\documentclass[iop,numberedappendix]{emulateapj}
\usepackage{chngcntr}
\usepackage{nicefrac}
\usepackage{sidecap}
\usepackage{multirow}
\usepackage{subfigure}
\usepackage{natbib}
\usepackage{mathtools}
\usepackage{MnSymbol}
\begin{document}

\received{receipt date} 
\revised{revision date} 
\accepted{acceptance date}
\slugcomment{draft version}

\title{Thermal and radiative AGN feedback have a limited impact on star formation \\ in high-redshift galaxies} 

\author{Orianne Roos, St\'ephanie Juneau, Fr\'ed\'eric Bournaud and  Jared M. Gabor} 
\affil{CEA-Saclay, 91190 Gif-sur-Yvette, France} 
\email{orianne.roos@cea.fr}

\shorttitle{Impact of AGN feedback on star formation} 
\shortauthors{Roos et al.}

\begin{abstract}
{The effects of Active Galactic Nuclei (AGNs) on their host-galaxies depend on the coupling between the injected energy and the interstellar medium (ISM)}. Here, we model and quantify the impact of long-range AGN ionizing radiation -- in addition to the often considered small-scale energy deposition -- on the physical state of the multi-phase ISM of the host-galaxy, and on its total Star Formation Rate (SFR). We formulate an AGN Spectral Energy Distribution {matched with observations}, which we use with the radiative transfer (RT) code Cloudy to compute {AGN} ionization in a simulated high-redshift disk galaxy. We use a high-resolution ($\sim6$~pc) simulation including standard thermal AGN feedback and calculate RT in post-processing. Surprisingly, while these models produce significant AGN-driven outflows, we find that AGN ionizing radiation and heating reduce the SFR by a few percent at most for a quasar luminosity ($L_{bol}=10^{46.5}$~erg~s$^{-1}$). {Although the circum-galactic gaseous halo can be kept almost entirely ionized by the AGN, most star-forming clouds ($n\gtrsim10^{2-3}$~cm$^{-3}$) and even the reservoirs of cool atomic gas ($n\sim0.3-10$~cm$^{-3}$) -- which are the sites of future star formation (100~-~200 Myrs), are generally too dense to be significantly affected}. {Our analysis ignores any absorption from a putative torus, making our results upper limits on the effects of ionizing radiation.} Therefore, while the AGN-driven outflows can remove substantial amounts of gas in the long term, the impact of AGN feedback on the star formation efficiency in the interstellar gas in high-redshift galaxies is marginal, even when long-range radiative effects are accounted for. 
\end{abstract}


\section{Introduction}
\label{Section:intro}
 
The impact of Active Galactic Nucleus (AGN) feedback on Star-Forming Galaxies (SFGs) remains an open question: contradictory answers are found in both simulations and observations. The amount of energy released by AGNs is theoretically high enough to blow all the gas out of their host-galaxies, or to maintain surrounding gas at high temperatures \citep{Croton2006,Ciotti1997,Matsuoka2012a}. {\citet{Curran2012} also show that there is always a finite ultra-violet (UV) luminosity above which all the gas in a radio galaxy or quasar host is ionized, up to redshift $\gtrsim3$}. This makes AGNs good candidates to quench star formation (SF) through outflows and ionization (``quasar mode"), or through jets (``maintenance" or ``radio" mode), which occurs in many simulations \citep[e.g.][]{Sijacki2007,Martizzi2012,Dubois2012,Dubois2013}.  Simulations also predict that galaxy mergers create starbursts and feed quasi-stellar objects (QSOs), which produce shock waves, expel the interstellar medium (ISM), and prevent it from falling back on the galaxy \citep[e.g.][]{DiMatteo2005,Hopkins2006,Li2007}.  However, mergers are rare and QSO phases are extreme events, and the gas expulsion depends on the coupling between the AGN and the ISM \citep[e.g.][]{DeBuhr2011}.  In contrast, other simulations show that AGN jets can trigger large-scale star formation by creating blast waves that compress the  gaseous clouds of the interstellar/intergalactic medium \citep{Gaibler2012,Dugan2014}.  
 
On the observational side, most X-ray selected AGNs up to redshift 3 are located in normal star-forming (main sequence) disk galaxies \citep{Mullaney2012} and normal SFGs frequently host an AGN up to redshift~$\sim 1~-~2$ \citep{Mullaney2012b,Juneau2013,Rosario2013c}, which suggests limited AGN impact on star formation, under the assumption that their star formation history is steady \citep{Elbaz2011}. 
Nonetheless, there is evidence for AGN quenching local elliptical galaxies \citep{Schawinski2007}, or suppressing star formation without necessarily quenching it \citep{Karouzos2014}. At higher AGN luminosities, molecular outflows have been observed in local quasars \citep{Feruglio2010} or Ultra-Luminous Infra-Red Galaxies (ULIRGs) \citep{Cicone2014,Veilleux2013}, however without necessarily affecting the Star Formation Rate (SFR) \citep{Spoon2013}. Finally, \citet{Keel2012} observed giant AGN-ionized clouds in low-redshift (mostly) interacting or merging galaxies, which could also impact the SFR of the hosts. On the other hand, there is observational evidence for AGNs triggering SF \citep[e.g.][]{Begelman1989,Graham1998,Klamer2004,Croft2006,Feain2007,Elbaz2009}, {and some studies show that the hosts of more powerful AGNs have higher nuclear SFR but similar global SFR as those of less powerful AGNs, up to intermediate redshift \citep{Diamond-Stanic2012,LaMassa2013,Esquej2014}. }
{Finally, \citet{Hickox2014} take into account the rapid variability of AGNs and find that, averaged over a period of $\sim100$~Myr, $every$ star-forming galaxy hosts at least one active episode and long-term black hole accretion rate (BHAR) is perfectly correlated with the SFR of the host, reproducing the observed weak correlation between the AGN luminosity and the global SFR.} A plausible explanation to this apparent discrepancy (both in simulations and observations) in the role of AGN feedback in star formation would be that AGN feedback works both ways, depending on the current accretion mode \citep{Zinn2013}, and on the timescale considered.

To determine if AGN feedback can reduce star formation, we need to know whether the gas expelled and/or ionized would have formed stars in the absence of AGN. In this paper, we will focus on the impact of thermal and radiative AGN feedback on the physical state of the gas, and therefore on its ability to form stars in a simulation representative of a high-redshift star-forming disk galaxy from \citet{Gabor2013b}.

AGN feedback is often implemented in simulations in the form of local (at the resolution scale) deposition of energy \citep[see][for a comparison of five models of AGN feedback]{Wurster2013a}. It has recently been shown that such AGN feedback creates outflows without impacting the disk \citep[e.g.][]{Gabor2014}. However, due to the lack of resolved small-scale observations of those mechanisms at high-redshift, the recipes remain quite arbitrary and do not have strong constraints on the scale at which energy should be re-injected.

  Until now, long-range effects of AGN radiation are rarely included because simulations cannot afford both high resolution and a complete treatment of the radiative transfer (RT), due to the cost in terms of computational time and memory.  For instance, {\citet{Vogelsberger2013,Vogelsberger2014c}, in their simulation Illustris, use the moving-mesh code AREPO to treat cosmological simulations including standard AGN and stellar feedback, plus a prescription for radiative electromagnetic AGN feedback.} Also, \citet{Rosdahl2013} implemented RT into RAMSES (RAMSES-RT). However, the resolution of the ISM structure of a galaxy is degraded, and effects of ionization on small-scale structures such as Giant Molecular Clouds (GMCs) cannot be considered. {The effect of long-range AGN radiation may however have a great impact on the structure of the ISM since most of AGN radiation is emitted in the optical, UV and X-ray wavelengths and is able to heat and/or ionize surrounding ISM, which could change the physical properties of the clouds \citep{Proga2014}. Furthermore, the fraction of ionizing photons emitted by the AGN that remains trapped in the ISM depends on the distribution of the gas into clumps \citep{Yajima2014}, and could in return change the properties of the ISM \citep{Maloney1999} and of the AGN-driven winds \citep{Dove2000}. Many authors already attempted to predict such effects with simple models (see Section \ref{subSection:simple_models}), but, due to the complexity of the ISM in a real galaxy and the broad wavelength range of observed AGN spectra, such models are not sufficient and RT calculations need to be performed.}
  
To do that, we post-process the results of our high-resolution simulation with a complete treatment of the RT and study the large-scale effect of AGN ionization on star formation. Therefore, we are not able to treat coupling to longer-term dynamical effects, but the ISM structure of the high-redshift disk galaxy allows us to probe which ISM phases can be impacted as a function of the AGN luminosity. {While it is known that very diffuse gas ($n\lesssim10^{-3}$~cm$^{-3}$ ; e.g. in the circum-galactic medium) can be entirely ionized by an AGN, and dense gas ($n\gtrsim 10^{2-3}$~cm$^{-3}$) is self-shielding \citep{Liu2013}, the impact of AGN radiation on other gas phases, such as atomic gas ($n\sim0.3-10$~cm$^{-3}$ ; which will form stars on a timescale of a few hundreds of Myrs), has not been studied extensively. Moreover, the effects of a clumpy distribution of gas inside a dense galactic disk on the propagation of AGN radiation are hard to determine with a simple model. } In Section~\ref{Section:method}, we describe the galaxy simulation and our method to predict the distribution of the gas photoionized by the AGN in the galaxy. Section~\ref{Section:results} shows maps of heated/ionized gas, and star-forming regions, total reduction of SFR and fractions of the total mass and volume of the gas that is heated/ionized by the AGN. Finally, in Section~\ref{Section:discussion} {we study the dependence of the SFR reduction on the structure of the ISM, give clues about long-term effects on star formation and the typical size and density of ionized regions. Finally, we try to account for other sources of ionization (e.g. stars, UV background). Our conclusions are in Section~\ref{Section:ccl}. \\

  \section{Method}
  \label{Section:method}

Using radiative tranfser calculations applied in post-processing to a high-resolution simulation of an isolated disk galaxy, we study the effects of photoionization by an AGN on the ISM.  Our procedure allows us to quantify the impact of AGN radiation on the SFR of well-resolved star-forming clouds in the galactic disk. In this section we describe the hydrodynamic galaxy simulation, the AGN SED used as the {only} ionizing source {(hypothesis discussed in Section \ref{subSection:transparent_lops})}, and the radiation transfer procedure.

\subsection{Sample of simulated galaxy snapshots}
\label{subSection:simu}

The simulation, described fully by \citet{Gabor2013b},  is an unstable clumpy disk representing an isolated $z\sim2$ disk galaxy with a well-resolved ISM \citep[see e.g.][for a comparison of such clumps with observations]{Elmegreen2008}. Such clumpy galaxies are predicted to efficiently fuel their central black hole (BH), resulting in frequent active phases \citep{Bournaud2011}, which is observationally supported up to intermediate redshift \citep{Bournaud2012}. The simulation is a cube of 50~kpc length using the adaptative mesh refinement (AMR) code RAMSES  \citep{Teyssier2002}. 

Stars, gas and dark matter are included. The total baryonic mass is $4\times10^{10}$ M$_\odot$, and the gas fraction is initially set to $\sim$~50~\%.  A cell is refined if its mass exceeds $6\times10^5$ M$_\odot$, if there are more than 30 dark matter particles in the cell, or  if the Jeans length is not resolved by at least 4 cell widths \citep{Truelove1997}. The maximal resolution of the simulation, taken as the size of the smallest cells, is about 6 pc. Thus, GMCs, where stars are supposed to form \citep{Waller1987}, are resolved. Star formation is allowed for cells denser than 100~m$_{H}$~cm$^{-3}$ (see Appendix~\ref{subSection:threshold} for discussion) and colder than $\sim10^4$~K \citep{Renaud2012}. {A thermal model for supernovae (SNe) feedback is included \citep[see][for details]{Gabor2013b}. The distribution of clouds in the simulation depends on the numerical noise and is therefore stochastic. Furthermore, star formation contains a random module for the mass distribution of new stars, which will affect the mass, shape, size and movements of the clumps.  Nevertheless, the simulations used all have the same statistical behaviour since they have the same probability density function (PDF) and power spectrum density (PSD) for gas density.}

AGN feedback is twofold: gas is heated, diluted and pushed away by the AGN directly in the simulation (thermal AGN feedback) ; and RT is added afterwards in 6 successive snapshots of the simulation (AGN photoionization), using version 13.02 of Cloudy \citep[last described by][]{Ferland2013} to study the large-scale effect of AGN ionization on star formation.  Thermal AGN feedback is based on \citet{Booth2009} and consists of local (10 pc-scale) energy re-deposition at a uniform pressure if it is sufficient to heat the gas to an average temperature of $10^7$ K, else energy is stored until the next time-step. Corrections to this method were added to prevent too high energy storage when the BH is embedded in very dense clumps. {The maximum temperature allowed in the region where energy is re-deposited by the AGN is $5\times10^9$ K. In the very infrequent case when this maximum temperature is reached, the radius of the re-deposition region is enlarged by a factor 1.25 until the temperature drops below the maximum temperature. When this happens, the radius is multiplied by a factor $\lesssim$ 2 on average. These prescriptions only apply to the very central region (10 pc-scale) of the galaxy and do not directly impact the large-scale gas structure and star formation in the intermediate and outer regions of the disk \citep[see Section 2.2.2 of][for details]{Gabor2013b}. }

The simulation shows high-velocity AGN-driven outflows, with mass outflow rates between about 10 and 100~\% of the SFR of the galaxy ($\sim$ 30 M$_\odot$ yr$^{-1}$). These outflows are mostly hot and diffuse, and do not impact large scale ($>100$~pc) star formation within short (10~-~20~Myrs) time-scales \citep{Gabor2014}.

We used two runs of the simulation with the same initial conditions, which develop a clumpy irregular ISM structure. The runs are identical until AGN feedback is shut down in one of them. This moment is defined as $t=0$. After that, both runs evolved separately. {SNe feedback remains in both runs.}  A series of 6 pairs of successive snapshots was studied, ranging from $t = 0$ to 88 Myrs. {Among all the snapshots studied, the maximum number density of hydrogen is $\sim10^{6}$~cm$^{-3}$, due to both the spatial resolution and the temperature floor.} Ionization was added to the snapshots of the simulation including AGN feedback, and the corresponding snapshots of the simulation with no {AGN} feedback were used to measure an uncertainty on the SFR. The ionization calculation is done on static snapshots of the simulation, and therefore the effect we see is instantaneous and indicates in which phases of the ISM the AGN radiation is absorbed, and whether it reaches high-density star-forming regions.\\

\subsection{Realistic Seyfert SED}
\label{subSection:sed}

  \begin{figure}[htp]
\includegraphics[trim=2.4cm 0cm 0cm 0cm, clip=true,width=\linewidth]{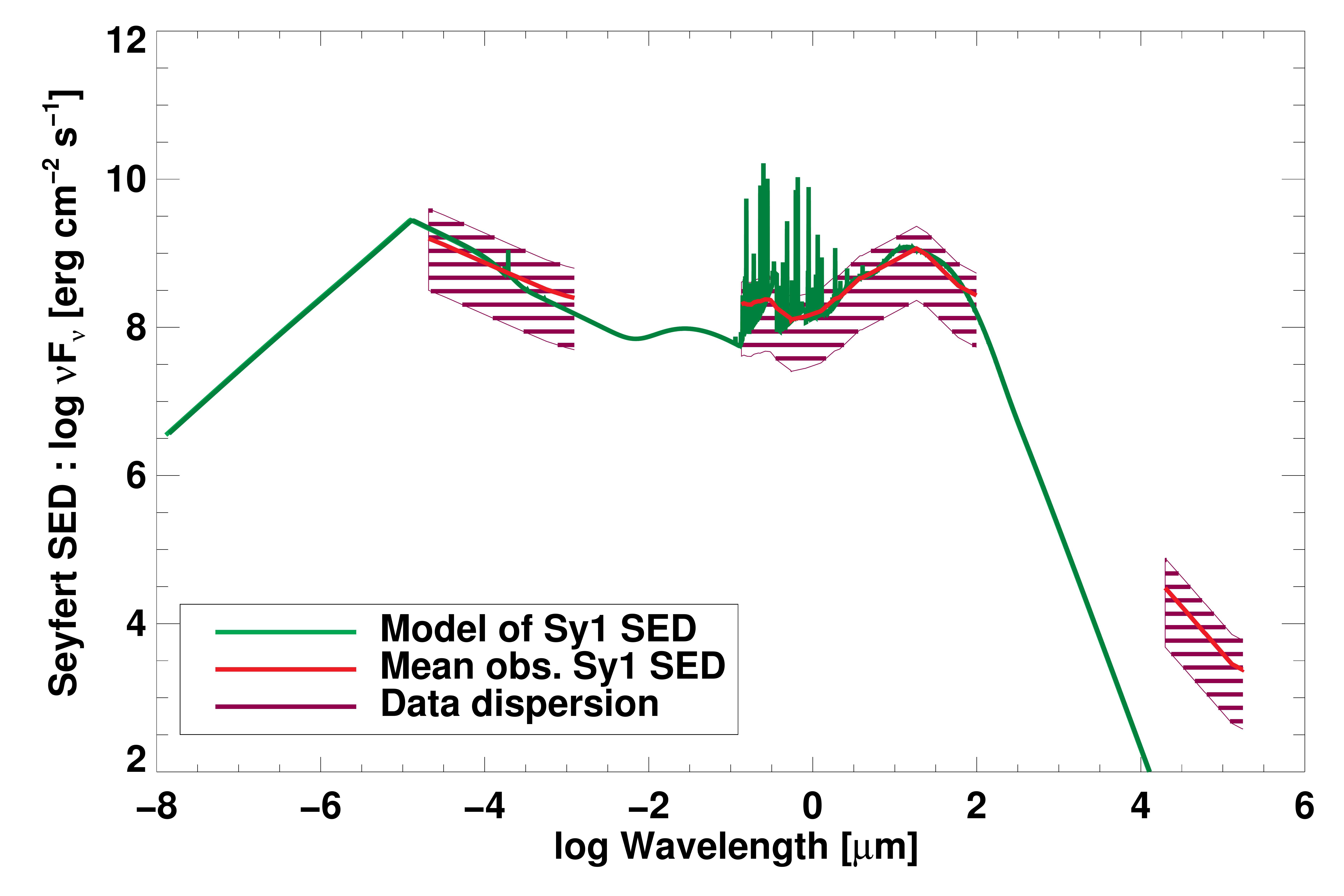} 
\caption{AGN SED model of the emission arising from the vicinity of a Seyfert 1 nucleus (i.e. AGN, dust and BLR clumps; in green). The red curve is the mean observational Seyfert 1 SED taken from \citet{Prieto2010}, and the dark hatched area is the data dispersion.\\} 
\label{fig:agn_sed}
\end{figure}

To study the effects of AGN radiation on the surrounding gas, we first require a realistic source spectral energy distribution (SED).  {In this work, we span a large range of wavelengths (from far infra-red (FIR) to X-rays), and match observational data in the wavelength domains of interest for this study. Hence, several bands of ionizing photons are taken into account (X-rays, extreme UV (EUV), UV), rather than a simple power-law model.} We formulated such an AGN SED for use in Cloudy, as shown in Figure~\ref{fig:agn_sed}.  {The so-called ``AGN spectrum'' used in the study is composed of radiation from the AGN itself (accretion disk and corona) but also takes into account the emission of dust and clumps surrounding the AGN, which are not resolved in our simulation.  The input spectrum is described in more detail in Appendix \ref{appendix:spectra}}. It was adjusted to the observational mean SED of the inner region of Seyfert 1 galaxies described by \citet{Prieto2010}.  

We chose to model the AGN spectrum of an unobscured (Type 1) Seyfert galaxy, i.e. the spectrum coming out of the central region seen face-on. The AGN radiation is propagated isotropically {-- we neglect the partial absorption of UV and X-ray photons by a dusty torus that would occur according to the Unified Model \citep{Urry1995}.  We make this assumption because the geometry and orientation of any such torus is unknown, and there are currently not enough resolved observations to be confident about them  \citep{Sales2014, RicciC2014,RicciTV2014b}. Consequently, our results are upper limits to the instantaneous impact of AGN radiation on star-formation in high-z disk galaxies -- including an obscuring torus would only decrease the amount of ionizing radiation emitted into the ISM.}

We note here that the FIR part of the spectrum does not match the observations. Nonetheless, those photons are not of a great interest in this study because they cannot ionize gas. We also verified that the presence of emission lines in the Broad-Line Region (BLR) spectrum did not affect our further results. 

\begin{deluxetable}{ccccc}
\tablecolumns{5} 
\tablewidth{\linewidth}
\tablecaption{Properties of AGN luminosity regimes\label{table.regimes}}
\tablehead{\colhead{Regime} &\colhead{Luminosity} & \colhead{Frequency}& \colhead{BHAR/SFR} & \colhead{$r_{ion}$}}

\startdata
Typical AGN & $10^{44.5}$ & $\sim30$ \% & $1.7\times10^{-3}$& 20~-~60\\
Strong AGN & $10^{45.5}$ & $\sim3$ \% & $ 1.7\times10^{-2}$&50~-~120\\
Typical QSO & $10^{46.5}$ & Rare & $1.7\times10^{-1}$&100~-~260
\enddata
\tablecomments{Bolometric luminosity in erg~s$^{-1}$. Values are given for $10^{10}~<~M_*~<~10^{11}$ M$_\odot$ typical SFGs. BH accretion rates (BHARs) are derived from the formula given by \citet{Mullaney2012b}. $r_{ion}$ is the predicted Str\"{o}mgren radius in parsecs and was derived from the calculation used by \citet{Curran2012} for gas at $10^3$~cm$^{-3}$ and 20 or 2,000 K respectively.}
\end{deluxetable}

{The normalization of the SED we used to perform the analysis is set to explore three luminosity regimes, as shown in Table~\ref{table.regimes}}. The bolometric luminosities are not related to the AGN luminosity inferred from BHAR in the simulation and we analyze the effects of each luminosity separately. The lowest luminosity regime corresponds to the typical bolometric luminosity observed for an AGN in a $10^{10}<M_*<10^{11}$ M$_\odot$ normal SFG, which is of a few $10^{44}$~erg~s$^{-1}$ at a redshift of $\sim2$ \citep{Mullaney2012b}. Such AGNs are hosted by roughly 30~\% of the standard main-sequence galaxies in the same mass range \citep{Mullaney2012,Juneau2013}. The second luminosity we used is reached by $\sim$~10~\% of all AGNs \citep{Mullaney2012}, which are hosted by $\sim$~3~\% of the SFGs. The last luminosity we studied corresponds to a QSO, which is quite uncommon in normal star-forming disks.

 {\citet{Mullaney2012b} show that typical AGNs up to redshift $\sim$ 2 have an average BHAR to SFR ratio of $\lesssim 10^{-3}$. Considering the same definition of the BHAR:
\begin{equation}
\dot{M}_{BH}=(1-\epsilon)\frac{L_{bol}}{\epsilon c^2},
\end{equation}•
where $L_{bol}$ is the bolometric luminosity of the AGN, $c$ is the speed of light and $\epsilon$ is the radiative conversion efficiency, set to 0.1 \citep{Merloni2004,Marconi2004}; and an average SFR of 30 $M_\odot$~yr$^{-1}$ for the simulation, we find that our lower-luminosity regime has a BHAR to SFR ratio roughly corresponding to the observed average. The middle and higher-luminosity regimes are respectively 10 and 100 times above (see Table~\ref{table.regimes}), which is consistent with them being less frequent.}

As QSOs and lower-luminosity AGNs do not have the same SED shape, a radio-loud quasar-matched spectral energy distribution was also formulated, using the \citet{Elvis1994} mean SED as a reference. 
The results are insensitive to the choice of the SED, and therefore we used the Seyfert spectrum in all luminosity regimes. We note here that even more luminous quasars have been observed \citep[e.g.][]{Stern2014} but since they are even rarer than the QSOs we studied, they might not impact  normal star-forming disk galaxies in general.

\subsection{Expectations from simple models}
\label{subSection:simple_models}

{Several simple models have been introduced by many authors to infer the effects of AGN radiation on their host galaxies. Some of them are addressed here. For instance,  according to \citet{Proga2014}, in optically thin clouds dominated by absorption opacity, radiation propagating through gas at constant density and pressure will uniformly heat the cloud, which will uniformly expand but will not be accelerated ; whereas for optically thin clouds dominated by scattering opacity, the radiation will uniformly accelerate the cloud away from the emitting source, without changing its size or shape, and inducing no mass loss. On the other hand, when a cloud is optically thick and is exposed to weak radiation, only a thin layer on the irradiated part of the cloud will be heated, inducing a slow mass loss. They find that BLR clouds -- which are very dense \citep[$\sim10^9$~cm$^{-3}$ ;][]{Mattews1985}, typically become optically thin in less than a sound-crossing time, are weakly accelerated, and that their structure, shape, and size change before they can travel a significant distance. A realistic ISM made of dense clumps and diffuse interclump medium would be a complex combination of the above trivial cases and RT needs to be treated in order to know the impact of AGN irradiation on such a realistic multi-phase ISM.}

{As another example, \citet{Curran2012} have shown that a UV luminosity of $L(1216\textrm{\AA})\sim 10^{23} $~W~Hz$^{-1} = 10^{30}$~erg~s$^{-1}$~Hz$^{-1}$ is able to ionize the cold neutral medium (CNM) with gas densities typical of GMCs ($n\sim10^3$ cm$^{-3}$), up to $\sim 50-100$~pc in the galactic disk, around the BH. For our three AGN luminosity regimes, we find $L(1216\textrm{\AA})~\sim~10^{29}, 10^{30}\textrm{, and }10^{31}$~erg~s$^{-1}$~Hz$^{-1}$ respectively. If we were using a constant density of $10^3$~cm$^{-3}$, we would thus expect the ionization front to be located at sub-kpc scale around the BH for gas temperatures between 20 and 2,000 K (see Table~\ref{table.regimes}). }

{Finally, a simple calculation of the optical depth in the UV gives an even lower value of this radius:  if the UV opacity is $\sim10^3$ cm$^2$ g$^{-1}$ \citep{Cayatte1994,Schaye2001}, the critical value of the column density above which the gas becomes self-shielding is 10$^{-3}$ g cm$^{-2}$. In this case, if the inner region close to the BH is at a uniform density of $10^3$ cm$^{-3}$, the UV emission is expected to be blocked within the first $\sim0.2$ pc.}
 
{However, these models are valid in idealized conditions such as smoothed galactic profiles or single-band AGN SEDs. In this study, we demonstrate the effect of a broad observationally-matched AGN spectrum, coupled to AGN-driven winds, on a realistic multi-phase distribution of the ISM. Applied to a disk galaxy with turbulent multi-phase ISM, these predictions {of the ionization radius} can vary by an order of magnitude, since most of the gas volume in our high-resolution clumpy ISM has a density low enough not to be opaque to AGN radiation, while a few dense clumps are able to completely block the radiation -- which can significantly change the escape fraction of photons emitted by the AGN. }

{Nonetheless, to account for dynamical effects such as mass loss or radiative pressure pushing clouds away, RT calculations need to be performed during the simulation, which is not the purpose of our post-processing treatment. However, the comparison between complete radiative treatment in post-processing and simplified but dynamical RT could help improve both methods, by developping accurate subgrid models.}

\subsection{Ionizing the simulated galaxy}
\label{subSection:process}

We estimate the effects of {AGN} photoionization on the gas in the simulated galaxy under the assumption that our AGN SED emerges isotropically from the location of the black hole.  We cast about 3,000 lines-of-propagation (LOPs) in all directions outward from the black hole, and calculate the radiative transfer along each LOP independently using  Cloudy{, a code designed to compute the radiative transfer and the atomic and molecular chemistry along 1-D lines. }

  \subsubsection{LOP building}
  
    \begin{figure}[htp] 
\includegraphics[width=\linewidth]{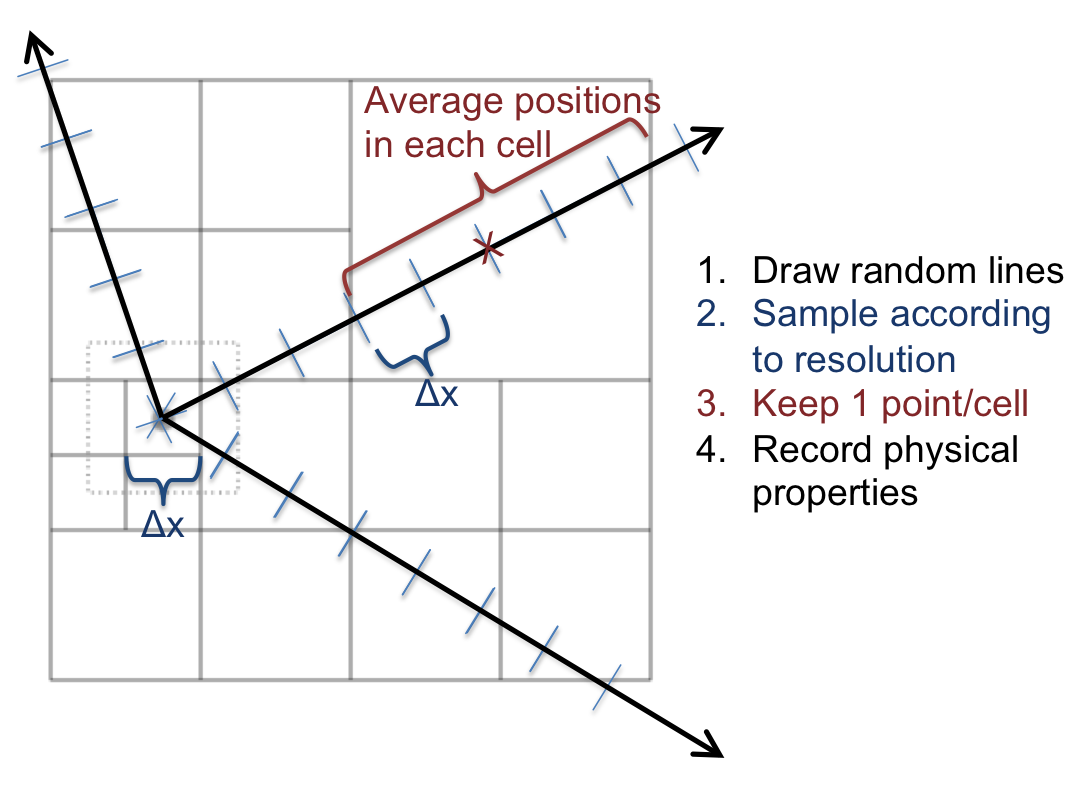} 
\caption{Sketch illustrating the procedure to create the LOPs in the AMR box. First, random lines are cast in the simulation box, from the location of the BH. They are then sampled with points separated by the size of the smallest cells. The positions of all points inside a given cell are averaged in order to keep only one point per cell and per LOP. For each point, relevant physical properties (gas density, temperature, etc.) are recorded. Based on the sketch in FLASH Users' Guide (version 4), from J.B. Gallagher.\\} 

\label{fig:sketch_sampling}
\end{figure}

  {The first step of our analysis is to build the LOPs, as illustrated by the sketch in Figure \ref{fig:sketch_sampling}. For each snapshot, LOPs are distributed as follows. The simulation box is sampled with 512 LOPs randomly cast into the entire box, plus 512 randomly cast LOPs restrained into the 30 degree half-opening angle cone of revolution perpendicular to the disk. In addition to that, the plane of the galactic disk is sampled  with 512 randomly cast LOPs (see Figure~\ref{fig:cell_lops}a, right). Finally, two arbitrary planes of the simulation box, perpendicular to the galactic disk, are sampled with 768 LOPs each (see Figure~\ref{fig:cell_lops}a, left, for one of these planes with its 768 LOPs). A third of these lines are cast randomly into the entire plane, another third in the projection of the cone on the plane, and the remaining into the galactic disk. All points of a given LOP are initially separated by the size of the smallest cells (see Figure~\ref{fig:sketch_sampling}). As shown on the sketch, the positions of all points inside a given cell are then averaged so that only one LOP point is kept per cell and per LOP. For each LOP point, all relevant physical properties (gas density, temperature, etc.) are recorded. {The hydrogen number density (hereafter, density) of the cell and its distance to the black hole are used to build the density profile along each LOP (see Figure~\ref{fig:lop_example} for two examples).}}

\begin{figure*} 
\subfigure[][]{%
\begin{minipage} [b]{.385\linewidth}
\includegraphics[width=\linewidth]{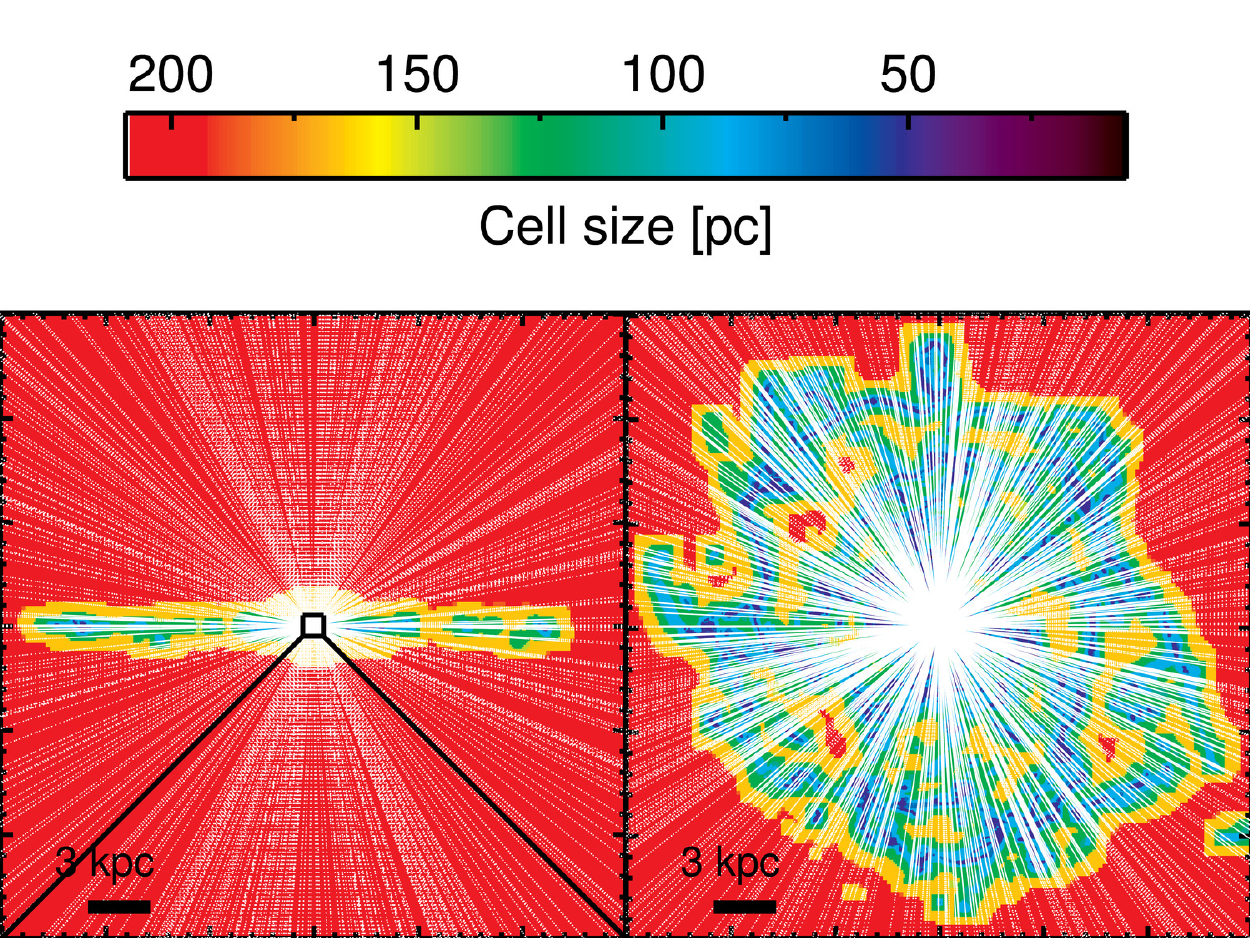}
\includegraphics[trim=0cm 0cm 0cm 3.1cm, clip=true,width=\linewidth]{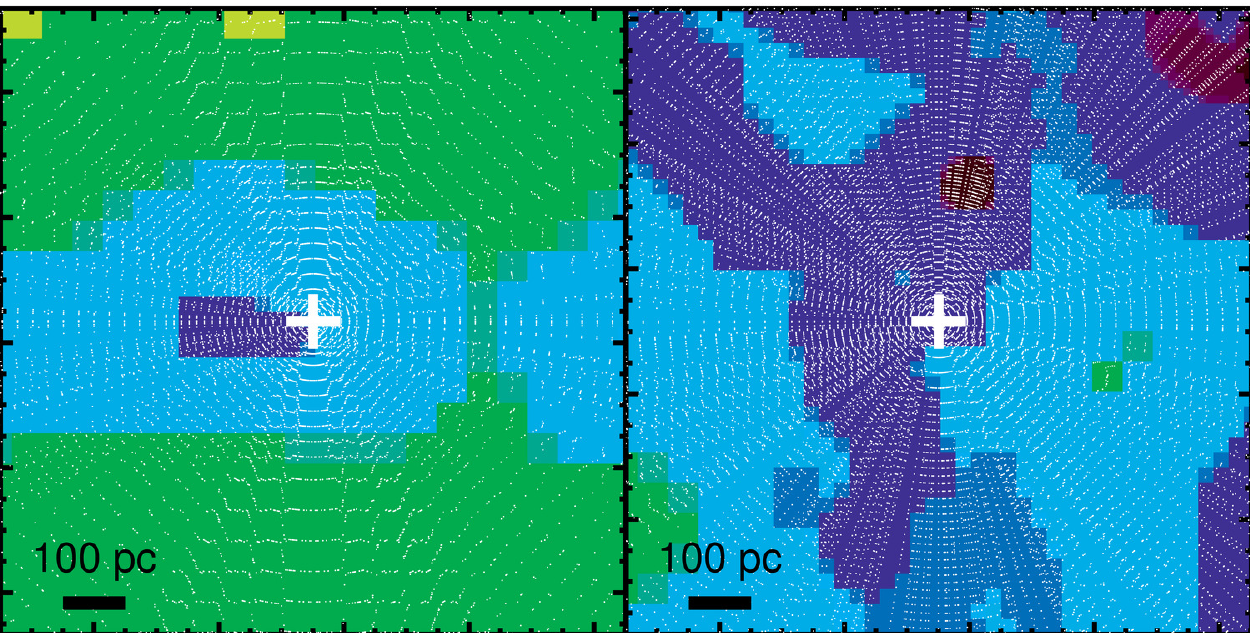}
\end{minipage}%
}%
\subfigure[][]{%
 \begin{minipage} [b]{.1\linewidth}
  \includegraphics[trim=9.4cm 7.2cm 1.65cm 0cm, clip=true,width=\linewidth]{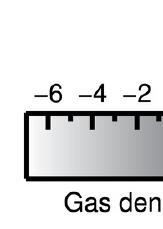}
 \includegraphics[trim=9.4cm 0cm 0cm 3.1cm, clip=true,width=\linewidth]{plot_lops_den_00100_all_z=2460right_light-eps-converted-to.pdf}
\includegraphics[trim=9.4cm 0cm 0cm 3.1cm, clip=true,width=\linewidth]{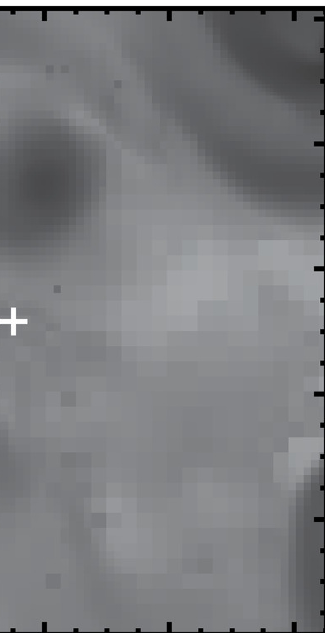}
\end{minipage}%
\hspace{-0.2cm}
    \begin{minipage} [b]{.1\linewidth}
      \includegraphics[trim=11.05cm 7.2cm 0cm 0cm, clip=true,width=\linewidth]{plot_lops_den_00100_all_z=2460right_light-eps-converted-to.pdf}
 \includegraphics[trim=0cm 0cm 9.4cm 3.1cm, clip=true,width=\linewidth]{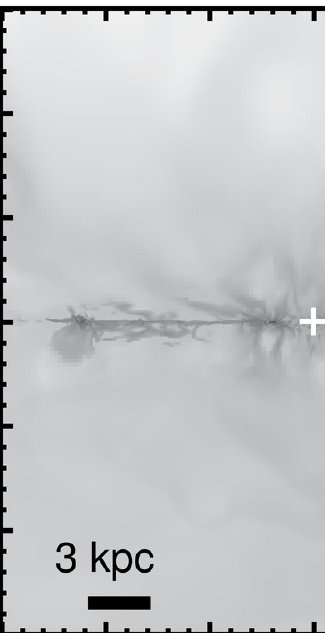}
\includegraphics[trim=0cm 0cm 9.4cm 3.1cm, clip=true,width=\linewidth]{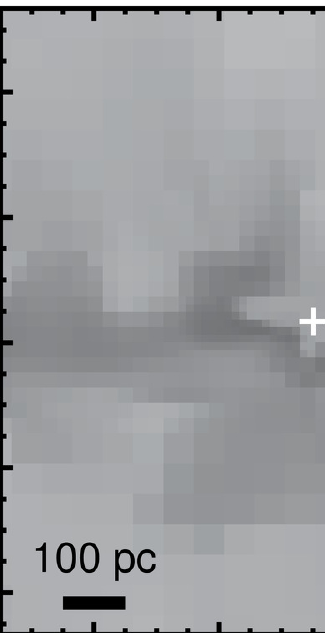}
\end{minipage}%
}%
\subfigure[][]{%
\begin{minipage}[b]{.385\linewidth}
\includegraphics[width=\linewidth]{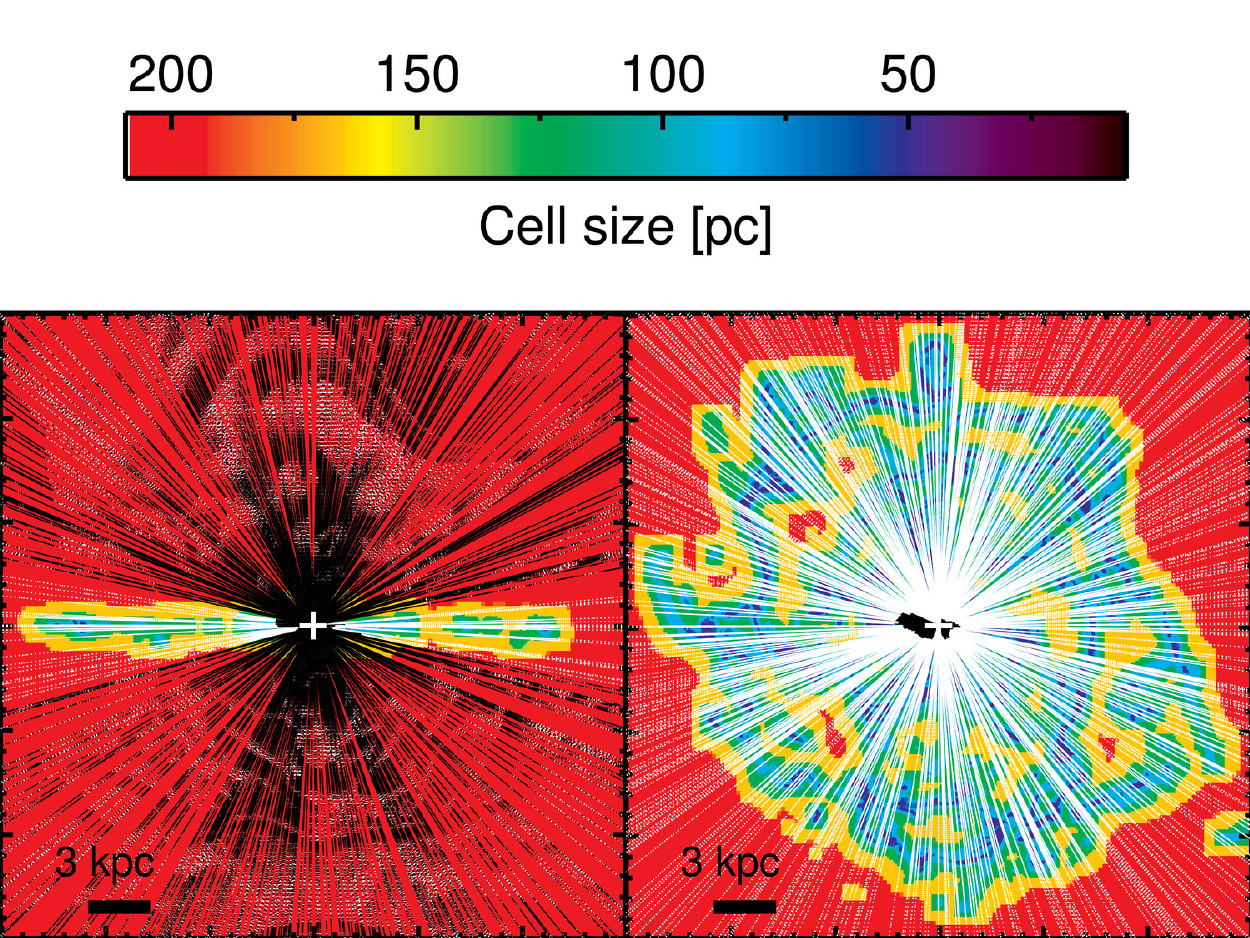}
\includegraphics[trim= 0cm 0cm 0cm 3.1cm,clip=true,width=\linewidth]{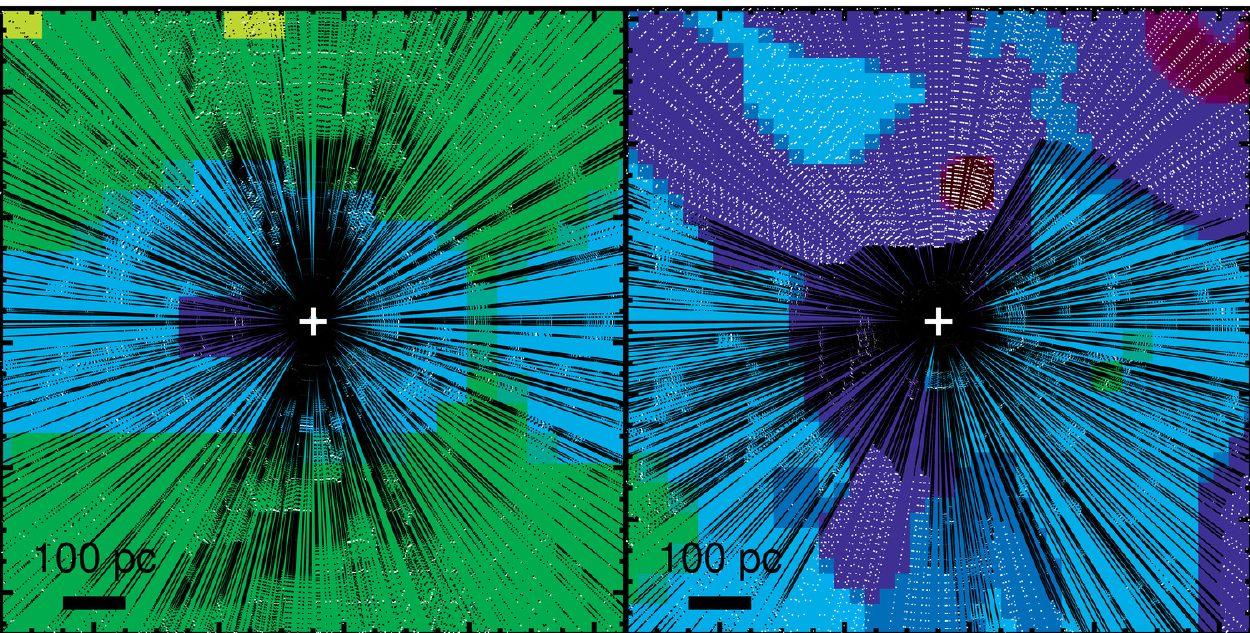}
\end{minipage}%
}%
\caption{{(a)} Large (\textit{top row}) and zoomed-in (\textit{bottom row}) edge-on (\textit{left column}) and face-on (\textit{right column}) views of the simulated galaxy showing the size of the cells. The points (\textit{white dots}) of the LOPs located in each plane are superimposed. The sampling is that before propagating the AGN radiation with Cloudy. The white cross shows the location of the BH. Thanks to the mass refinement criterion, substructures are well sampled (see Figure~\ref{fig:den_cells}b). \label{fig:cell_lops} \\ {(b)} Comparison with the mass-weighted density maps. The views and zooms are the same, but only half of the maps is displayed, for comparison with the maps located on the cropped side. Small AMR cells trace substructures (see Section \ref{subSection:simu}). \label{fig:den_cells} \\ {(c)} Same key as Figure~\ref{fig:cell_lops}a. During the computation, Cloudy resamples the LOPs according to LTE zones. The original AMR LOP is sampled with lower density where the physical conditions are alike, whereas it is sampled with higher density where AGN radiation induces significant and rapid changes. The computation stops when the temperature drops under 4,000 K (roughly where there are no more ionizing photons left), or at the end of the line. The Cloudy sampling (\textit{black}) is superimposed to the original one (\textit{white}). For the LOP regions located past the point where Cloudy stopped, the original sampling is used. The Cloudy sampling depends on the AGN luminosity. Here, we show that of the typical AGN regime ($L_{bol}=10^{44.5}$~erg~s$^{-1}$). \label{fig:cell_cloudy}} 
\end{figure*}

    \begin{figure}[htp] 
\includegraphics[width=\linewidth]{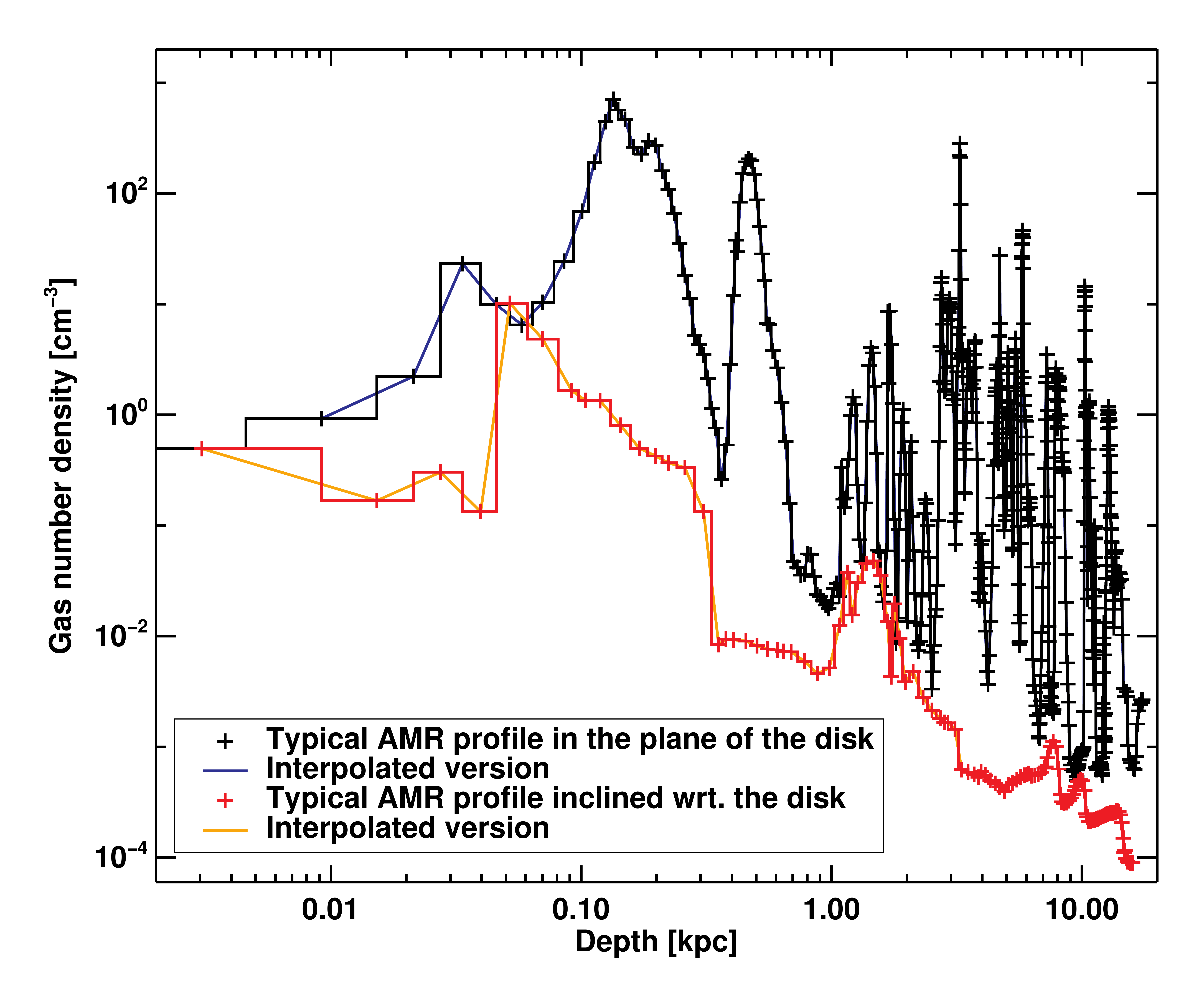} 
\caption{Example of a typical density profile in the plane of the disk (\textit{black and dark blue}) and outside the plane of the disk (\textit{red and orange}). The histogram lines \textit{(black and red)} show the discrete AMR profiles. The lighter \textit{(dark blue and orange)} curves are the interpolated profiles used in Cloudy. Typical profiles show high density contrasts between the clumps and the interclump medium.\\} 

\label{fig:lop_example}
\end{figure}
  
  \subsubsection{Cloudy computations}

{Before computing the RT, Cloudy interpolates the discrete density profile coming from the AMR simulation to get a continuous profile (see Figure~\ref{fig:lop_example}). By comparing the size of the cells (Figure~\ref{fig:cell_lops}a) to their density (Figure~\ref{fig:den_cells}b) and from the typical LOP profiles (Figure~\ref{fig:lop_example}), we know that substructures, namely GMCs, are well sampled along the lines.}

The Cloudy computation occurs on a static snapshot and the density profile does not change during the radiative transfer computation. The calculation stops either when the computed temperature along the line drops under 4,000~K, meaning that the rest of the line is not affected by the ionizing source ; or at the end of the line, in which case the line is totally ionized. 

{When calculating the one-dimensional RT along a single LOP, Cloudy treats the system as a spherically symmetric set of concentric shells (or zones), centered on the ionization source. These zones are dynamically determined with a local thermal equilibrium (LTE) criterion. Thus, the sampling of each line changes during the process, and depends on the AGN luminosity (see Figure~\ref{fig:cell_cloudy}c for the sampling at the end of the process for the typical AGN regime): the LOPs are sampled with higher density in regions where AGN radiation induces significant changes, while they are sampled with lower density where the physical properties are similar. The LOP regions where the Cloudy calculation did not occur (because the equilibrium temperature is below 4,000~K) keep the initial sampling.}

  {In each zone, ionization processes (photo-ionization, collisions, charge transfer, etc.) and recombination processes (radiative and charge transfers, etc.) are balanced to compute the propagation of radiation along a given LOP, according to the input hydrogen number density profile, {filling factor} and radiation source, and gives the resulting SEDs and physical conditions \citep{Ferland2013}. The filling factor is set to 0.2 and accounts for the multi-phase gas distribution along the 2 other spatial dimensions in the sphere. In each sphere, the flux of the ionizing source is proportional to the inverse square of the distance to the source. To rebuild the galaxy, all computations are assembled using only the radial dependence, and interpolated. This pseudo-3D ray-tracing approach is similar to that used in Cloudy\_3D \citep{Morisset2006}, now re-written in Python and renamed pyCloudy \citep{pyCloudy}. The reflection and scattering between two neighbouring LOPs are not considered explicitly. We assume that those phenomena are reproduced on average for the whole galaxy, since reflection and scattering are taken into account inside each individual sphere.}

 Lastly, the initial temperature along the lines is not used for the calculation, and the gas is considered initially neutral. {We do not include the SNe and the UV background in the ionization process. However, they are implemented in the simulation, and from the value of the initial temperature in the cells, we know that they heat the halo and some regions in the disk: gas at T $>10^6$~K is likely to be ionized and some of the $10^{4-5}$~K gas is possibly ionized. Such regions could then be transparent to AGN radiation.  A test study on a series of LOPs shows that the transparency of such diffuse gas that could be already ionized by other sources has no impact on the effect of AGN radiation on star formation (see Section~\ref{subSection:transparent_lops}). Thus, transparent gas is conservatively neglected in the main study. Considering that the gas is initially neutral allows us to study the effect of AGN photo-ionization alone and confront our model of thermal feedback to the ability of the AGN to ionize gas. Here, as the AGN is able to ionize the cavities created by winds entirely, we assume that our model of AGN feedback (thermal + RT) is self-consistent.}
 
\subsubsection{Output parameters for a given cell}

\paragraph{Temperature}
The final temperature of a given cell containing at least one Cloudy point is defined as the maximum between the average equilibrium temperature of the Cloudy points in the cell (corresponding to the temperature gas would reach if it was not externally re-heated), and the initial temperature of the cell\footnote{If the RT calculation returns a lower temperature, it means that a mechanism other than AGN photoionization (e.g.  UV background, SNe, compressive motions, etc.) is already heating the gas at a temperature higher than what the AGN ionization can provide and so these other sources will dominate.}. For the points where the computation did not occur (because the temperature on the line dropped under 4,000 K), the assumed temperature is the initial temperature. The latter is not necessarily equal to the equilibrium temperature given by Cloudy because gas is constantly heated by thermal AGN and stellar feedback in the simulation. {The {instantaneous relative temperature change} is defined as:
\begin{equation}
\textrm{RTC} = \frac{T_{final}-T_{initial}}{T_{initial}},
\end{equation}
where $T_{initial}$ is the initial temperature of the cell in the simulation (before RT), and $T_{final}$ is the final temperature of the cell defined above. Regions with $RTC<1$ \% (or $T_{initial} \leq T_{final}< 1.01\times T_{initial}$) are considered not heated.}

 \paragraph{Star Formation Rate} 
 A cell is star-forming and has a non-zero value of the SFR if its temperature is below $10^4$~K and its density is above a threshold of 10 m$_H$ cm$^{-3}$ (hereafter, m$_H$ will be implicitly assumed). This value distinguishes diffuse gas and dense star-forming clumps and is chosen arbitrarily and independently of the density threshold in the simulation. It is varied in Appendix~\ref{subSection:threshold}. The SFR in each cell $i$ is computed according to the Schmidt-Kennicutt law \citep{Kennicutt1998} :
  \begin{equation}
  \textrm{SFR}_i =  \epsilon ~\sqrt{\frac{32\textrm{G}}{3\pi}} ~\rho_i^{\textrm{N}} V_i~\textrm{if~} T_i<T_{\textrm{thr}} \textrm{ and } \rho_i>\rho_{\textrm{thr}},
  \end{equation}
  where $\rho_i$ is the mass density of cell $i$, $V_i$ its volume and $T_i$ its temperature before or after RT. The Kennicutt index N is equal to 1.5 and the efficiency $\epsilon$ is 1 \%. $T_{thr}$ and $\rho_{thr}$ are respectively the temperature and density thresholds for star formation defined above. The SFR per cell is limited to a maximum value, computed to account for the fact that molecular clouds generally do not turn more than 30 \% of their gas into stars during their $\sim$ 10~-~100~Myrs lifetime \citep{Matzner2000,Elmegreen2002,Renaud2012}. 
  The final SFR of the simulated galaxy is the sum of the individual values -- interpolated as described below,  and is compared: (1) to the initial SFR of the simulation including AGN feedback, (2) to the SFR of the reference simulation, where AGN feedback is shut down. Option~(1) allows to measure the impact of ionization feedback alone, while option (2) allows to measure the impact of all types of AGN feedback on the SFR.
  
{   \paragraph{Neutral fraction of hydrogen}
 As for the temperature, the fraction of neutral hydrogen in a given cell is defined by averaging the values of all LOP points inside the cell. LOP regions where RT did not occur are assumed to be neutral.
 A cell is considered ionized if the neutral fraction of hydrogen is smaller than 10~\%. The fraction of neutral hydrogen, unlike temperature, does not take into account the other heating/ionizing sources (such as stars).}

\subsubsection{Interpolation onto the simulation box}
About 3,000 lines-of-propagation were cast in the simulation box. Even though the box is relatively well covered with LOPs, and due to the Cloudy resampling in LTE zones,  all AMR cells in the snapshots do not necessarily contain a LOP point. Thus, values of the physical properties output by Cloudy (temperature, ionization fraction, etc.) need to be interpolated in order to apply to the whole simulation box.

  We use the fact that the Jeans length has to be resolved by at least 4 neighbouring cells in the simulation (see Section~\ref{subSection:simu}) -- and therefore their properties are alike -- and assume all cells containing no LOP point neighbouring a processed cell\footnote{A cell containing either one (or more) LOP point where the Cloudy computation occured, or one point from LOP regions where the Cloudy computation stopped before the end (because there was no further effect of the ionizing radiation). Two percent of the cells (in mass and in number) contain at least 1 Cloudy point. } within a radius of 4 times the cell size have the same output parameters. With this method, 20~\% of the total number of cells for each snapshot -- equivalent to 20~\% of the total gas mass in the simulation box (galaxy and gaseous halo) -- have known post-RT physical properties. {We call these cells ``4-neighbouring cells''}.
  
  To get the values of the physical properties for the entire grid of the simulation, the properties of the ``4-neighbouring cells''  are corrected with respect to the joint histogram of density and cell size. We assume that cells at a given density and a given radius in the galactic disk or gaseous halo are similarly affected by the AGN. {Due to the geometry of the simulated galaxy (a disk with a vertical exponential density profile in a diffuse gaseous halo) and the mass criterion of the refinement, the size of the cells (see Figure~\ref{fig:cell_lops}a) increases when going away from the galactic disk. This naturally ensures a good sampling of the outer gaseous halo, since fewer LOPs per unit angle are needed to cross larger cells.} In a few cases, some density bins are not sampled by the ``4-neighbouring cells" criterion, which is enlarged to 8 neighbouring cells. \\

\section{Results}
\label{Section:results}

The aim of this work is to switch on the AGN located at the center\footnote{As the central BH of the galaxy slightly moves during the simulation \citep{Gabor2013b}, we use the exact position of the BH particle of the simulation, not the geometrical center of the simulation box.} of the simulated galaxy, probe how far gas is ionized {by the AGN} and whether many star-forming regions are heated and/or ionized and prevented from forming stars. In this section, we compare the ionization, temperature, and SFR of the gas before our RT calculations to those after the RT calculations for the three AGN luminosity regimes presented in Section~\ref{subSection:sed}.\\

\subsection{Maps of the ionized galaxy}
\label{subSection:maps}

\begin{figure*}
\centering
\includegraphics[trim=1.0cm 0cm 3.1cm 0cm, clip=true,width=0.85\textwidth]{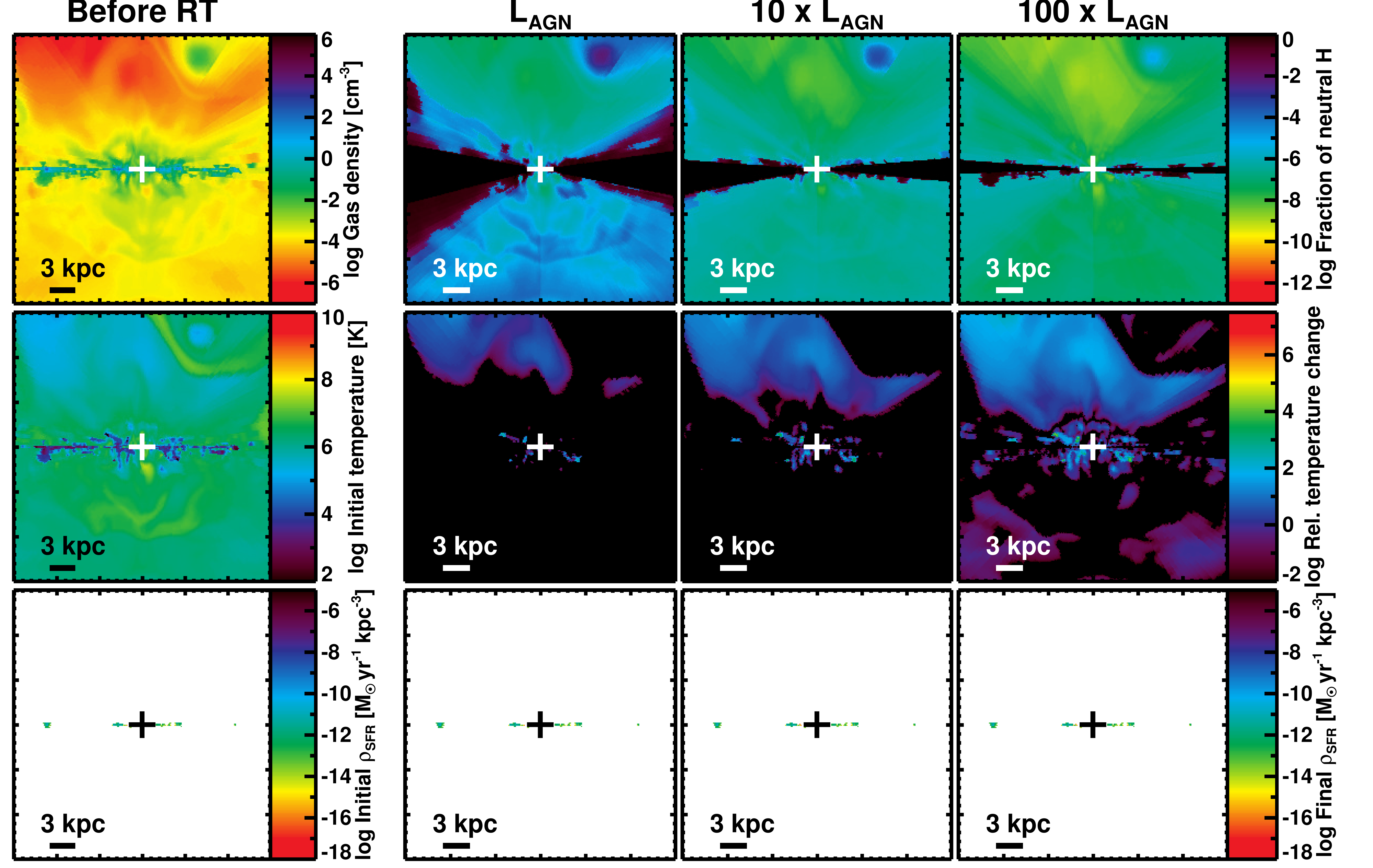}
\caption{Large edge-on view of the simulated galaxy (snapshot \#2). \textit{Top row}: Hydrogen density, Fraction of neutral hydrogen after RT. \textit{Middle row}: Temperature before RT, Relative temperature change. \textit{Bottom row}: $\rho_{SFR}$ before RT,  $\rho_{SFR}$ after RT. The `+' sign shows the location of the BH and the density threshold for SF is 10 cm$^{-3}$. Parameters after RT are given for the three AGN luminosities (L$_{\textrm{AGN}}=10^{44.5}$~erg~s$^{-1}$). AGN ionization and heating are more visible in the halo and in diffuse regions surounding the galactic disk, as AGN luminosity increases.} 
\label{fig:den_temp_sfr_large_edge-on_10Hcc_00100}
\end{figure*}
\begin{figure*}
\centering
\includegraphics[trim=1.0cm 0cm 3.1cm 0cm, clip=true,width=0.85\textwidth]{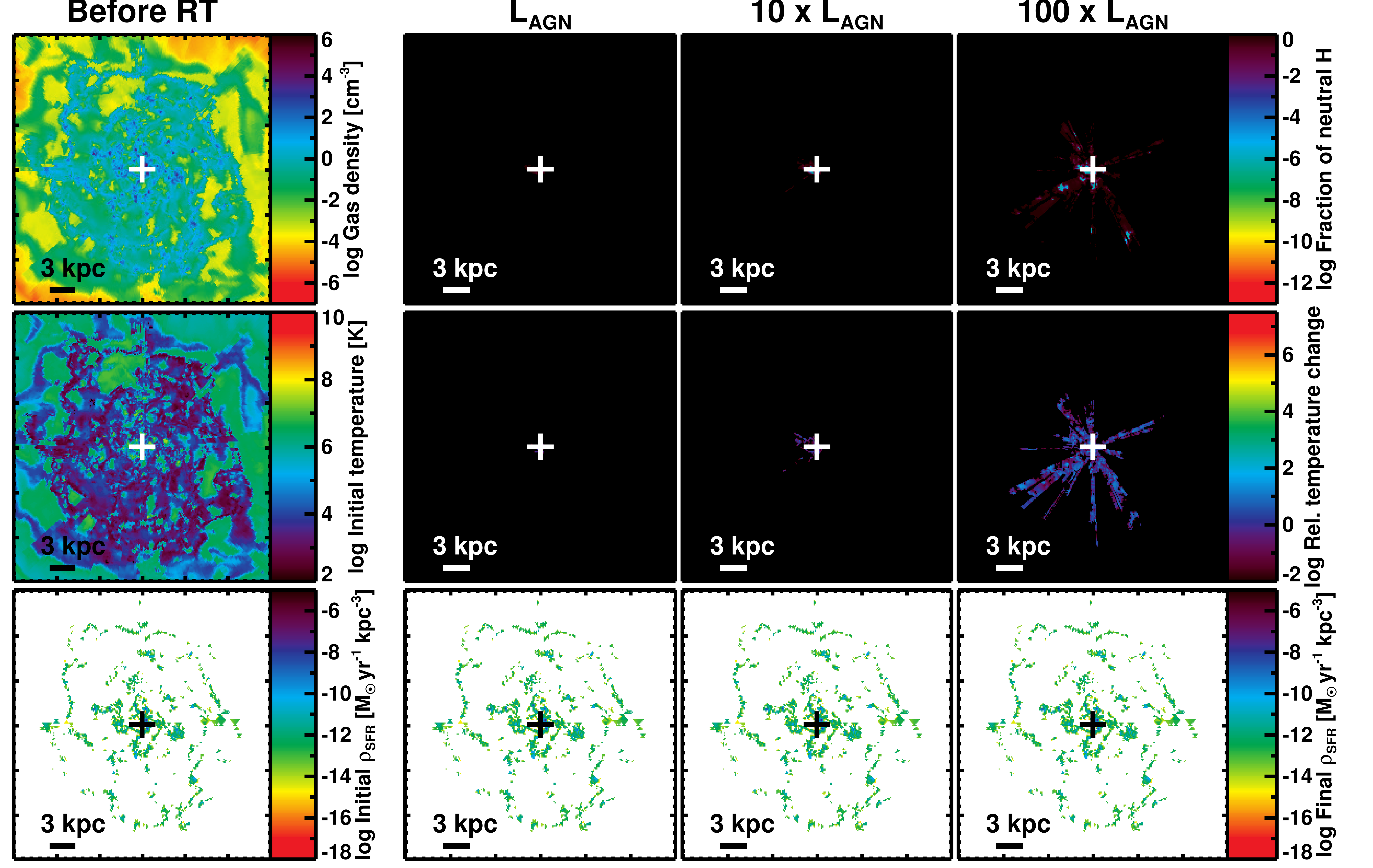}
\caption{Large face-on view of the simulated galaxy (snapshot \#2).   \textit{Top row}: Hydrogen density, Fraction of neutral hydrogen after RT. \textit{Middle row}: Temperature before RT, Relative temperature change. \textit{Bottom row}: $\rho_{SFR}$ before RT,  $\rho_{SFR}$ after RT. The `+' sign shows the location of the BH and the density threshold for SF is 10 cm$^{-3}$. Parameters after RT are given for the three AGN luminosities (L$_{\textrm{AGN}}=10^{44.5}$~erg~s$^{-1}$).  Ionization and heating spots of diffuse gas are visible at large scale in the disk for the highest AGN luminosity.} 
\label{fig:den_temp_sfr_large_face-on_10Hcc_00100}
\end{figure*}

\begin{figure*}
\centering
\includegraphics[trim=1.0cm 0cm 3.1cm 0cm, clip=true,width=0.85\textwidth]{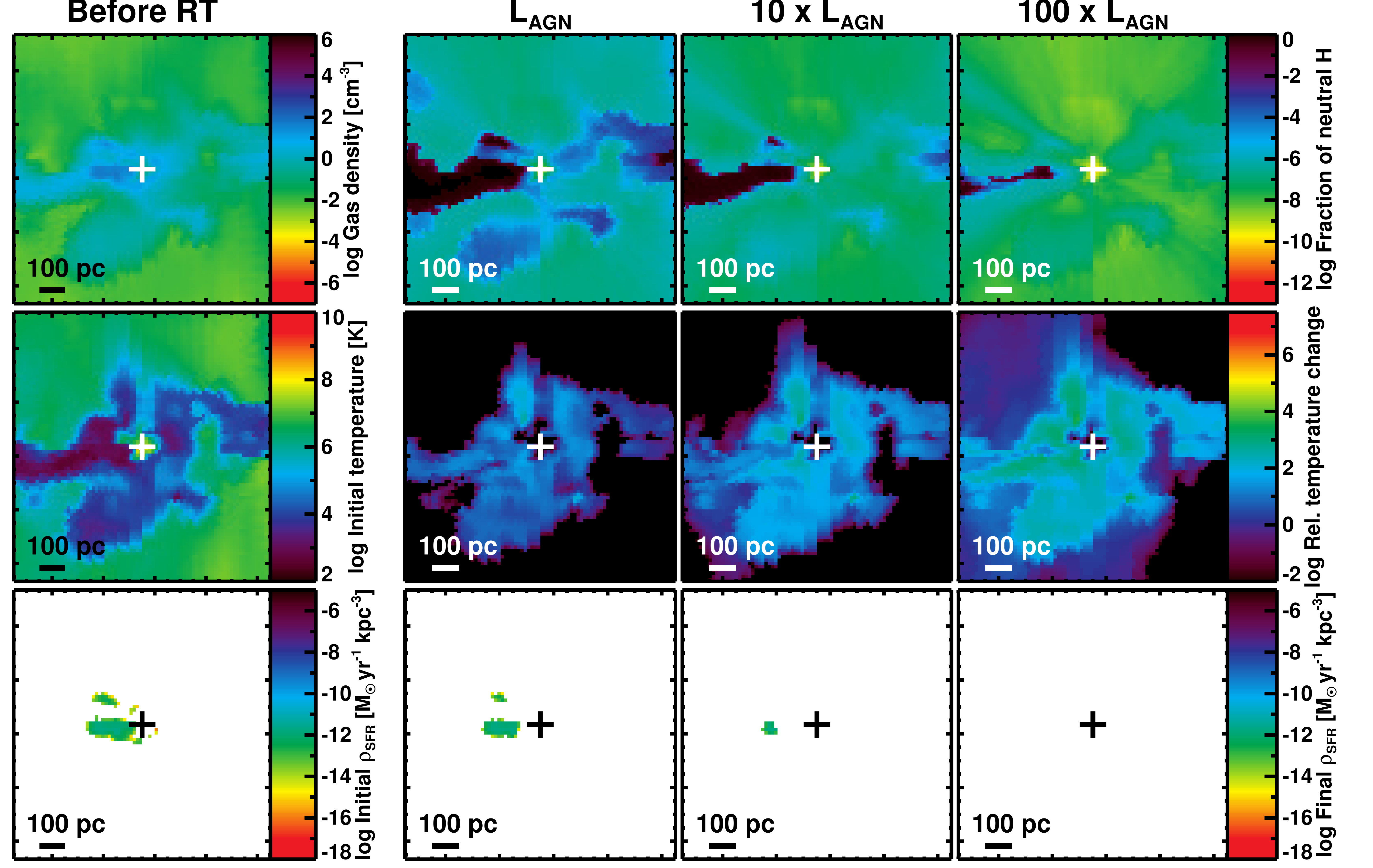}
\caption{Zoomed edge-on view of the simulated galaxy (snapshot \#2).  \textit{Top row}: Hydrogen density, Fraction of neutral hydrogen after RT. \textit{Middle row}: Temperature before RT, Relative temperature change. \textit{Bottom row}: $\rho_{SFR}$ before RT,  $\rho_{SFR}$ after RT. The `+' sign shows the location of the BH and the density threshold for SF is 10 cm$^{-3}$. Parameters after RT are given for the three AGN luminosities (L$_{\textrm{AGN}}=10^{44.5}$~erg~s$^{-1}$). At small scale, diffuse regions are heated above the temperature threshold for SF.} 
\label{fig:den_temp_sfr_zoom_edge-on_10Hcc_00100}
\end{figure*}
\begin{figure*}
\centering
\includegraphics[trim=1.0cm 0cm 3.1cm 0cm, clip=true,width=0.85\textwidth]{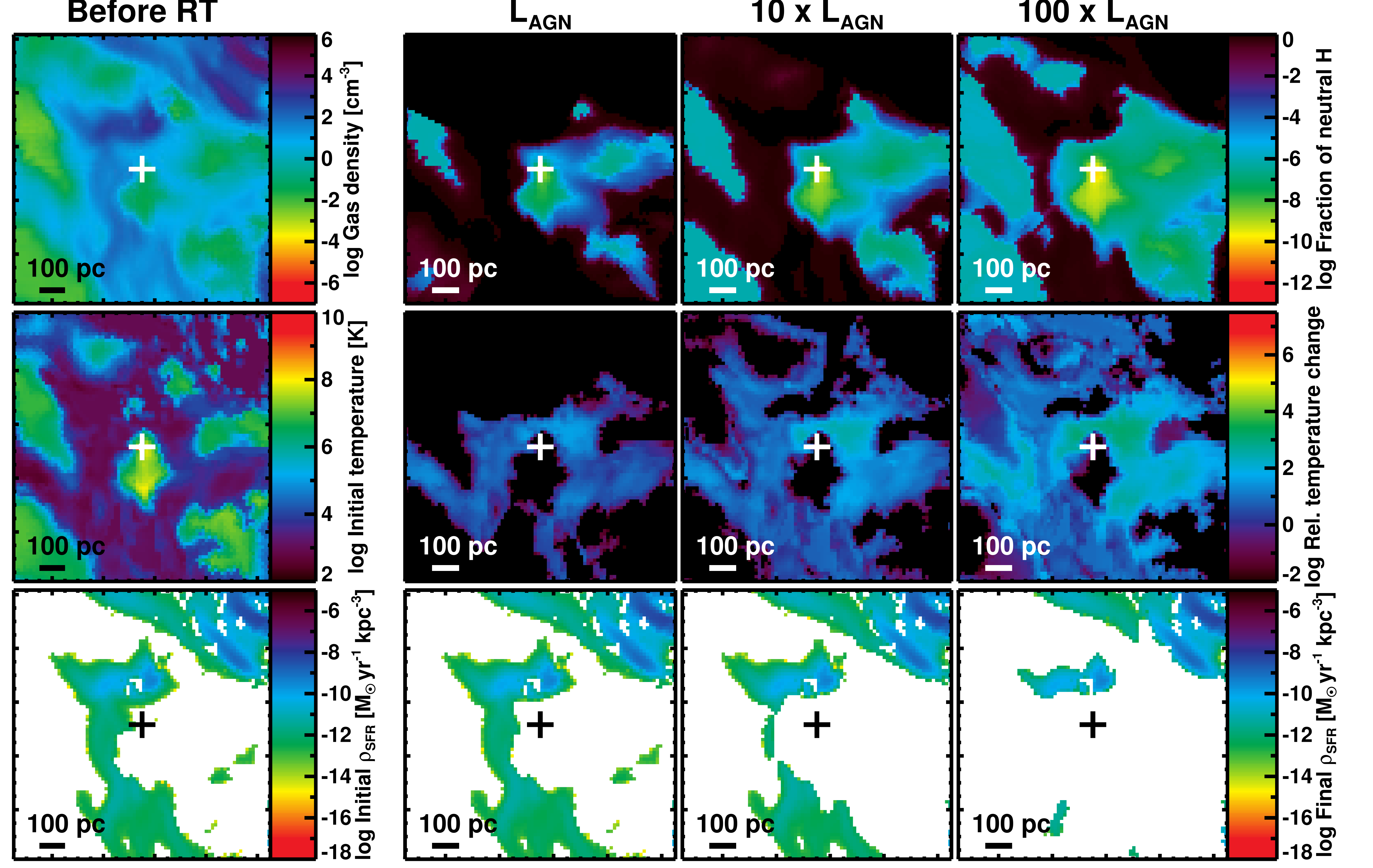}
\caption{Zoomed face-on view of the simulated galaxy (snapshot \#2).   \textit{Top row}: Hydrogen density, Fraction of neutral hydrogen after RT. \textit{Middle row}: Temperature before RT, Relative temperature change. \textit{Bottom row}: $\rho_{SFR}$ before RT,  $\rho_{SFR}$ after RT. The `+' sign shows the location of the BH and the density threshold for SF is 10 cm$^{-3}$. Parameters after RT are given for the three AGN luminosities (L$_{\textrm{AGN}}=10^{44.5}$~erg~s$^{-1}$). The densest SF-cores are not affected, even with the highest AGN luminosity.} 
\label{fig:den_temp_sfr_zoom_face-on_10Hcc_00100}
\end{figure*}

After propagating  AGN radiation throughout the galaxy, the ionization state of the gas and its new temperature are known, and the new SFRs are computed for all cells. Gas density is the same before and after RT (see Section~\ref{subSection:process}). Figures \ref{fig:den_temp_sfr_large_edge-on_10Hcc_00100} and \ref{fig:den_temp_sfr_large_face-on_10Hcc_00100} show edge-on and face-on views of a representative snapshot, and zooms on the central regions are displayed in Figures \ref{fig:den_temp_sfr_zoom_edge-on_10Hcc_00100} and \ref{fig:den_temp_sfr_zoom_face-on_10Hcc_00100}. Each map is a thin slice of the simulation box, centered on the BH. The maps for some other snapshots are available in Appendix~\ref{appendix:maps}. 

We will first focus on the effect of the lowest AGN luminosity on the galaxy, namely the first two columns of each figure. The impact of the higher luminosities are discussed in next section. The ionization panels of the disk seen edge-on at both large (Figure~\ref{fig:den_temp_sfr_large_edge-on_10Hcc_00100}) and small (Figure~\ref{fig:den_temp_sfr_zoom_edge-on_10Hcc_00100}) scales show that the AGN radiation escaping the galaxy has a typical bi-conical shape, as observed by \citet{Muller-Sanchez2011} (see Section~\ref{discussion:cones} for discussion). 

In the central region of the disk, gas at densities of about 10$^{1-2}$ cm$^{-3}$ is heated by roughly a factor of ten, though the effect is invisible at kpc-scale. Dense clumps are only slightly heated on the illuminated side but shield themselves and the material behind them. At approximately 500 pc around the BH in the disk, even  directly illuminated diffuse gas is not impacted. {In the upper outer part of the gaseous halo, and in some diffuse spots above and under the disk, very diffuse gas ($10^{-3}-10^{-6}$ cm$^{-3})$ is relatively inefficiently heated ($RTC<10~\%$) by the AGN. These outer heated regions relatively resemble the upper outer part of the ionization cone.} It is however important to note that a complete ``temperature cone'' similar to the ionization cone does not appear here since the measurement of the final temperature takes into account the initial temperature in the simulation (in which gas is already heated by thermal AGN and SNe feedback, etc.), whereas ionization is only based on the radiative transfer due to the AGN radiation (which assumes the gas was previously neutral, see Section~\ref{subSection:process}). {We therefore see that the effect of AGN heating  dominates the other kinds of feedback only in the outer parts of the halo, or in diffuse regions around the galactic disk, near the BH.}

Finally, only a thin layer (up to $\sim$ 40~pc for 10$^{1-2}$~cm$^{-3}$~; $\sim$ 15~pc for 10$^{2-3}$~cm$^{-3}$) on the illuminated side of the star-forming regions is heated above the temperature threshold, and most of the ionized regions were not initially forming stars anyway. The remaining star-forming regions shield themselves from the radiation starting approximately at 10$^3$ cm$^{-3}$. Around 100 pc away from the BH, even the diffuse star-forming regions remain. The density threshold for SFR is 10 cm$^{-3}$, but more diffuse clumps would not be self-shielding from the AGN radiation, and would contribute negligibly to the total SFR of the galaxy (see discussion in Appendix~\ref{subSection:threshold}).

All snapshots (see Appendix \ref{appendix:maps}) show roughly the same behaviour, except \#6, where the BH is embedded into a very dense clump of gas (n $\sim$ 10$^{4-5}$ cm$^{-3}$). In this case, AGN radiation is blocked within the central clump and regions that could be affected by the AGN are below the resolution limit. Snapshot \#3 shows both behaviours, since a dense clump just above the BH blocks the radiation above the disk, but not on the other side. Thus, the distribution of the gas into clumps has an important impact on the propagation of the AGN radiation, sometimes preventing it from escaping at all.\\

\subsection{Dependence on AGN luminosity}
\label{subSection:comp_lum}

In this section, we investigate the effect of increasing AGN luminosities (right two columns of Figures \ref{fig:den_temp_sfr_large_edge-on_10Hcc_00100}, \ref{fig:den_temp_sfr_large_face-on_10Hcc_00100}, \ref{fig:den_temp_sfr_zoom_edge-on_10Hcc_00100} and \ref{fig:den_temp_sfr_zoom_face-on_10Hcc_00100}). As expected, the higher the luminosity, the more extended the ionized and/or heated regions, and the more star-forming regions are suppressed.

In the halo, diffuse gas at 10$^{-6}$ cm$^{-3}$ and 10$^5$ K is heated at 10$^6$ K in the standard AGN regime, and at 10$^{6.5}$ and 10$^7$ K in the strong AGN and QSO regimes.  At small scale {in the disk}, 10$^{1-2}$ cm$^{-3}$ gas is heated from 10$^{3-4}$ to 10$^{4-5}$~K in the normal AGN regime, at approximately 500 pc around the BH. Denser gas or gas located further in the disk is not heated.  In the strong AGN regime, gas with the same density and temperature is also heated, but to a higher degree (10$^{4-5.5}$ K) and further away from the BH (up to $\sim$ 1 kpc). In the QSO regime, {the effect is even more important: diffuse} 10$^{3-4}$ K gas is heated to 10$^{4-6.5}$ K, within up to $\sim$ 10 kpc around the BH, {thanks to the diffuse interclump medium allowing the QSO radiation to go past the inner kiloparsec}. However, the densest clouds ($> 10^{4}$ cm$^{-3}$) are not heated and shield the gas behind them. {The clumpy distribution of gas in the ISM is responsible for the high variability of the maximal radius at which gas is heated by the AGN in the disk, which cannot be probed with smooth density profiles.}

As AGN luminosity increases, star-forming regions are suppressed further from the BH. In the strong AGN regime, diffuse star-forming regions close to the BH and not shielded by dense clumps are destroyed up to 300 pc away in the disk. In the QSO regime, this distance increases to 1 kpc and only the $n> 10^3$ cm$^{-3}$ star-forming regions survive at the center.
However, most of the SFR lies in the densest star-forming clouds, which are not affected by the AGN radiation.

Snapshot \#6 (AGN embedded in a dense clump, see Appendix \ref{appendix:maps}) shows a more extreme behaviour. Even if AGN radiation escapes the central clump and an ionization cone is visible in the strong AGN and QSO regimes, the reduction of SFR is nearly zero whatever the luminosity of the AGN. This, once again, shows that gas ionized is mainly not initially star-forming, even at high AGN luminosity.

\begin{figure}[htp]
\includegraphics[trim=0cm 0cm 0cm 0cm, clip=true,width=\linewidth]{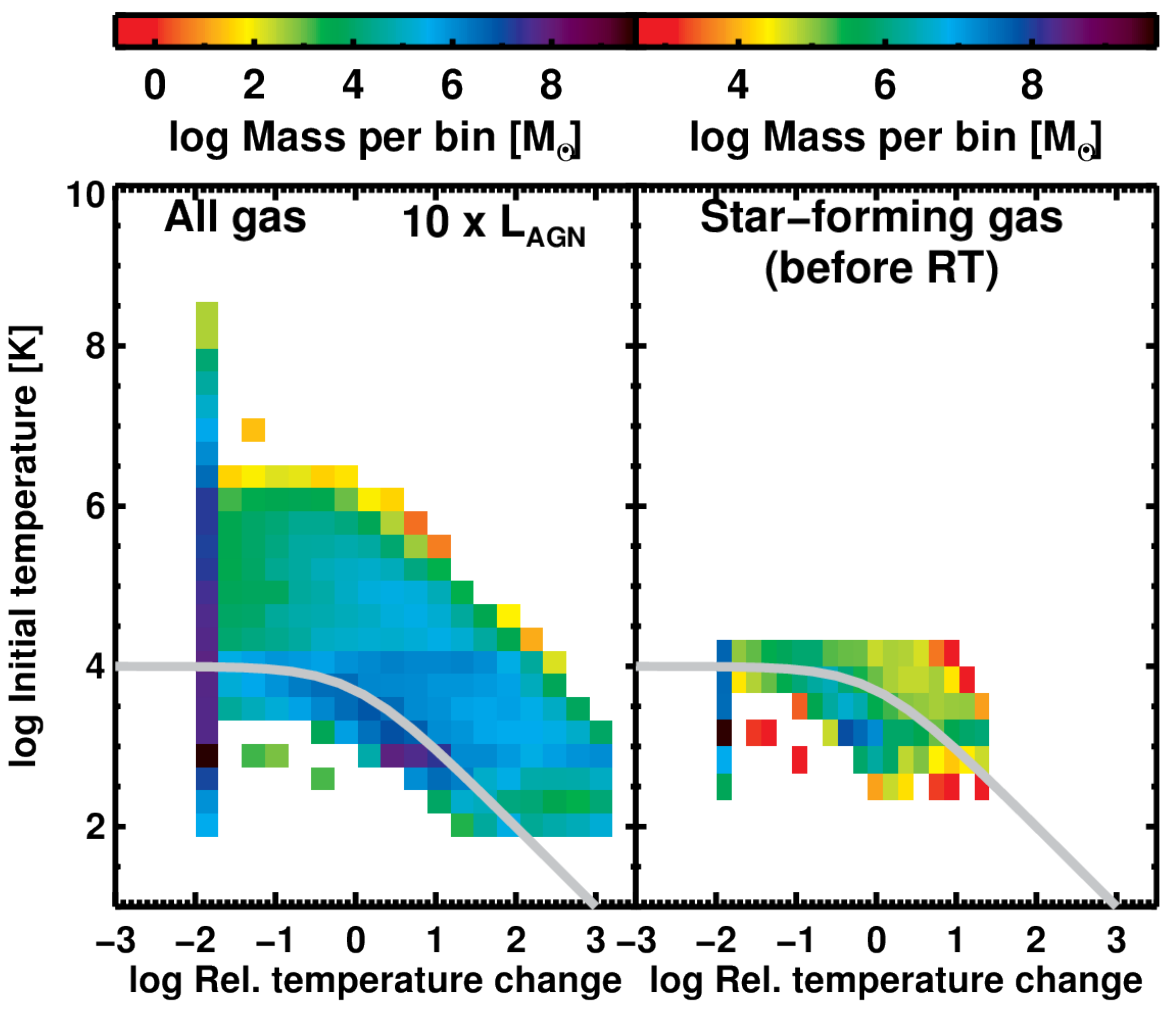}
\includegraphics[trim=0cm 0cm 0cm 0cm, clip=true,width=\linewidth]{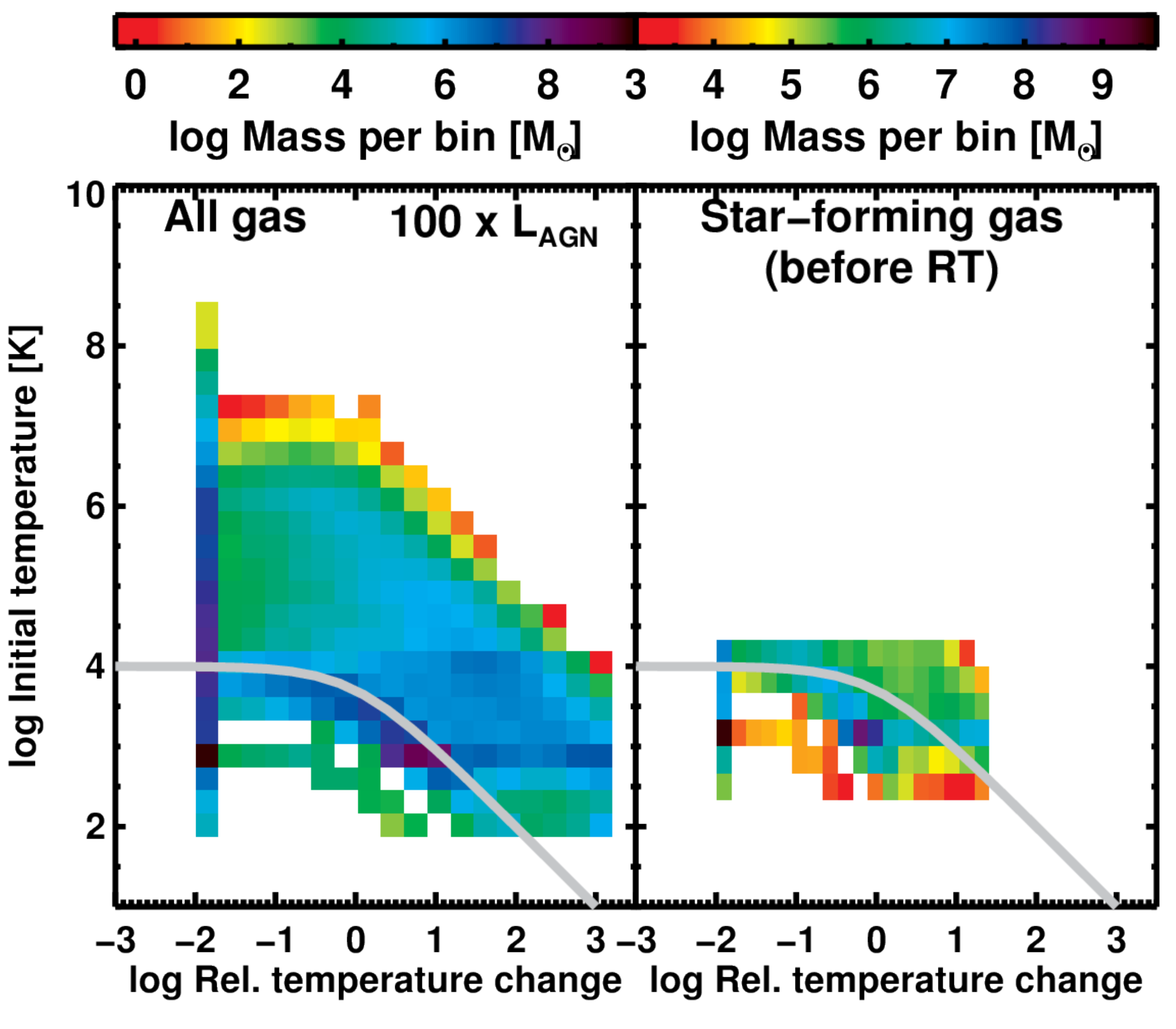}
\caption{\textit{Top:} Initial temperature of the gas as a function of the relative temperature change for the strong AGN regime. \textit{Bottom:} Same plot for the QSO regime. The left panels show all the gas in the central region of the disk (a cylinder of 2 kpc radius and 4 kpc height, centered on the BH). The right panels show only the gas that form stars before applying the RT process (in the same region). Each 2-D bin is color-coded with its total gas mass. All cells with $RTC<1$~\% are shown in the lowest $RTC$ bin. Gas above the gray dividing line stops forming stars after the RT process because its temperature crosses the temperature threshold for star formation. The bulk of the gas which is star-forming before RT remains star-forming after the RT process for all AGN luminosity regimes.\\}
\label{fig:RTC_T_diagrams}
\end{figure}

{The idea that star-forming gas is mostly left unaffected even in the strong AGN and QSO regimes is also well presented by the temperature versus relative temperature change diagrams (see Figure~\ref{fig:RTC_T_diagrams}). The temperature represented is the initial temperature (before RT) and only the gas in the central region of the box -- defined as the cylinder of radius 2 kpc and height 4 kpc centered on the BH, is shown. All cells with $RTC<1$~\% are considered not heated and are shown in the first bin ($\log RTC=-2$). The demarcation line between gas forming stars before RT and remaining below the temperature threshold for star formation after the RT process, and gas forming stars before RT but crossing the temperature threshold after RT is defined as:}
\begin{equation}
T_{initial}=\frac{T_{thr}}{RTC+1},
\label{eqn:demarcation}
\end{equation}
{where $T_{initial}$, $T_{thr}$ and $RTC$ are the quantities defined above. Star-forming gas above the demarcation line defined in Equation \ref{eqn:demarcation} is prevented from forming stars due to AGN ionization. Figure~\ref{fig:RTC_T_diagrams} clearly shows that, even though a greater amount of gas is heated to a higher degree when increasing the AGN luminosity, the bulk of the star-forming gas is not heated enough by AGN radiation to exceed the temperature threshold for star formation.}

In summary, increasing AGN luminosity indeed heats a greater amount of gas with densities reaching 10$^{3}$ cm$^{-3}$ above the temperature threshold for star formation. However, given the low density {of the gas in most affected regions}, we expect the decrease of the total SFR due to photoionization to be relatively small, as quantified below.\\

\subsection{Reduction of the total SFR}
\label{subSection:total}

\begin{SCfigure*}[][]
\epsscale{1.7} 
\plotone{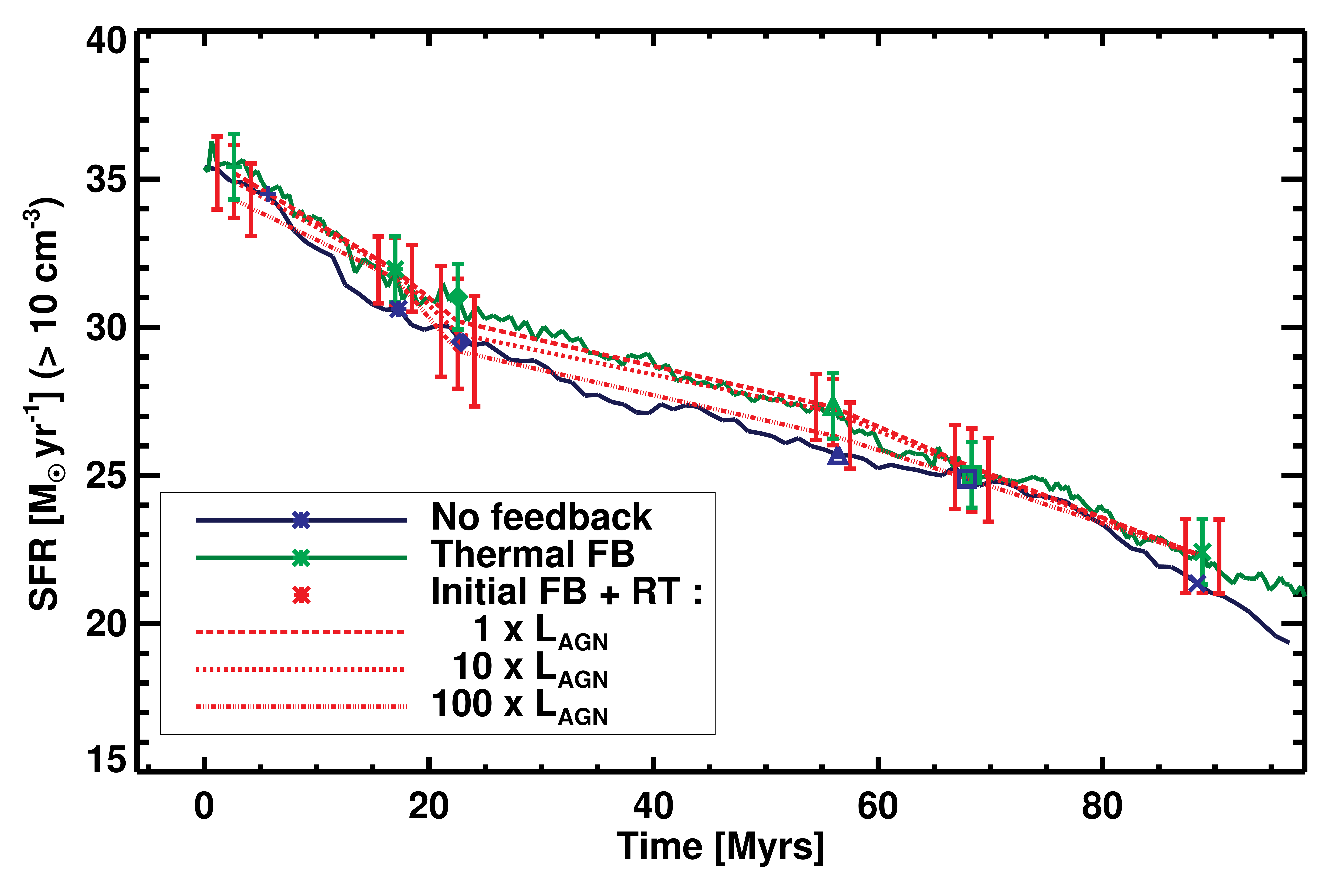} 
\caption{Star Formation Rate as a function of time for the simulation with no AGN feedback (blue) and the simulation including thermal AGN feedback before and after ionization (green and red respectively). Each line style corresponds to an AGN luminosity, as labelled. Each symbol corresponds to a given snapshot. For clarity, the error bars of the typical AGN regime have been shifted 1.5~Myr to the left, and those of the QSO regime 1.5~Myr to the right. The difference between the reference curve (blue) and the initial feedback one (green) gives the error on the value of the SFR (green error bars). The additional errors are due to the correction of the SFR obtained after the Cloudy computations  (described in Section~\ref{subSection:process}). The red error bars are the sum of the latter and the stochastic errors (green error bars). The AGN does not have a significant impact on the instantaneous SFR of the whole galaxy along time, even with the highest AGN luminosity.\\} 
\label{fig:sfr}
\end{SCfigure*}

\begin{figure}[htp] 
\epsscale{1.15} 
\plotone{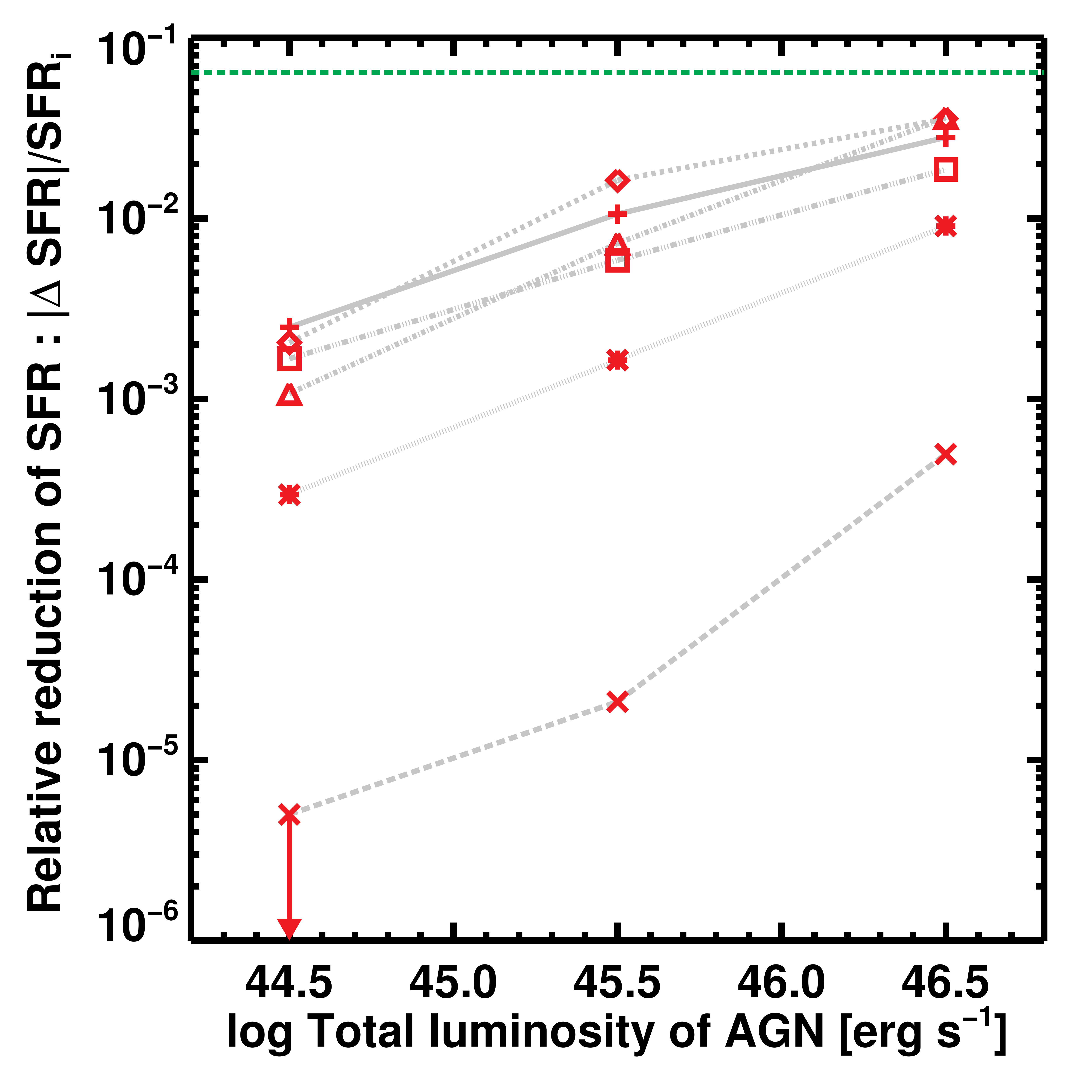} 
\caption{Relative reduction of the Star Formation Rate after radiative transfer as a function of AGN luminosity. Each symbol represents the same snapshot as in Figure~\ref{fig:sfr}. They are linked for clarity. The green line shows the maximal relative difference between the simulations with and without AGN feedback. The lower curve corresponds to snapshot \#6, where the BH is embeded in a very dense clump. The change in SFR due to the AGN is of a few percent at most.\\} 
\label{fig:sfr_ratio}
\end{figure}

We estimate the global effects of AGN photoionization by summing the SFRs of all cells in the simulation, and calculating total ionized/heated mass and volume fractions (see Section~\ref{subSection:heated-ionized}). Figure~\ref{fig:sfr} shows the evolution of the total SFR as a function of time for the simulation without AGN feedback and the simulation including AGN feedback before and after ionization. Here the density threshold for star formation is 10 cm$^{-3}$ (see Appendix~\ref{subSection:threshold}  for other thresholds). The slow decrease is due to gas consumption over time -- as no new gas is added to the simulation box. 

The difference between the SFR without AGN feedback and the SFR with only thermal AGN feedback (before RT) arises due to the fluctuations of the SFR in the simulation, which is highly dependent on the distribution of gas into clumps. As this distribution is stochastic (see Section \ref{subSection:simu}), short-term variation and a difference of a few percent between two simulation runs are not surprising {and AGN feedback -- if it does play a role in this change -- is probably not the main driving mechanism}. Similarly, the SFR with thermal AGN feedback being greater than the SFR without AGN feedback is most likely due to a random event, and is not necessarily a sign of positive AGN feedback (SF-triggering). 

Figure~\ref{fig:sfr} clearly shows that the impact of RT on the total SFR of the simulation with feedback is small at all luminosities. As the final SFR is based on the final post-RT temperature (defined as the maximum between the Cloudy temperature and the initial temperature, see Section~\ref{subSection:process}), it takes into account the effect of the thermal AGN and stellar feedback implemented in the simulation even though theses sources are not considered for the RT computation itself. Thus, we conclude that, not only does RT change the SFR marginally compared to other feedback models\footnote{All simulations include SNe feedback.} but also, in this particular simulation, the change in SFR due to all kinds of AGN feedback is not significant.

Figure~\ref{fig:sfr_ratio} shows the relative reduction of the SFR due to the radiative transfer, as a function of the AGN luminosity. It is defined as :
\begin{equation}
\label{eqn:delta}
\Delta_{\textrm{rel}} = \frac{\left| \textrm{SFR}_{final}~-~\textrm{SFR}_{initial} \right|}{\textrm{SFR}_{initial}},
\end{equation}
where $\textrm{SFR}_{initial}$ is the SFR of the simulated galaxy with thermal AGN feedback before computation of the RT and $\textrm{SFR}_{final}$ is the SFR after computation of the RT. The total values of the SFR -- both before and after RT -- were corrected the same way to account for the resampling induced by the Cloudy computation (see Section~\ref{subSection:process}).
We see that the effect is indeed very small, and though there is an increasing trend at higher luminosities, the overall effect is marginal: a maximum of a few percent in the QSO  regime, for the snapshots with the most diffuse inner regions. The lowest curve corresponds to snapshot \#6, where no effect is visible on the $\rho_{SFR} $ maps. In the standard AGN regime, the final SFR of this snapshot, which is a rare configuration, is not reduced at all. \\

We conclude that adding instantaneous AGN photoionization feedback to a simulation containing thermal AGN feedback and stellar feedback changes the SFR of the whole galaxy by only a few percent at most. Moreover, this reduction is much smaller than the difference in SFR between two runs of the same simulation with and without thermal AGN feedback -- which represents the fluctuations of the SFR due to the stochastic distribution of clouds -- and shows that AGN feedback does not have a significant impact on the SFR on short time-scales.

\subsection{Fractions of heated and/or ionized gas}
\label{subSection:heated-ionized}

\begin{figure*} 
\epsscale{0.95} 
\plottwo{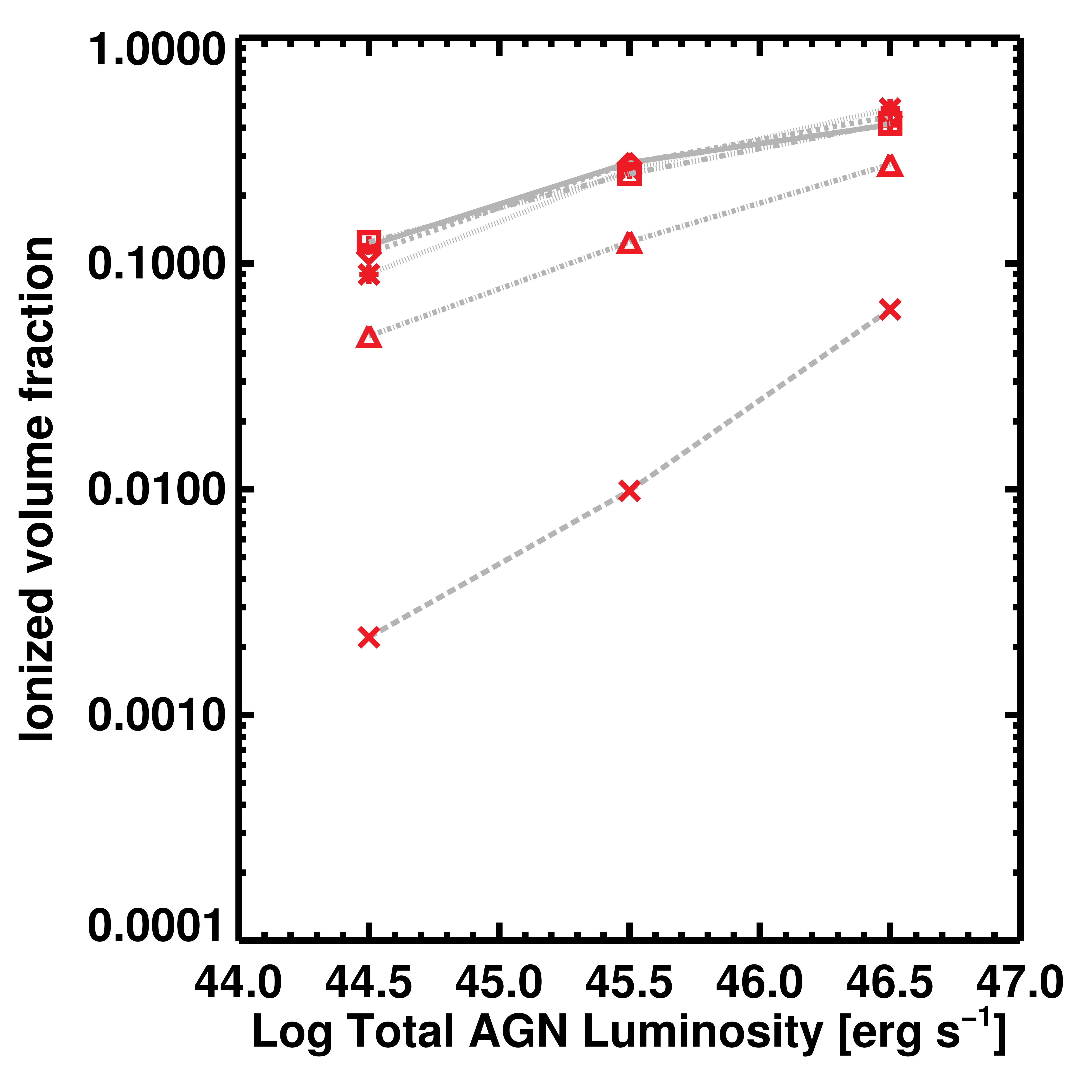}{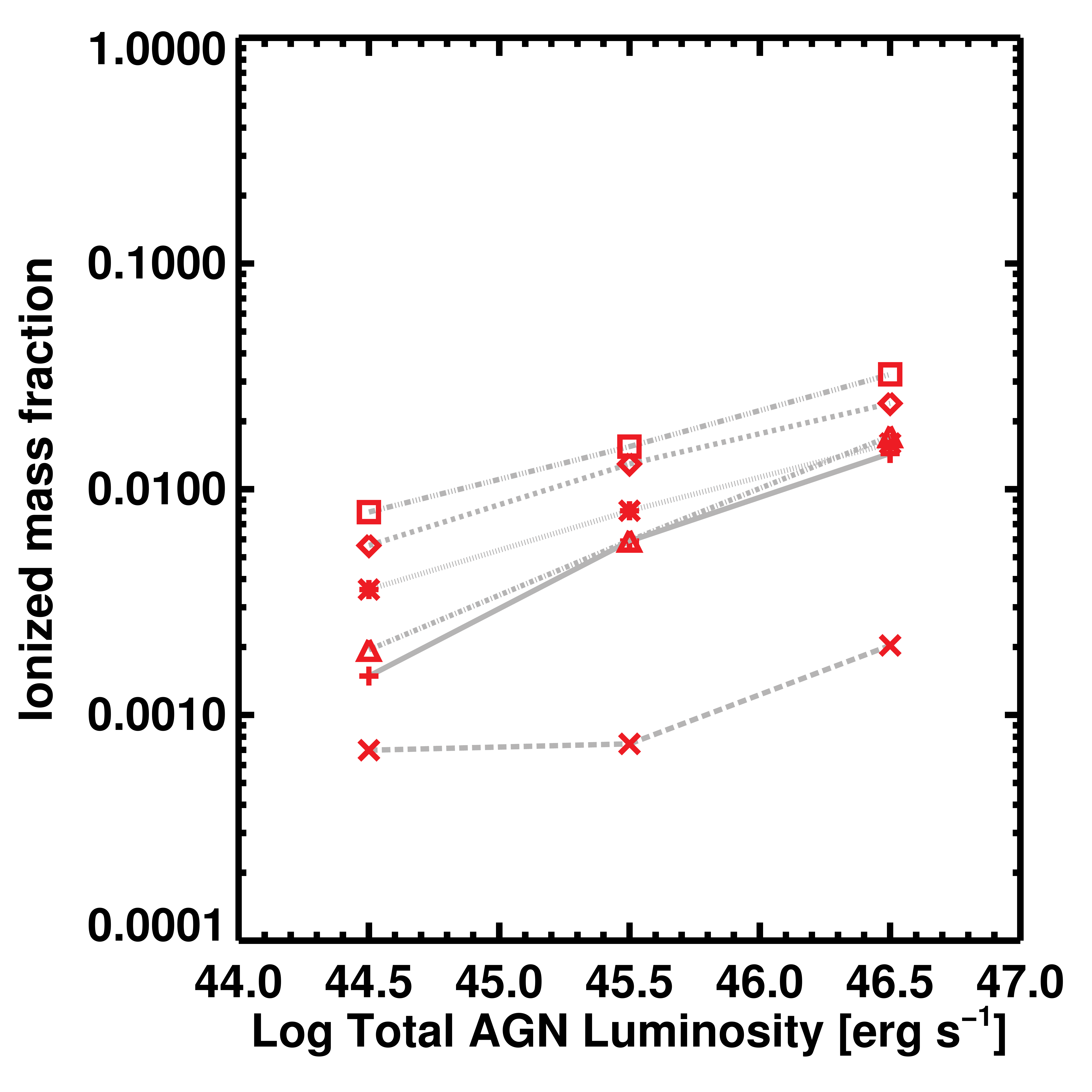} 
\caption{Ionized volume fraction (\textit{left}) and mass fraction (\textit{right}) of the gas in the galactic disk as a function of the AGN luminosity (same symbols as Figure~\ref{fig:sfr}). The word `ionized' refers to the gas ionized by the single AGN ionization feedback. Before RT, the whole galaxy is assumed to be neutral. Even if the volume fraction of ionized gas in the galactic disk is high, the corresponding mass fraction remains low, meaning that only diffuse gas is ionized by the AGN.} 
\label{fig:ionized_mass}
\end{figure*}

\begin{figure*} 
\epsscale{0.95} 
\plottwo{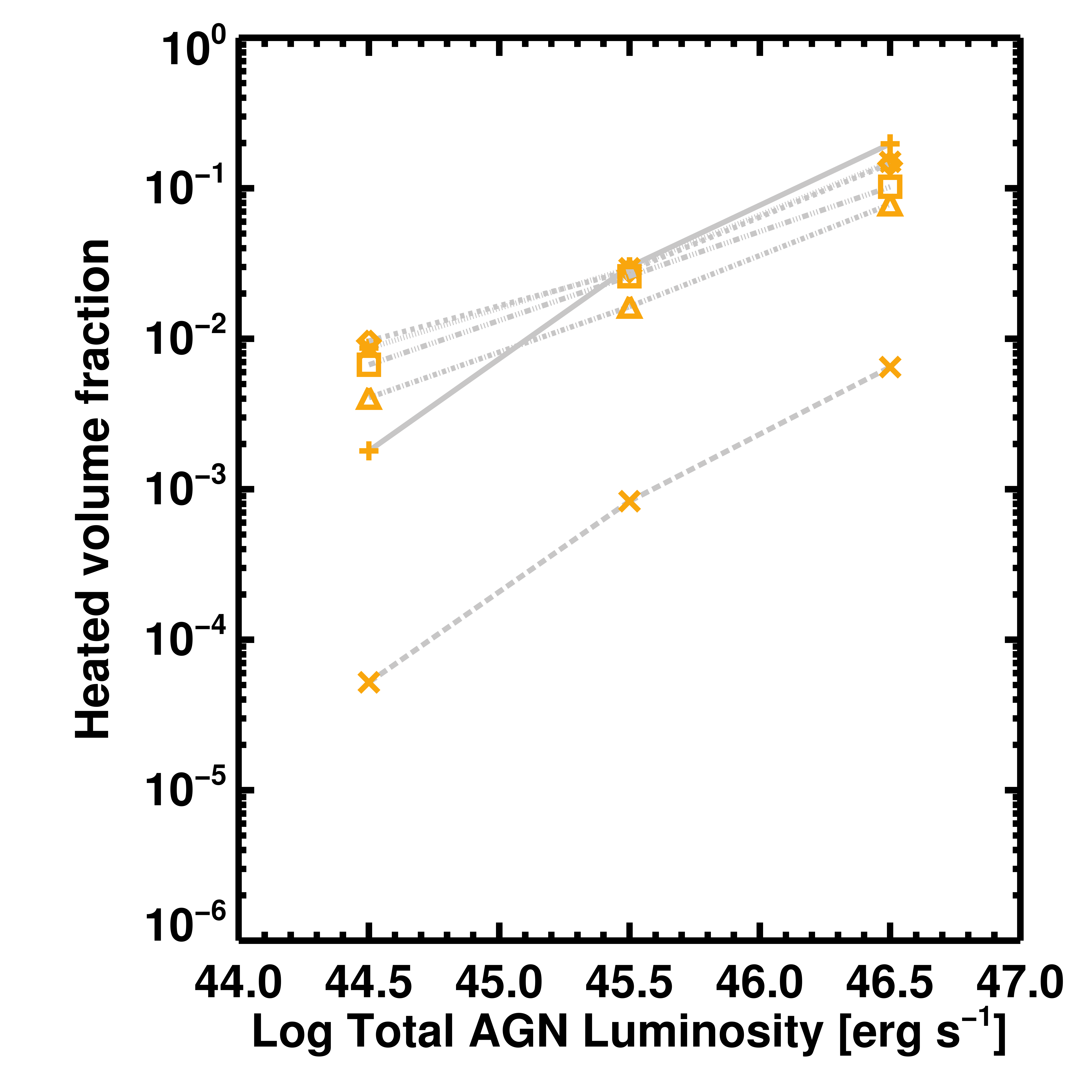}{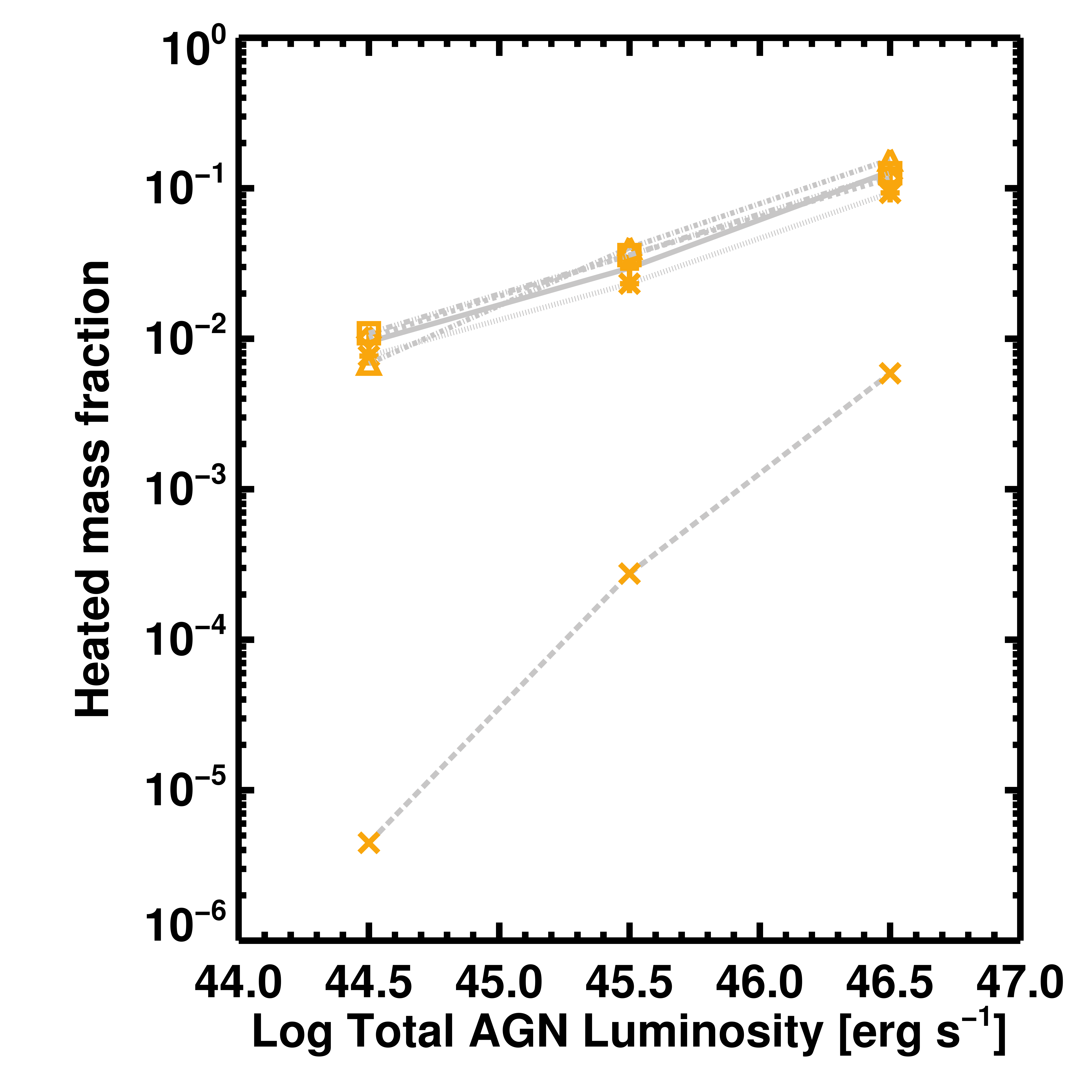} 
\caption{Heated volume fraction (\textit{left}) and mass fraction (\textit{right}) of the gas in the galactic disk as a function of the AGN luminosity (same symbols as Figure~\ref{fig:sfr}). The word `heated' refers to the gas heated by the single ionization feedback above the temperature at which it has been heated by all other kinds of feedback implemented in the simulation (thermal AGN FB, stellar FB). The volume and mass fraction of heated gas are low, which means that AGN ionization overwhelms all kinds of FB implemented in the simulation for only a small amount of the diffuse gas.\\} 
\label{fig:heated_mass}
\end{figure*}

From the small reduction of the SFR, we expect the gas {mass} fraction that is heated or ionized by the AGN to be small. Figure~\ref{fig:ionized_mass} shows the volume and mass fractions of the gas in the galactic disk (2 kpc-thick layers on each side of the median plane of the box) that is ionized by the AGN, as a function of the luminosity. The ratios are defined as follows:
\begin{equation}
\label{eqn:ion}
m_{\textrm{ion}} = \frac{M_{\textrm{ionized, corrected}}}{M_{\textrm{total, corrected}}}~;~v_{\textrm{ion}} = \frac{V_{\textrm{ionized, corrected}}}{V_{\textrm{total, corrected}}}
\end{equation}
As for Equation \ref{eqn:delta}, both the pre- and post-RT parameters are those of the simulation with AGN feedback respectively before and after photoionization, and were corrected the same way to account for the resampling due to the Cloudy computation.
We consider gas to be ionized if its neutral hydrogen fraction is below 10 \%. Contributions from OB stars and SNe or thermal AGN feedback are not included since Cloudy computes the RT as if the BH was the only ionizing source and the gas was initially neutral (see Section~\ref{subSection:process}).  Given that the halo is initially set in the simulation to have very diffuse gas ($10^{-5}$ cm$^{-3}$), the halo component is highly ionized by the AGN, but would be easily ionized by another source such as OB stars and SNe feedback or UV background (see Section~\ref{subSection:transparent_lops} for discussion). The simulation is not designed to study the gaseous halo properties and one would need to account for cosmological context or at least close environment and satellites. Restraining the study to the gas in the disk reduces this effect but does not cancel it entirely and thus this value shows an upper limit to the amount of neutral gas that could be ionized by an AGN.

We see that even though the fraction of volume ionized by the AGN is large (from 5 to 40 \% in all representative snapshots depending on the luminosity regime), the corresponding mass fraction is low: from 0.1 to 3 \% at most. This confirms the intuition that, even though some regions that are ionized have a large spatial extent, they are not significant contributors to the total mass of the galaxy, and therefore to the total SFR.\\

Figure~\ref{fig:heated_mass} shows the heated mass and volume fractions of gas in the disk. The ratios are similar to Equation~\ref{eqn:ion}. Gas is heated if the equilibrium temperature given by Cloudy is greater than the initial temperature in the simulation, meaning that the AGN ionization alone is able to heat the cell above the temperature at which it has been heated by all the other kinds of feedback (thermal AGN and stellar) in the simulation. Thus, heated gas takes into account the ``feedback history'' of the snapshot (contrarily to ionized gas) and traces the regions where heating due to the AGN photoionization outweighs the other kinds of feedback. However, the behaviour is very similar to that of ionized gas, and even when the photoionization has a stronger effect than the other forms of feedback together, the SFR is only slightly impacted because most of the affected gas is not initially star-forming.\\

\section{Discussion}
\label{Section:discussion}

In the following, we discuss the dependence of our results on the structure of the ISM. We also deduce a trend for the 100 Myr-scale effects on star-formation and develop our study of the ionization cones. Finally, we try to account for the gas that would already be ionized by other sources before applying RT.  \\

\subsection{Role of ISM structure}
\label{subSection:other}

The profiles used in the LOPs have complex structures with large contrasts between the clump and inter-clump densities (see Figure~\ref{fig:lop_example}). In order to study the role of the ISM structure on the propagation of AGN radiation,  we used Cloudy to calculate the propagation of the three AGN regimes we used before along homogeneous lines-of-propagation, using the inner and outer radii of a typical LOP in the plane of the disk, {and the same filling factor (see Figure~\ref{fig:unif})}. We compared the resulting neutral hydrogen fractions to that of a typical LOP in the plane of the disk (as shown in black in Figure~\ref{fig:lop_example}), whose mean density is $\sim$ 40 cm$^{-3}$. We only show the typical AGN and QSO luminosities, and note that the strong AGN case is intermediate, as expected.

\begin{figure}[htp]
\center
\includegraphics[trim=2cm 0cm 7.4cm 0cm, clip=true,width=0.7\linewidth]{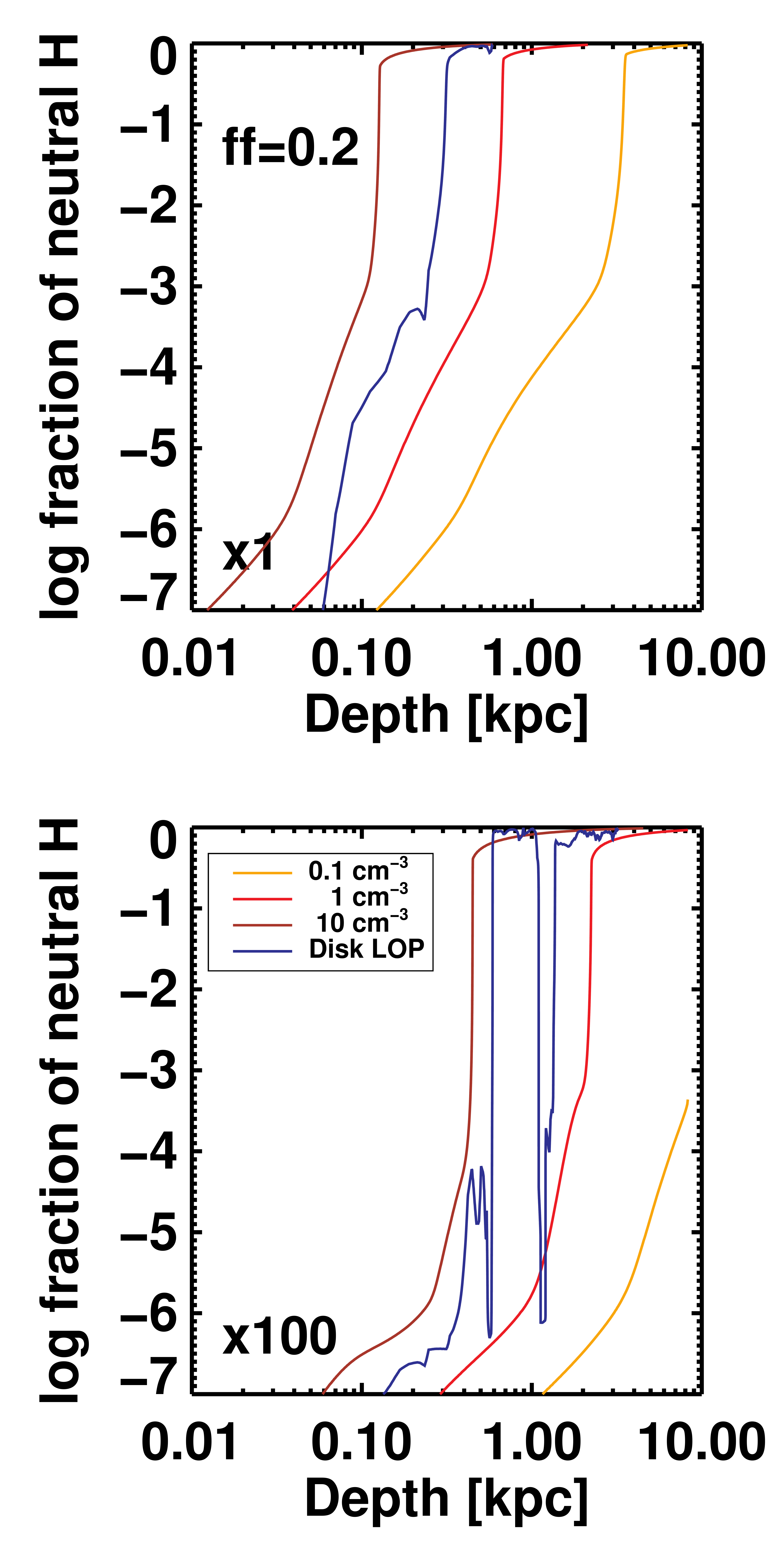}
\caption{Fraction of neutral hydrogen for three constant density profiles at 0.1, 1 and 10 cm$^{-3}$ and a typical LOP in the disk plane (mean density of 40 cm$^{-3}$). The top panel shows the typical AGN luminosity and the bottom panel shows the strong QSO luminosity. In a uniform galactic disk at 10~cm$^{-3}$, a QSO would only ionize the central few hundreds of  parsecs.\\} 
\label{fig:unif}
\end{figure}

The ionization profile of the LOP inside the plane of the disk used in the main study is located between those of the constant LOP at 10~cm$^{-3}$ and at 1~cm$^{-3}$, whereas its mean density is $\sim40$~cm$^{-3}$. For all constant profiles (except for the 10~cm$^{-3}$ in the typical AGN regime), the radius at which gas becomes less than 10 \% ionized is larger than for the main study LOP, showing that ionization and heating by the AGN goes deeper in the disk than for the typical disk plane LOP. However in the uniform-density case, the ionization fraction decreases smoothly whereas that of the main study LOP is not monotonic and regions of highly ionized gas are found far in the disk in the strong QSO regime (see Section~\ref{discussion:cones}). {Such spikes correspond to diffuse interclump regions that are ionized by the AGN. In the QSO regime, they are located at radii up to $\sim8$ kpc in the disk (see Figure~\ref{fig:den_temp_sfr_large_face-on_10Hcc_00100}).} This suggests that typical high-redshift disk galaxies are on average dense enough to screen AGN radiation, and that holes between the dense clumps are necessary for radiation to go past the inner kiloparsecs. 

A smooth exponential density profile enclosing the same mass as the simulated galaxy gave similar results.\\

We conclude that the ISM structure plays a major role in the propagation of AGN radiation since holes explain that ionization often reaches large distances in the disk or in the halo, whereas dense clumps are not necessary to explain the low ionized mass fraction, since the average density is high enough for the gas to be self-shielding.

\subsection{Long-term effects on star formation}
\label{discussion:longterm}

\begin{deluxetable}{ccc}
\tablecolumns{3} 
\tablewidth{\linewidth}
\tablecaption{Effect of AGN on future (100~-~200 Myrs) star formation.\label{table.reservoir}}
\tablehead{\colhead{Regime} &\colhead{Heated mass rate} & \colhead{Ionized mass rate}}

\startdata
Typical AGN &  $0^*-4$ \% & $0^*-2$ \%\\
Strong AGN &  $0.2-9$ \% & $0.01-3$ \%\\
Typical QSO & $2-30$ \% & $0.1-8$ \%
\enddata
\tablecomments{Rates are given for atomic gas ($0.3-10$ cm$^{-3}$). $^*$Rate is zero for snapshot \#6.}
\end{deluxetable}

Long-term effects on star formation can be deduced from our instantaneous study by looking at the HI reservoirs in the envelope of GMCs {and in the atomic clouds around them (``proto-GMCs''), which are} composed of 0.3~-~10 cm$^{-3}$ ISM \citep{Dobbs2008}.    

{Indeed, the first step to launch star formation is to form a dense cloud of molecular gas (a GMC) out of diffuse ISM. Such GMCs are believed to be created by spiral-wave-induced shocks \citep{Heyer1998,Dobbs2008b}, and live about 15~-~40 Myrs \citep{Murray2011}, until they are disrupted by the stars that formed inside them and their envelope is dispersed \citep{Elmegreen2007a}.  In the absence of external heating, those 0.3~-~10~cm$^{-3}$ regions form proto-GMCs which gradually fall under $10^4$~K. These are likely to collapse due to shocks and create new GMCs, which in turn induce the formation of new stars in the next 100~-~200~Myrs, and the cycle continues. }

However, if this gas phase is maintained hot or ionized by the AGN, {it cannot cool down and collapse and} future star formation is suppressed {on a time-scale shorter than that necessary to refuel the interclump medium with cold infalling gas}. The instantaneous effect of the three luminosity regimes on {atomic gas} is computed the same way as before (see Section~\ref{subSection:heated-ionized}) but accounts only for the gas at densities 0.3~-~10 cm$^{-3}$ and is displayed in Table~\ref{table.reservoir}.

From Table~\ref{table.regimes}, AGN radiation is emitted at $10^{44.5}$~erg~s$^{-1}$ about 30 \% of the time, and at $10^{45.5}$~erg~s$^{-1}$ about 3 \% of the time, considering an AGN duty cycle of \nicefrac{1}{3}. 

Instantaneously, the heating/ionization of the GMCs due to the AGN in such regimes is negligible (see Table~\ref{table.reservoir}), and therefore it is highly unlikely that cumulative effects will become important in the following 100~-~200~Myrs and thus no star formation quenching is expected. Even if the AGN were emitting 100 \% of the time, the results would not change for both the typical and strong AGNs. However, in the case where a QSO would be emitting for an extended period of time {because of, e.g., a merger}, GMCs would more likely be impacted and cumulative effects could significantly reduce future star formation.

{The longer-term ($\gtrsim 200$ Myrs~-~1 Gyr) SFR evolution depends on the ability of the AGN to keep the gaseous halo warm or ionized. Indeed, keeping the halo hot over an extended period of time could prevent inflows from reaching the disk and starve the galaxy by suppressing its gas supplies \citep{Dubois2012}. However, the simulation we used is an isolated galaxy and its initial gaseous halo is not designed to be realistic and thus we cannot predict whether the AGN is able to quench the galaxy on a time-scale of a few Gyrs.}

\subsection{Distribution of ionized gas}
\label{discussion:cones}

{This section focuses on the ionization maps in Figures \ref{fig:den_temp_sfr_large_edge-on_10Hcc_00100}, \ref{fig:den_temp_sfr_large_face-on_10Hcc_00100}, \ref{fig:den_temp_sfr_zoom_edge-on_10Hcc_00100} and \ref{fig:den_temp_sfr_zoom_face-on_10Hcc_00100}.} In the typical AGN regime, the inner part of the disk surrounding the AGN is ionized up to 50~-~700 pc, depending on the location of the nearest dense clumps ($n> 10^3$ cm$^{-3}$), which shield themselves, block the AGN radiation and protect the diffuse material behind them. The galactic disk remains neutral at larger scale. The LOPs that do not cross dense clumps are ionized until the end, meaning that the material in the halo is not able to stop the radiation. However, this simulation was not designed to have a realistic gaseous halo and is not in its cosmological context.

At small scales around the black hole, the limit between neutral and ionized gas goes from 1~-~10 cm$^{-3}$ in the typical AGN regime, to 10 and 100 cm$^{-3}$ in the strong AGN and QSO regimes respectively. With increasing AGN luminosity, the distance at which clumps are able to shield the diffuse regions behind them is larger, allowing ionization to go further within the disk. Though, it does not affect the densest clumps or the disk itself, which remains neutral. 

In the strong AGN and QSO regimes, not only are the ionized regions more extended, but also the fraction of remaining neutral hydrogen is smaller by a factor $\sim$~10~-~100 in the regions that were already ionized in the typical AGN regime. In the QSO regime, ionized spots are visible at large scale within the disk (up to 8 kpc from the BH), though SFR is not impacted since those regions are mostly not initially star-forming.

{Clearly, the impact of AGN photo-ionization is greater than expected for a simple model (see Section \ref{subSection:simple_models}), showing that the multi-phase distribution of the gas plays a key role in the propagation of AGN radiation: while dense clumps can block the ionizing radiation at a very small scale-length depending on their distance to the BH, AGN radiation is allowed to propagate past the inner kiloparsecs thanks to  the diffuse interclump medium. The morphology of the galaxy may also be of great importance, since the calculation done by \citet{Curran2012} reproduces observations of (most likely) elliptical radio galaxies and quasar hosts \citep{Curran2006a,Curran2008a} up to redshift 3, but fails to reproduce the propagation of AGN radiation in a simulated star-forming disk at redshift $\sim2$.}

Ionized gas (AGN ionization only) and heated gas (AGN ionization stronger than thermal AGN and SNe feedback) have distinct distributions, showing that, in the simulation presented here, AGN ionization itself does not overwhelm all other forms of feedback and ionization from other sources -- at least instantaneously -- for the three luminosity regimes. 

Yet, our study reproduces the observed biconical shape of AGN emission (see Figure~\ref{fig:den_temp_sfr_large_edge-on_10Hcc_00100}), even though the propagation of AGN light is isotropic. This shows that the simulated ISM is able to collimate the AGN radiation to some degree. These ionization cones may be larger-scale analogs to those predicted by the AGN Unified Model of \citet{Urry1995}. {Furthermore, as the AGN is the only ionizing source, we show that other sources of ionization such as stars are not needed for AGN radiation to escape the galaxy,  which is consistent with AGNs being the main drivers of ionization cones. Accounting for these other sources of ionization would only favor the escape of photons emitted by the AGN.}

The bases of the cones are not circular and are not necessarily centered on the BH . Their shape depends a lot on the cloud distribution.
With a higher AGN luminosity, the ionization cones are wider and their basis is larger, {which is in broad agreement with \citet{Hainline2013} in the sense that the size of the Narrow-Line Regions (NLRs) increases with AGN luminosity}. Finally, the inclination of the cones with respect to the galaxy spin axis decreases for a higher AGN luminosity. {The diffuse and almost entirely ionized gaseous halo may also be consistent with the observations of nearly circular NLRs in radio-quiet quasars by \citet{Liu2013b}}.

\subsection{Other sources of ionization}
\label{subSection:transparent_lops}

\begin{deluxetable*}{lrrrr}
\tablecolumns{5} 
\tablewidth{0pt}
\tablecaption{Mean mass and volume fractions of hot gas and mean relative change of $\Delta_{rel}$ due to hot gas.  \label{table.hot_gas}}
\tablehead{Transparency criterion &T $> 5\times10^4$ K & T $>10^5$ K & T $> 10^6$ K & T $> 10^7$ K }

\startdata
Mean mass fraction$^1$ & $4.00 ~(\pm0.38)$ \% & $2.80 ~(\pm0.42)$ \% & $1.23~ (\pm0.42) $\% & $0.15~ (\pm0.15)$ \%  \\
Mean volume fraction$^1$ & $98.57 ~(\pm0.54)$ \% & $98.22~ (\pm0.80)$ \% & $74.71 ~(\pm7.06)$ \% & $6.13 ~(\pm5.30)$ \% \\
\tableline\\
Mean $\Delta_{C-T}$ for the typical AGN$^2$ &  $-6.00~(\pm16.90)$ \% & $-7.79~(\pm18.50)$ \%   &  $-2.20~(\pm22.47)$ \% & $3.65~(\pm14.30)$ \%\\
Mean $\Delta_{C-T}$ for the strong AGN$^2$ &  $3.81~(\pm6.70)$ \% & $-5.10~(\pm20.43)$ \% &  $5.66~ (\pm9.53)$ \% & $2.01~ (\pm5.82)$ \%\\
Mean $\Delta_{C-T}$ for the typical QSO$^2$ &  $-25.00~(\pm55.90)$ \%& $-18.20~(\pm41.92)$ \% &  $13.57~ (\pm33.25)$ \% & $-0.61 ~(\pm1.72)$ \%\\

\tablecomments{1. Mass and volume fractions take into account the entire galaxy (disk and gaseous halo). The mean is done on the 6 snapshots studied. 2. $\Delta_{C-T}$ is the relative change between the $\Delta _{rel}$ of  the semi-transparent line ($T$) and the original $\Delta_{rel}$ used for comparison ($C$), where $\Delta_{rel}$ is the relative reduction of SFR defined in Equation~\ref{eqn:delta}. A negative value indicates enhanced SFR suppression compared to the original LOP ; a positive value indicates that the relative reduction of SFR is smaller for the semi-transparent line. As SF-triggering cannot be probed with our method (see Footnote~5), the latter is only due to the resampling of the lines. The mean is done on the LOPs of the sample that are star-forming before the RT process. Values between parentheses correspond to the standard deviation.\\}
\end{deluxetable*}

\begin{figure*}[htp]
\center
\includegraphics[trim=0.5cm 1.4cm 0.4cm 2.8cm, clip=true,width=0.8\linewidth]{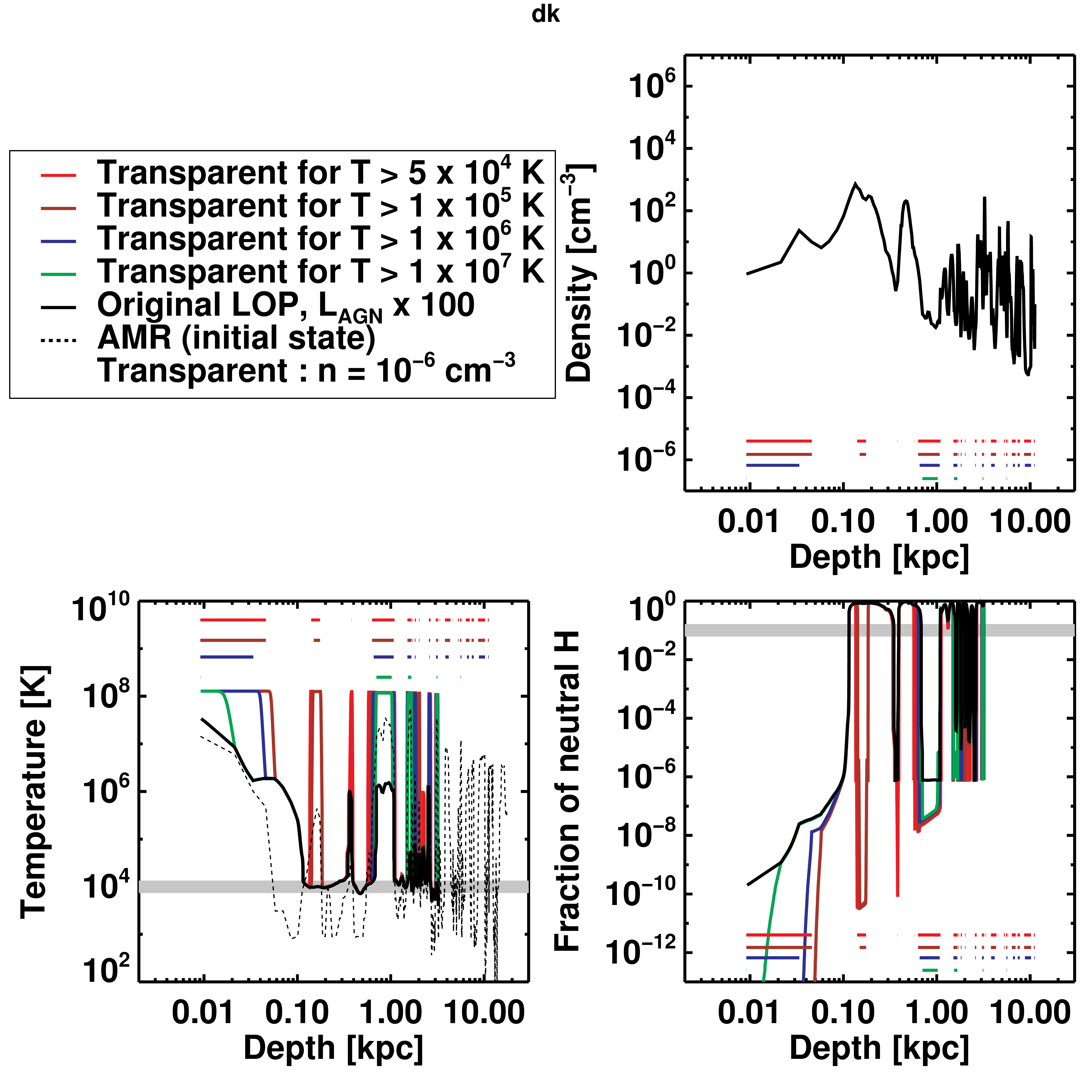}

\caption{Density (\textit{top}), temperature (\textit{left}) and neutral hydrogen fraction (\textit{right}) profiles for the original LOP (\textit{black}) and the transparent LOPs, with transparency criteria as labelled. Transparency criteria are shifted for the sake of visibility on the density panel, and reproduced on the two other panels to make the plots easier to read. The gray lines show the temperature threshold for star formation and the ionized gas demarcation used throughout the paper. This LOP is a typical LOP of the galactic disk. Accounting for gas ionized by other sources does not change the physical properties in the star-forming regions.\\} 
\label{fig:transp_lops_disk}
\end{figure*}

{As the simulation contains several heating sources (stars, UV background, thermal AGN feedback, etc. ; see Section~\ref{subSection:process}), a large volume fraction of the gas in the ISM or in the halo is already very hot before the RT process (see Table~\ref{table.hot_gas}), and is  likely to be ionized. This gas could then be transparent to the AGN radiation during the RT process, whereas in our model, all the gas is supposed to be initially neutral. Some of the AGN photons are then artificially absorbed because they interact with gas they should not encounter.} 

{To see if the AGN radiation would have a bigger impact if encountering transparent gas, we ran a series of LOPs, in which the number density is set to an arbitrarily low value ($10^{-6}$ cm$^{-3}$) in every cell whose initial temperature is above a given threshold ($5\times10^4, 10^5, 10^6 \textrm{ and } 10^7$ K). These values are set in order to probe different transparency levels, since the ionization state is not explicitly computed in the simulation: at our resolution, gas at T $> 10^6$ K is very likely to be ionized, while gas at a few $10^4$ K could be ionized, depending on its density, or due to shocks.}

{Among the series of LOPs studied, we present the results for the propagation of the QSO radiation along a typical LOP in the plane of the galactic disk (see Figure~\ref{fig:transp_lops_disk}). We compare ionization and temperature along such ``semi-transparent'' lines to those along the original density profile used in the main study.}

{Transparent regions -- which are essentially not forming stars, even before RT -- are more heated, and more ionized than their original counterparts. However, such differences may not be physically relevant, since (1) such regions are heated to $10^8$ K, a value that entirely depends on the value of the density we chose for the transparent regions, and (2) the fraction of neutral H is very low ($\lesssim 10^{-4}$) in both cases. In the semi-transparent configuration, a few more ionized spots exist at intermediate radius in the disk compared to the typical LOP, but dense (``opaque'') clumps do not seem to be more affected, whatever the AGN luminosity. Furthermore, most star-forming regions are located beyond the range of ionization computed with Cloudy. As accounting for transparent regions does not allow radiation to go further in the disk compared to the original LOPs (because it does not increase the radius at which the temperature drops under 4,000 K), taking transparent gas into account has no major effect on star formation. }

{Quantitatively, LOPs that are not star-forming before the RT process stay non-star-forming after the RT process for all transparency criteria (i.e. $SFR_f=SFR_i=0$). For LOPs that are star-forming before the RT process, we compute the relative change of the SFR reduction $\Delta_{rel}$ (see Equation~\ref{eqn:delta}), defined as:}
\begin{equation}
\Delta_{C-T}=\frac{\Delta_{rel,C}-\Delta_{rel,T}}{\Delta_{rel,C}},
\end{equation}•
  {where $\Delta_{rel,T}$ is the relative reduction of SFR for the different transparency criteria and $\Delta_{rel,C}$ is that of the original profile. $\Delta_{C-T}$ can be seen as an error bar on the relative reduction of SFR shown in Figure~\ref{fig:sfr_ratio}. The values for the different AGN regimes are gathered in Table \ref{table.hot_gas}. The standard deviations are large due to the small sample, and there is a high discrepancy between enhancement of the SFR reduction ($\Delta_{C-T}>0$) and attenuation of the SFR reduction (which is only due to resampling errors since we cannot probe SF-triggering\footnote{During the RT process, the temperature either increases, or remains constant, and the density profile is constant as well (see Section~\ref{subSection:process}). Thus, a small decrease of the SFR can only be due to the resampling of density and temperature along the LOP.}; $\Delta_{C-T}<0$) for the different transparency criteria. From this, the change in the SFR reduction induced by transparent gas is consistent with zero at all AGN luminosities. Furthermore, a variation of 25~\% at most of the relative reduction $\Delta_{rel}$ (shown in Figure~\ref{fig:sfr_ratio}) is marginal. We therefore neglect all ionizing sources other than the AGN in the main study.\\}

\section{Conclusions}
\label{Section:ccl}

In order to determine the effect of AGN long-range photoionization and local thermal energy re-deposition on star formation, we used Cloudy to propagate AGN radiation from the BH located at the center of a simulated high-redshift disk galaxy through the ISM and the gaseous halo. We built a model of a Seyfert 1 SED {spanning a large range of wavelengths} and used it as ionizing source in Cloudy. Radiative transfer was computed through the whole galaxy using lines-of-propagation (LOPs) emerging from the central BH in all directions with a good spatial coverage. We tested three different AGN luminosities, corresponding to three ``AGN regimes'': typical AGN ($L_{bol}=10^{44.5} $~erg~s$^{-1}$), strong AGN ($L_{bol}=10^{45.5} $~erg~s$^{-1}$), and QSO ($L_{bol}=10^{46.5} $~erg~s$^{-1}$) and studied how AGN photoionization heats/ionizes the different gas phases in the galactic disk and halo, and how that temperature increase might impact the SFR of the whole galaxy. Our results are as follows :
 \begin{itemize}
 
  \item {The AGN mainly affects the diffuse phase of the ISM and the gaseous halo. Indeed, the AGN is able to ionize a large volume fraction of the gas in the galactic disk -- 5 to 40 \% depending on the AGN regime, and in the gaseous halo -- 30 to 90 \%. However, the corresponding mass fraction remains very low (0.1 to 3~\%), showing that most of the ionized and/or heated gas is diffuse, while GMCs are left unaffected.   }\\
 
 \item {As the bulk of star-forming gas occurs in dense clumps, the decrease of the total SFR due to local AGN heating and distant AGN radiative effects  is marginal, even if there is an increasing trend with luminosity. The maximum relative reduction of SFR is of a few percent in the most diffuse cases at a QSO luminosity.}\\
 
 \item Gas distribution in the simulated high-redshift disk plays a major role on the propagation of AGN radiation: dense star-forming clumps shield themselves and the diffuse material behind them against {AGN X-ray to UV ionizing radiation and optical heating radiation}, while holes allow it to propagate further in the disk or in the halo. {In the QSO regime, diffuse regions of gas in the interclump medium are heated and/or ionized up to $\sim8$ kpc around the BH. Furthermore, the clumpiness of the ISM induces a high variablity of the maximal radius at which the AGN is able to heat/ionize gas in the disk, which is not predicted by smooth distributions of gas.}\\
 
  \item {Atomic gas around the GMCs and in their envelopes -- thought to be the reservoir of future star-formation -- is not strongly impacted by a typical or strong AGN, whether through winds, or photoionization.} Taking into account a standard AGN duty cycle, we show that the cumulative effects of AGN feedback on star formation are small {on a timescale of a few hundreds of million years}. Thus, in a standard configuration, not only is AGN radiation unable to substantially affect the dense molecular regions that dominate instantaneous star formation, but it is also inefficient at destroying cool atomic reservoirs ($\sim$ 10 cm$^{-3}$) for the future sites of star formation.\\
  
 \end{itemize}

Our detailed calculations suggest that the coupling between AGN radiation and the star-forming ISM is very weak, at least in the most typical SFGs at high-redshift. {We showed that well-resolved dense star-forming clumps shield themselves against AGN UV radiation -- which has been hypothesized by e.g. \citet{Vogelsberger2013,Vogelsberger2014c} in their simulation Illustris at lower resolution, but also against X-ray emission ; while diffuse gas is affected by the AGN at large scale in the gaseous halo, and even at large scale inside the galactic disk for the QSO regime.}

This study, in association with the results of \citet{Gabor2014}, supports the idea that frequent AGNs in high-redshift SFGs are a promising mechanism to regulate or remove the mass of galaxies, without impacting star formation even on relatively long timescales of hundreds of Myrs, i.e. preserving the steady-state evolution with relatively constant star formation histories.

{Finally, due to the high variability of the ionization radius induced by the well-resolved ISM, we expect that simple Str\"{o}mgren spheres do not correctly model the impact of AGN ionizing radiation on the pysical state of the gas. We recommend a more complex subgrid model to treat RT without explicitly implementing it in the simulations, including at least non-smoothed gas density (spherical symmetry is not valid in a clumpy ISM), gas column density and distance from the AGN ; and AGN luminosity, inferred from the BH accretion rate.}\\

\acknowledgments

We acknowledge support from the EC through grants ERC-StG-257720 and the CosmoComp ITN. Simulations were performed at TGCC and IDRIS under GENCI allocations 2013-GEN2192 and 2014-GEN2192. We thank the anonymous referee for valuable comments, which improved the content and clarity of this paper. 

\appendix

{The appendices gather more detailed explanations about our study and describe some secondary tests that we did to check the consistency of our model. They are organized as follows: a detailed description of the AGN spectrum and its components is available in Appendix \ref{appendix:spectra}; maps of ionized/heated gas and $\rho_{SFR}$ for three other snapshots are displayed in Appendix \ref{appendix:maps}; we also present a study of the value of the density threshold for star formation (see Appendix \ref{subSection:threshold}) and of the filling factor (see Appendix \ref{subSection:fill_fact}). Finally, we compare the results of our analysis to a lower-resolution simulation (see Appendix \ref{subSection:resolution}).\\}

\section{Seyfert SED}
\label{appendix:spectra}

  \begin{figure}[htp]
\epsscale{1.1} 
\plottwo{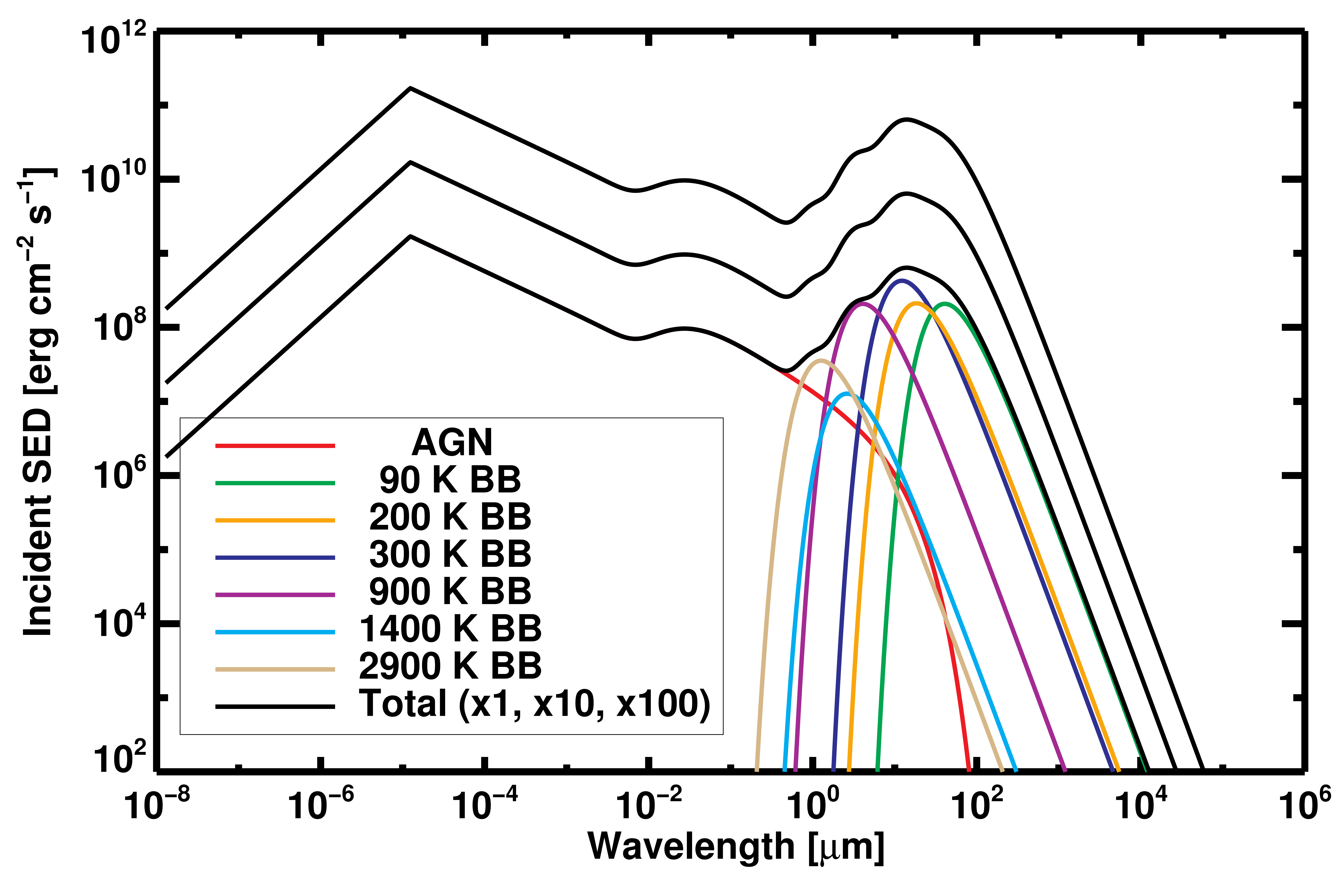}{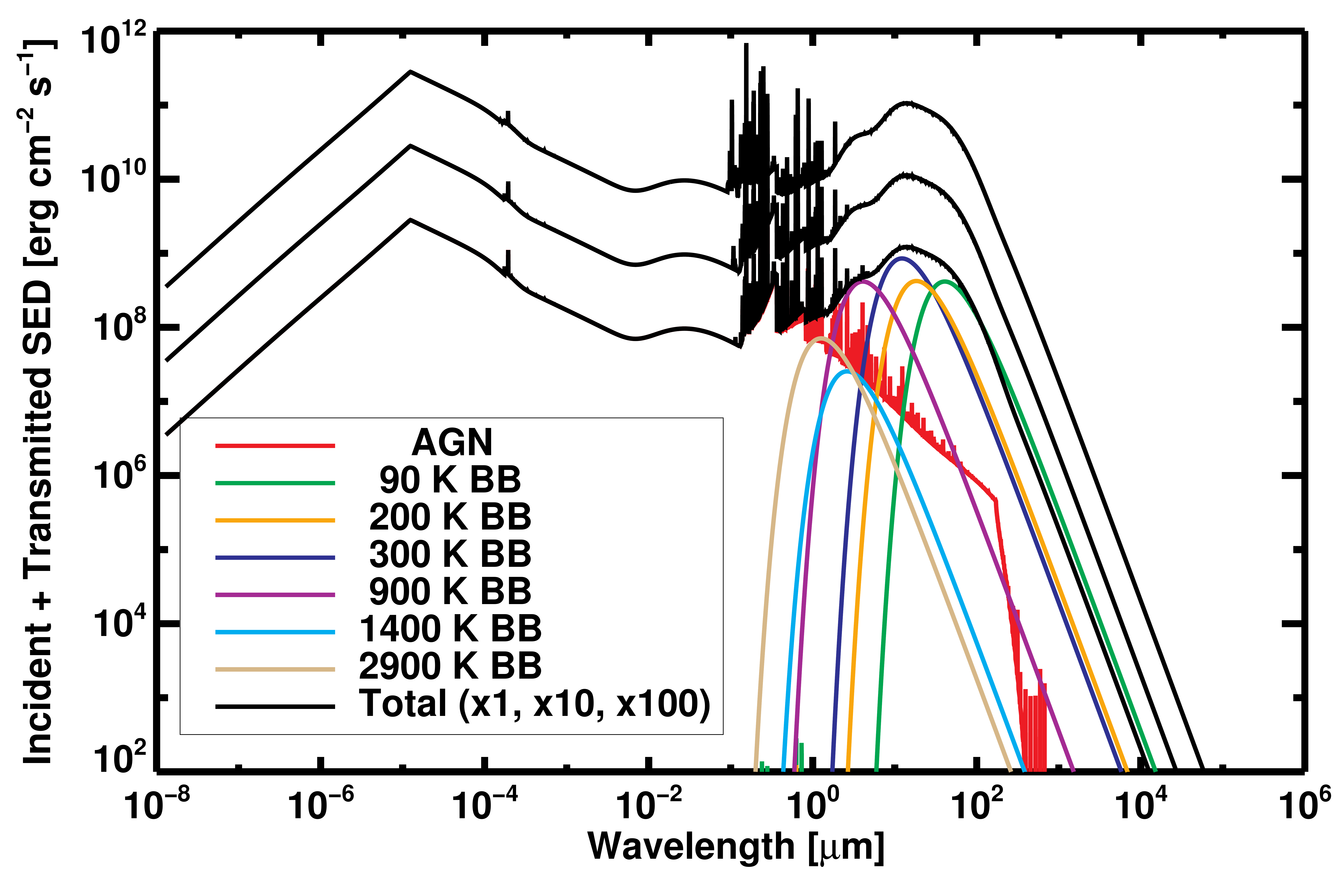} 
\caption{Left: Incident SED of the inner region of a Sy 1 galaxy, i.e. the sum (black) of the AGN (red) and the dust (blackbodies; other colors) incident emissions. Right: Emergent SED of the inner region  of a Sy 1 galaxy, i.e. the sum the incident SED and the radiation transmitted through the BLR. The total SEDs for the three luminosity regimes are shown (black); the components are those of the typical AGN regime.\\} 
\label{fig:components}
\end{figure}

The input SED can be interpreted in the framework of the Unified Model of AGN, according to which the supermassive black hole is surrounded by an accretion disk, which thickens to become a torus approximately at the sublimation radius of dust. Our AGN SED is composed of (1) UV radiation from the accretion disk, and X-rays from the accretion disk's corona, (2) a linear combination of blackbodies corresponding to the heated dusty torus, and (3) illuminated BLR clumps.  These different components are defined as follows: 
\begin{itemize}
\item  The AGN component was built using the $agn$ command in Cloudy, defining a multi-component continuum similar to that observed in typical AGNs: 
 \begin{equation}
F_\nu = \nu^{\alpha_{\textsc{UV}}}e^{-\frac{h\nu}{kT_{\textsc{BB}}}}e^{-\frac{kT_{\textsc{IR}}}{h\nu}} + A\nu^{\alpha_\textsc{X}}
    \end{equation}
where $T_{BB}$ is the cut-off temperature of the Blue Bump, so that it peaks at 10 microns and $T_{IR}$ is the IR cut-off temperature. $\alpha_{UV}$ and $\alpha_{X}$ are the exponents in the UV and in the X fields, respectively.  Coefficient A was adjusted so that the optical to X-ray spectral index $\alpha_{OX}$ had the specified value, defined by:
 \begin{equation}
\frac{F_\nu(2\textrm{ keV})}{F_\nu(2500\textrm{ \AA})} = \left(\frac{\nu_{2\textrm{ keV}}}{\nu_{2500\textrm{ \AA}}}\right)^{\alpha_{OX}}
   \end{equation}
 We used the following parameters: T $ = 5.5\times10^5$ K, $\alpha_{OX}=-0.7$, $\alpha_{UV}=-0.3$, $\alpha_{X}=-0.48$. These were initially based on the parameters taken by \citet{Ferguson1997} but then modified to match observed data. \\

\item Dust was assumed to have solar abundance and the ISM to be Milky-Way-like. {Cloudy computes all processes relevant for the treatment of dust grains: photons absorption and scattering, stochastic heating, photoelectric effect, Auger emissions in X-ray environments, electron and ion collisional charging, collisions between gas and dust and resulting energy exchange, etc. As the radiation field is propagated, gas and dust opacities affect photons, while photons change the properties of grains after absorption.} Dust sublimation and grain depletion were taken into account {\citep[see Section 2.5 of][and online Cloudy documentation Hazy for further details]{Ferland2013}}. Often, dust emission is described as a greybody (see \citet{Rathborne2010} for instance). Because greybodies cannot be implemented in Cloudy, we used a linear combination of blackbodies with temperatures ranging from 90 to 2,900 K. However, heated dust does not contribute to the ionization of gas and thus the accuracy on the IR part is not crucial for this study. The inner and outer radii of the computation were defined so that radiative transfer occurred through the dust sublimation zone --  roughly 0.1 pc when the temperature is about 1,400 K \citep{Honig2010}. \\

\item The transmitted spectrum of the BLR was computed by propagating the incident AGN SED with Cloudy, through clumps of gas representing a typical BLR, i.e. homogeneous clumps of gas with a hydrogen density of $10^9$ cm$^{-3}$ \citep{Mattews1985} and a filling factor of $10^{-3}$ \citep{AGN3}.
\end{itemize}

These components are shown in Figure~\ref{fig:components}, for the typical AGN regime. The strong AGN and QSO regime SEDs are also shown for comparison, but not their components. The input scripts that led to this Seyfert SED and the QSO SED (not shown) are available as online-only material.\\

\section{Maps for other snapshots}
\label{appendix:maps}

This section gathers the maps for three of the other snapshots studied in the simulation including AGN feedback, in order to illustrate how the distribution of clumps affects the propagation of AGN radiation. The density threshold for SF is 10 cm$^{-3}$. As the distribution of clouds evolves with time and the black hole moves \citep[see][]{Gabor2013b}, the snapshots look similar but are not exactly identical. 

The different configurations clearly modify the propagation of the ionizing radiation :
 \begin{itemize}   
 \item Figure~\ref{fig:den_temp_sfr_10Hcc_00075w_app} shows the maps corresponding to snapshot \#1, which is representative of the evolution of the galaxy. In this snapshot, the BH is located slightly above the galactic disk: the typical AGN radiation going downward is mainly blocked, and the lower ionization cone is small. However, in the two other regimes, AGN radiation is energetic enough to go through the center of the galactic disk. 
\item Figure~\ref{fig:den_temp_sfr_10Hcc_00210_app} displays the maps for snapshot \#6, that where the BH is embedded into a dense clump (n $>$ 10$^4$ cm$^{-3}$). This configuration is not frequent because the number of dense clumps in the disk is not large, though it boosts the accretion rate of the BH. In this snapshot, the typical AGN luminosity regime is too weak to ionize the gas. In the strong AGN and QSO regimes, the fraction of photons that escape the central clump increases and an ionization cone with a very small basis appears. The impact on SF is the lowest among all snapshots studied and the changes $\rho_{SFR}$ are below the resolution limit. 
\item Finally, Figure~\ref{fig:den_temp_sfr_10Hcc_00150_app} shows the maps of snapshot \#4, where a dense clump lies on the above edge of the BH. In all luminosity regimes, AGN radiation is blocked in the upper half of the simulation box, similarly to snapshot \#6, whereas the lower part behaves like snapshots \#1 and \#2.\\\\
 
\end{itemize}

\begin{figure*}
\includegraphics[trim=1.0cm 0cm 3.1cm 0cm, clip=true,width=\textwidth]{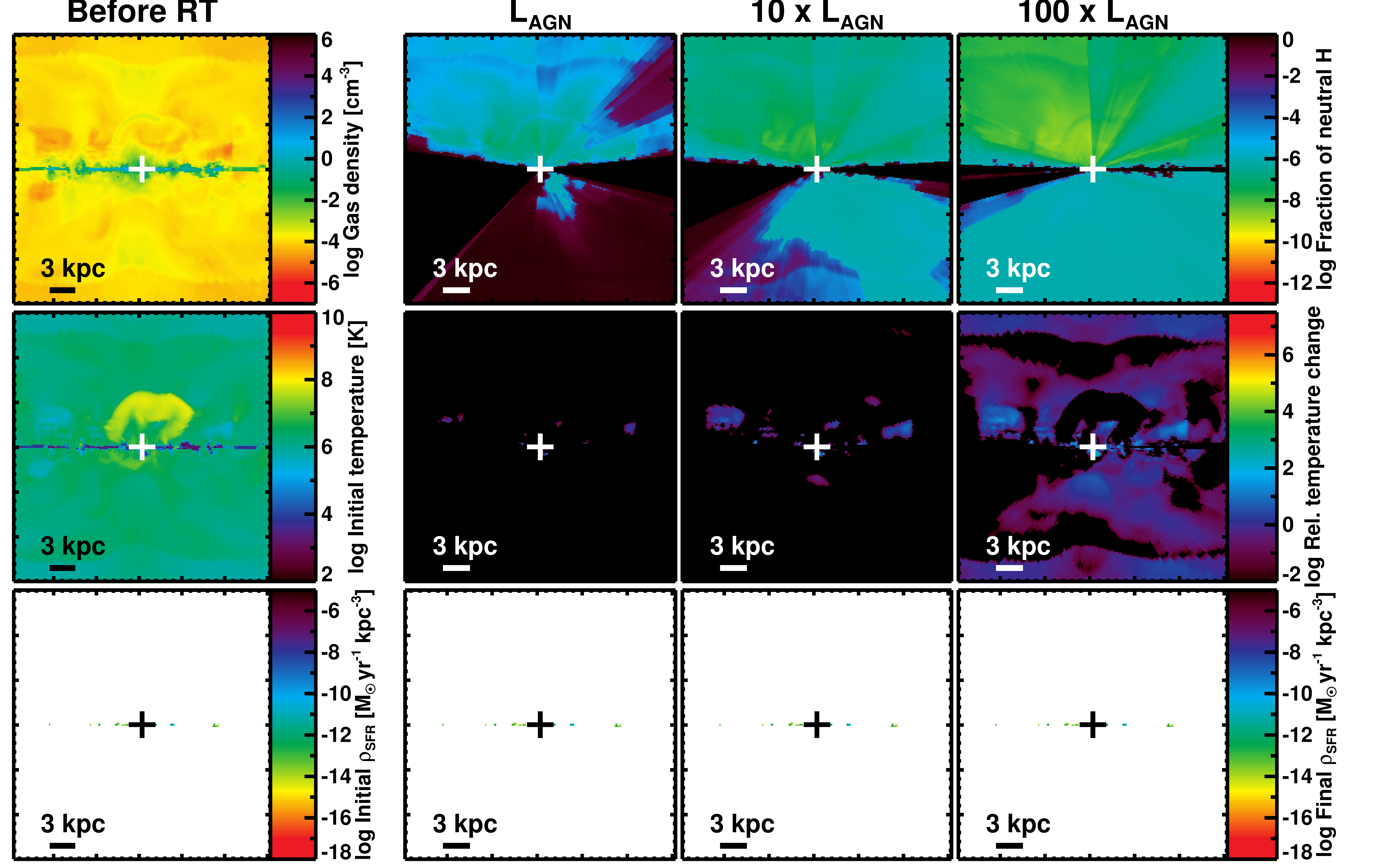}
\includegraphics[trim=1.0cm 0cm 3.1cm 0cm, clip=true,width=\textwidth]{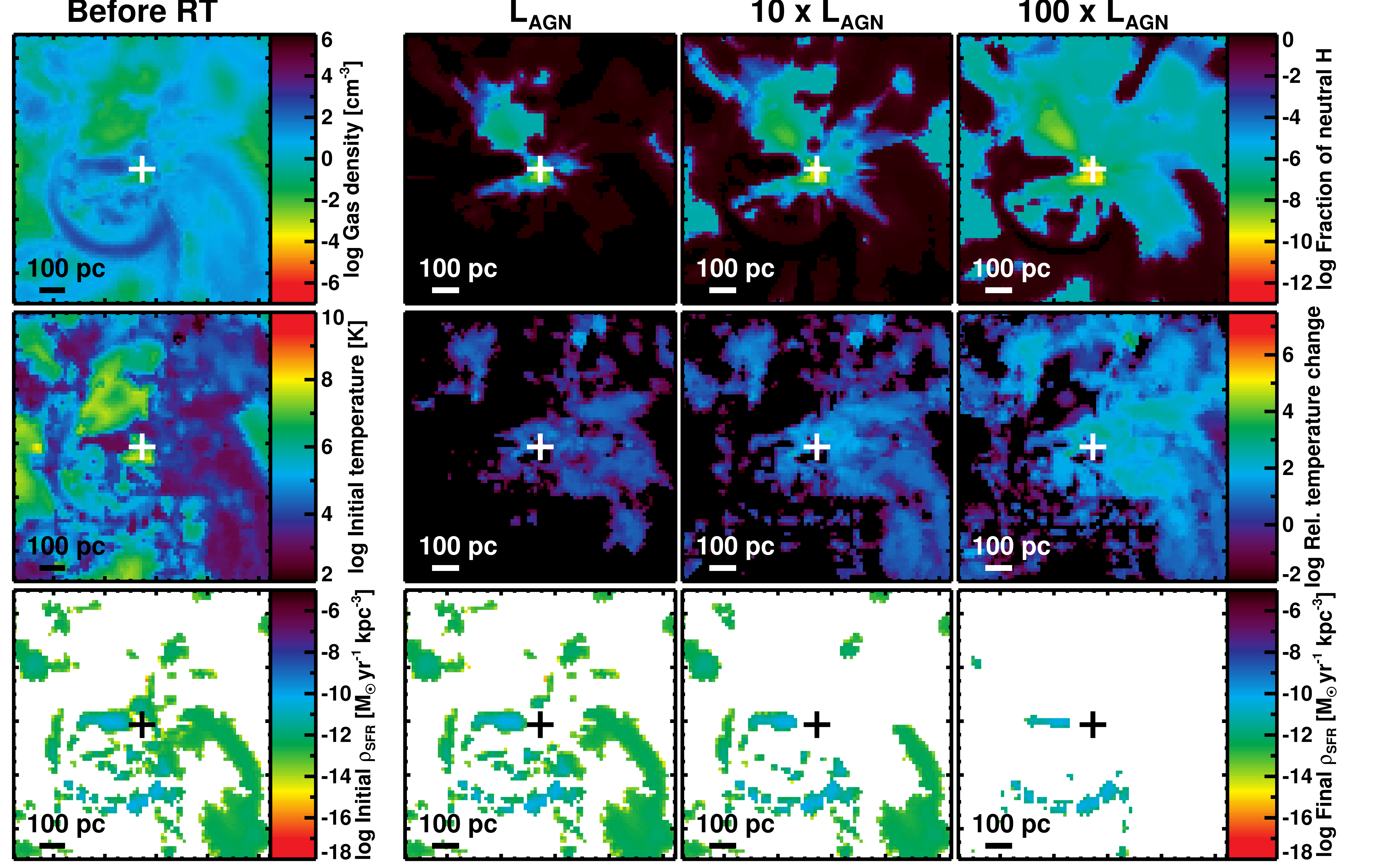}
\caption{Large edge-on and zoomed face-on views of the simulated galaxy (snapshot \#1).  \textit{Top row}: Hydrogen density, Fraction of neutral hydrogen after RT. \textit{Middle row}: Temperature before RT, Relative temperature change. \textit{Bottom row}: $\rho_{SFR}$ before RT,  $\rho_{SFR}$ after RT. The `+' sign shows the location of the black hole and the density threshold for SF is 10 cm$^{-3}$. Parameters after RT are given for the three AGN luminosities (L$_{\textrm{AGN}}=10^{44.5}$~erg~s$^{-1}$).  } 
\label{fig:den_temp_sfr_10Hcc_00075w_app}
\end{figure*}

\begin{figure*}
\includegraphics[trim=1.0cm 0cm 3.1cm 0cm, clip=true,width=\textwidth]{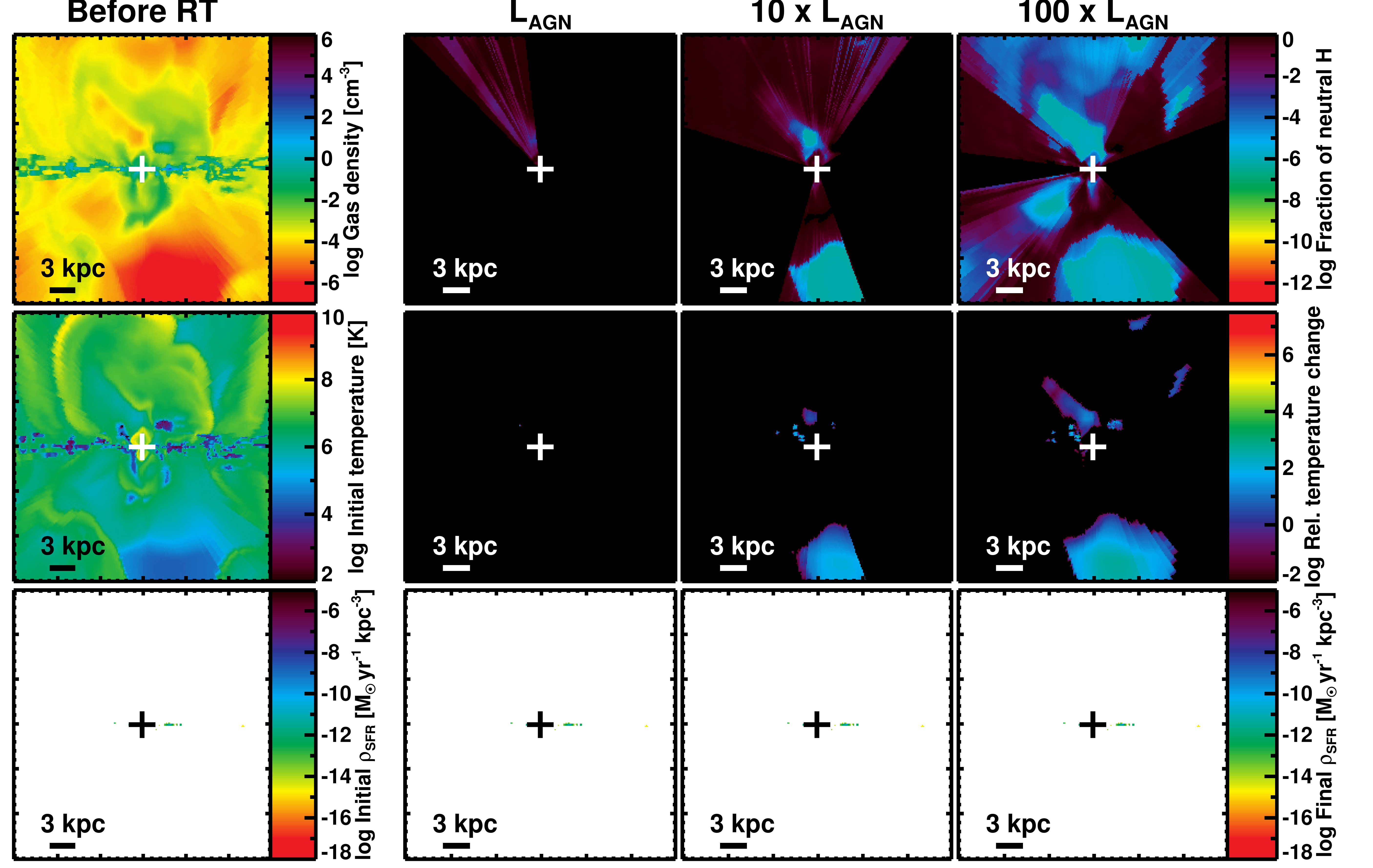}
\includegraphics[trim=1.0cm 0cm 3.1cm 0cm, clip=true,width=\textwidth]{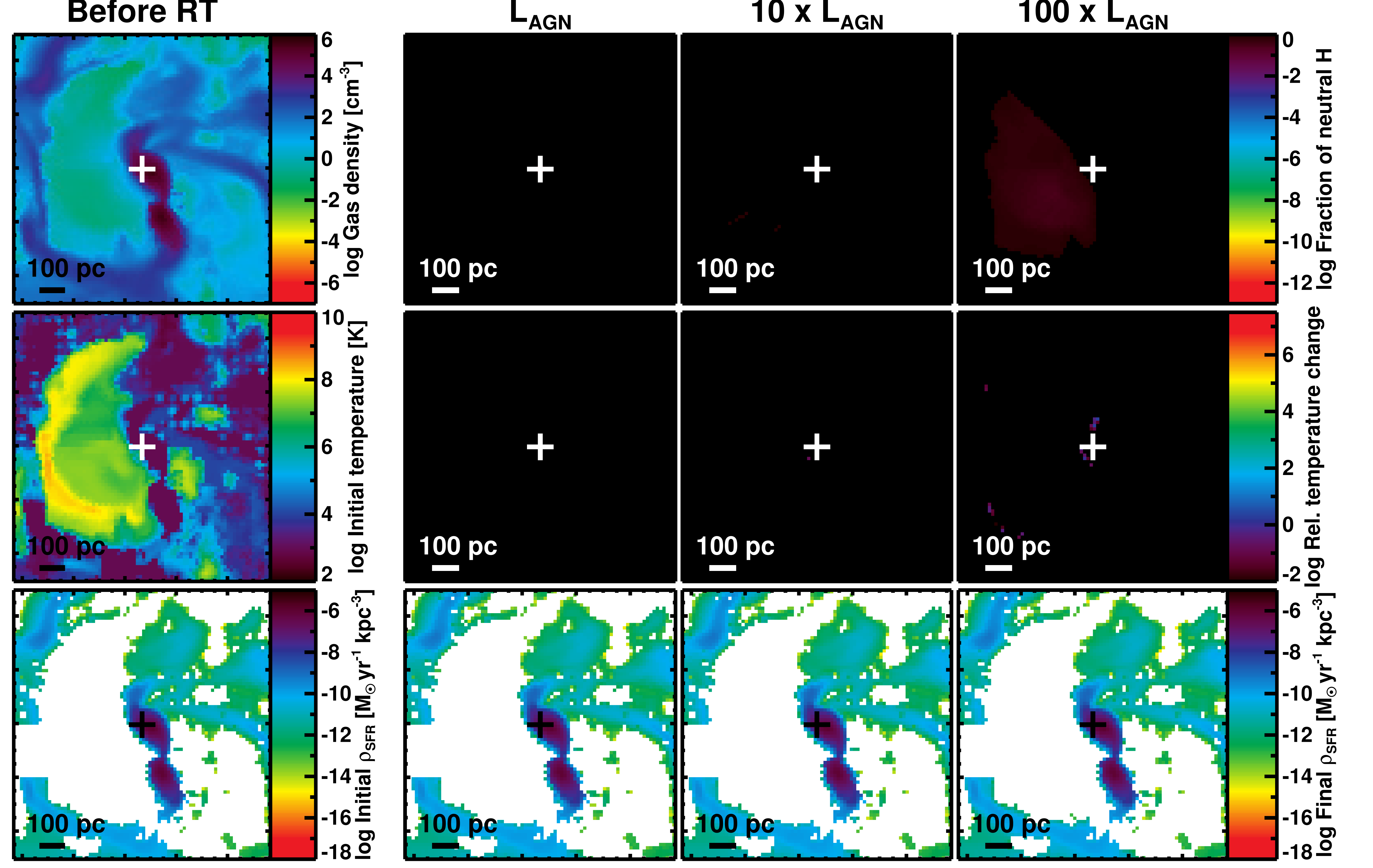}
\caption{Large edge-on and zoomed face-on views of the simulated galaxy (snapshot \#6).   \textit{Top row}: Hydrogen density, Fraction of neutral hydrogen after RT. \textit{Middle row}: Temperature before RT, Relative temperature change. \textit{Bottom row}: $\rho_{SFR}$ before RT,  $\rho_{SFR}$ after RT. The `+' sign shows the location of the black hole and the density threshold for SF is 10 cm$^{-3}$. Parameters after RT are given for the three AGN luminosities (L$_{\textrm{AGN}}=10^{44.5}$~erg~s$^{-1}$).  } 
\label{fig:den_temp_sfr_10Hcc_00210_app}
\end{figure*}

\begin{figure*}
\includegraphics[trim=1.0cm 0cm 3.1cm 0cm, clip=true,width=\textwidth]{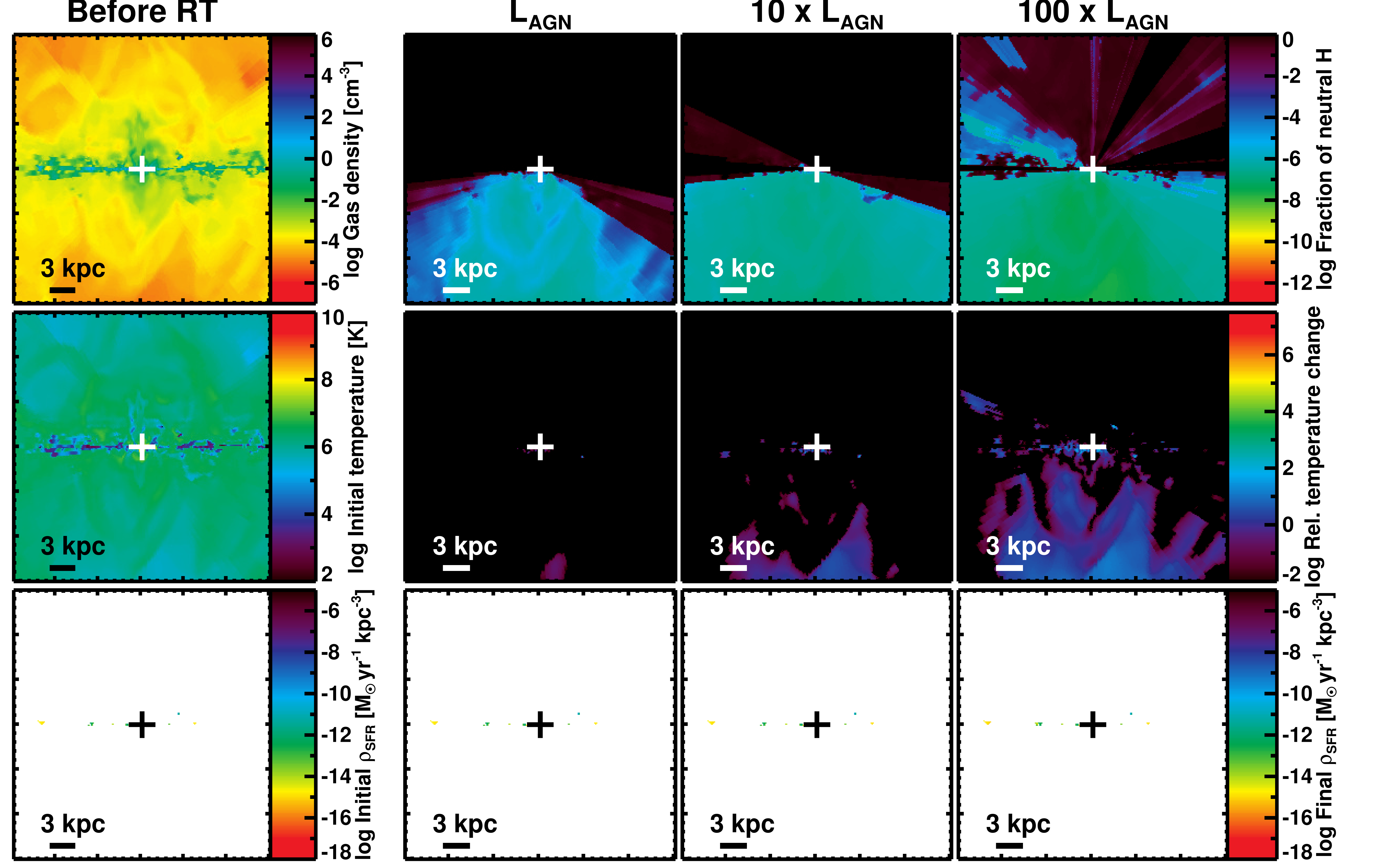}
\includegraphics[trim=1.0cm 0cm 3.1cm 0cm, clip=true,width=\textwidth]{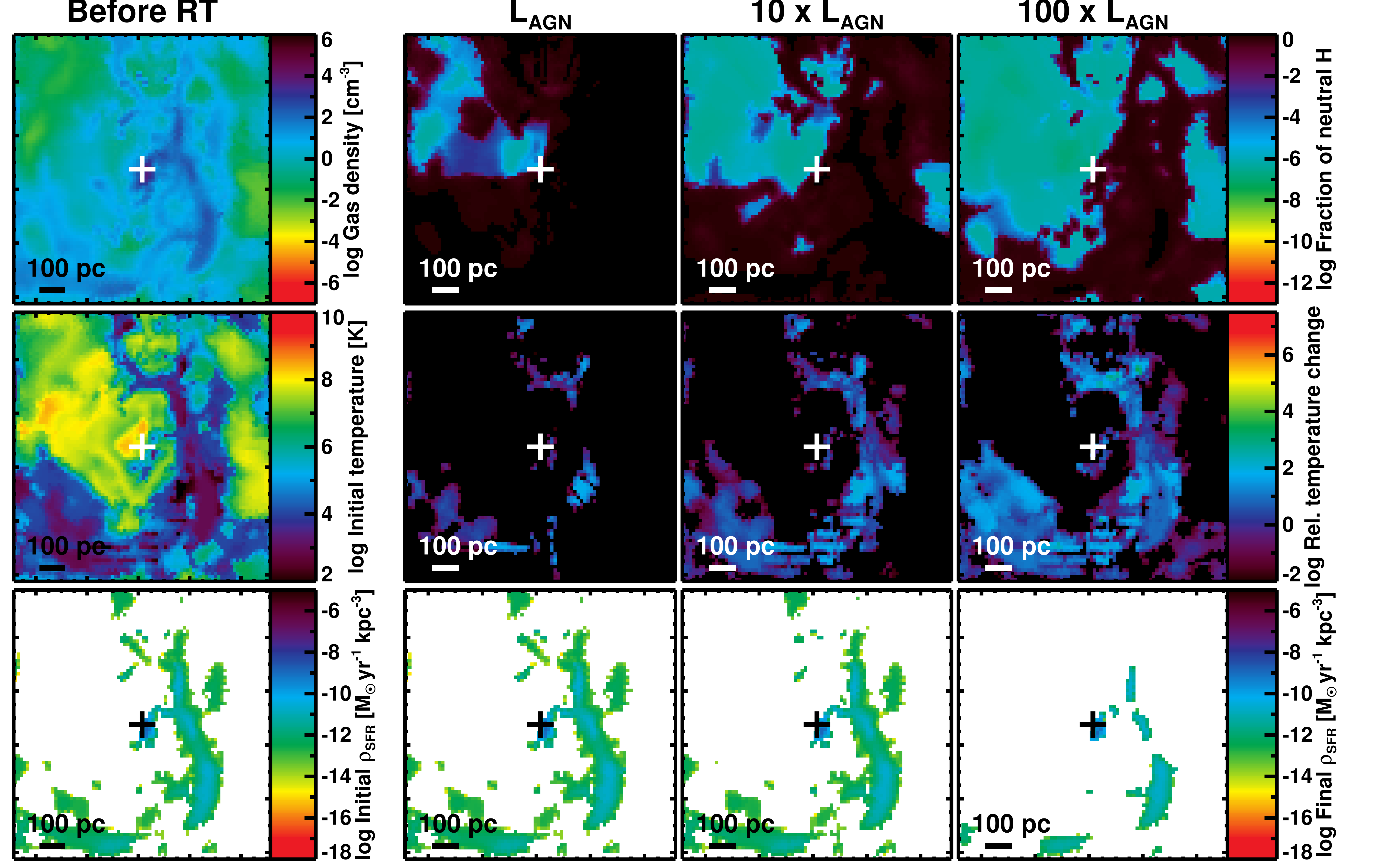}
\caption{Large edge-on and zoomed face-on views of the simulated galaxy (snapshot \#4).   \textit{Top row}: Hydrogen density, Fraction of neutral hydrogen after RT. \textit{Middle row}: Temperature before RT, Relative temperature change. \textit{Bottom row}: $\rho_{SFR}$ before RT,  $\rho_{SFR}$ after RT. The `+' sign shows the location of the black hole and the density threshold for SF is 10 cm$^{-3}$. Parameters after RT are given for the three AGN luminosities (L$_{\textrm{AGN}}=10^{44.5}$~erg~s$^{-1}$). } 
\label{fig:den_temp_sfr_10Hcc_00150_app}
\end{figure*}

\section{Density threshold for star formation}
\label{subSection:threshold}

\begin{figure*}
\includegraphics[trim=1.0cm 69.8cm 3.1cm 0cm, clip=true,width=\textwidth]{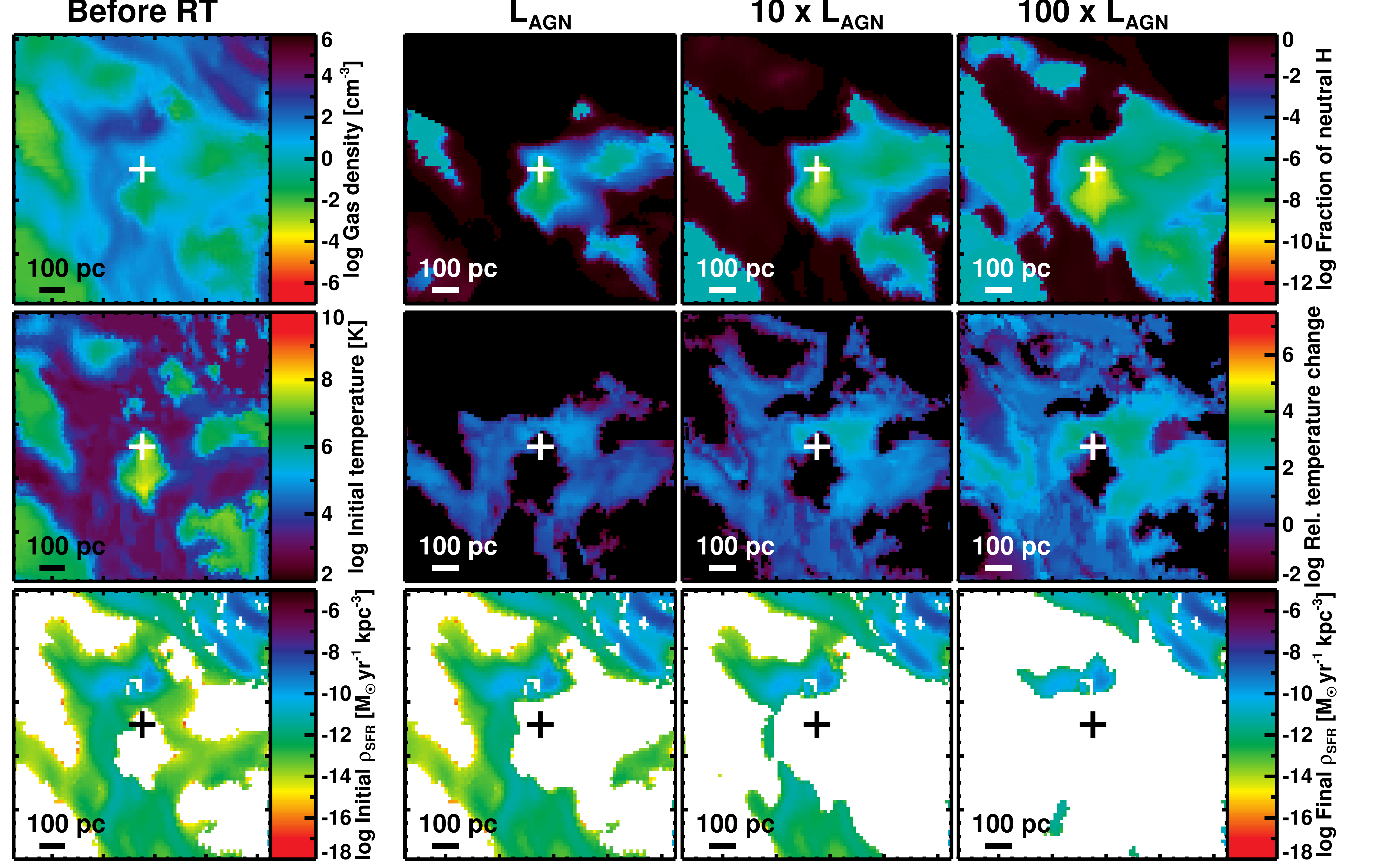}
\includegraphics[trim=1.0cm 0cm 3.1cm 49.0cm, clip=true,width=\textwidth]{./rel_temp_change_all_den_temp_sfr_1Hcc_00100dk_zoom=82_comp_interp-eps-converted-to.pdf}
\includegraphics[trim=1.0cm 0cm 3.1cm 49.1cm, clip=true,width=\textwidth]{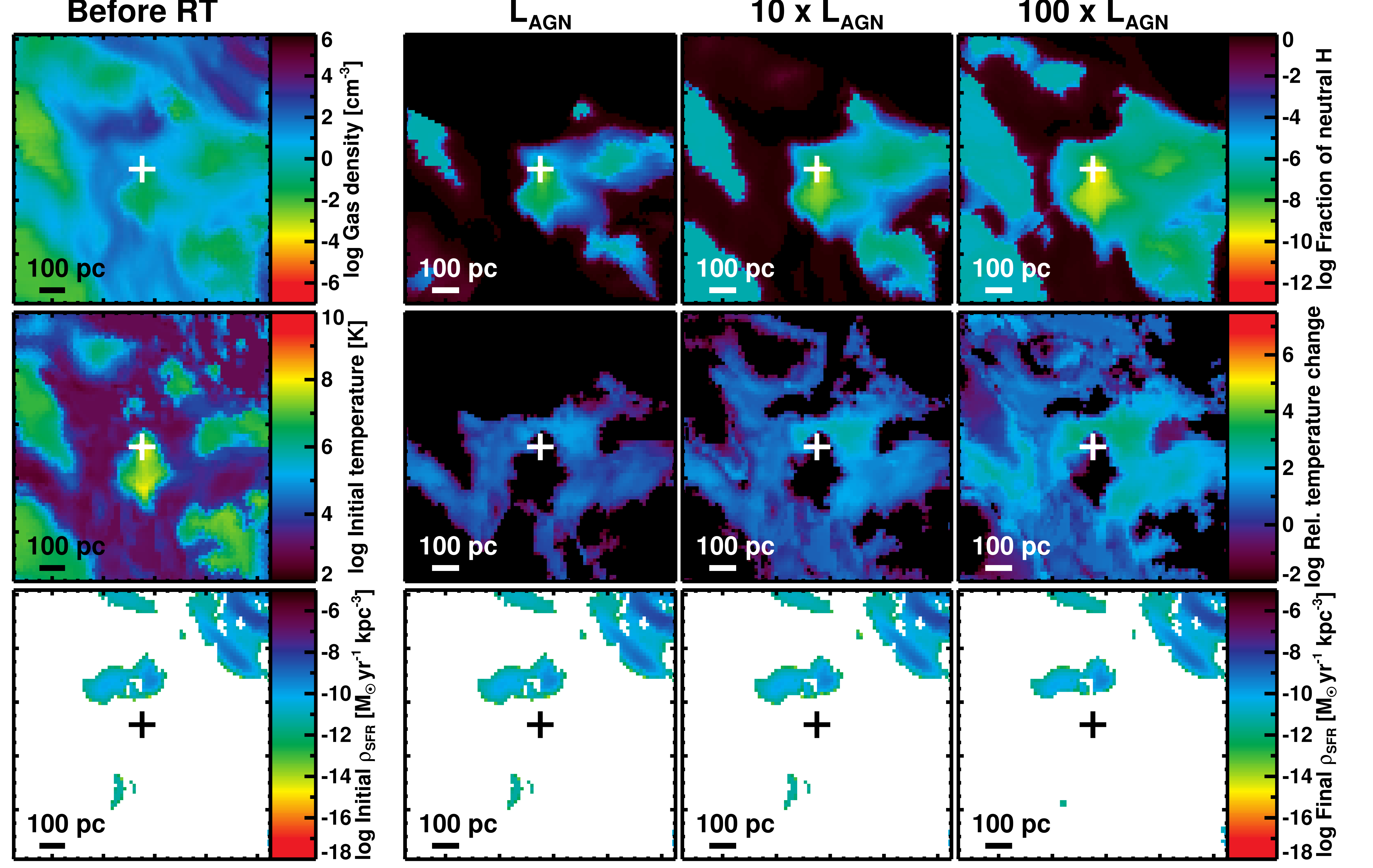}
\caption{SFR density maps before RT (\textit{left column}) and after RT (\textit{right three columns}) for the three AGN luminosities. Top line shows a density threshold for star formation of 1 cm$^{-3}$, bottom line of 100 cm$^{-3}$. L$_{\textrm{AGN}}=10^{44.5}$~erg~s$^{-1}$. With a lower threshold, star-forming regions are more extended but the extra contribution to the total SFR is small (see Figure~\ref{fig:sfr1-100Hcc}). } 
\label{fig:threshold}
\end{figure*}

The density threshold above which a cell is able to form stars is 100 cm$^{-3}$ in the simulation but the value used for the analysis is 10 cm$^{-3}$. Such a change does not affect the efficiency of star formation, but plays the role of a delimiter between star-forming and non-star-forming regions \citep{Kraljic2014}.  In the post-processing, we tested three values of this threshold: 1, 10 and 100 cm$^{-3}$. With a low threshold, the star-forming regions are much more extended than with a higher threshold (see Figure~\ref{fig:threshold}, for 1 and 100 cm$^{-3}$), though the additional star-forming clouds are so diffuse that the total SFR of the entire galaxy remains comparable (see Figure~\ref{fig:sfr1-100Hcc}). For QSO luminosities (see right column of Figure~\ref{fig:sfr1-100Hcc}), diffuse regions are ionized and the spatial extent of the remaining star-forming regions in the very center of the galaxy hardly depends on the value of the threshold we varied. 

\begin{SCfigure*} 
\epsscale{0.9} 

\plotone{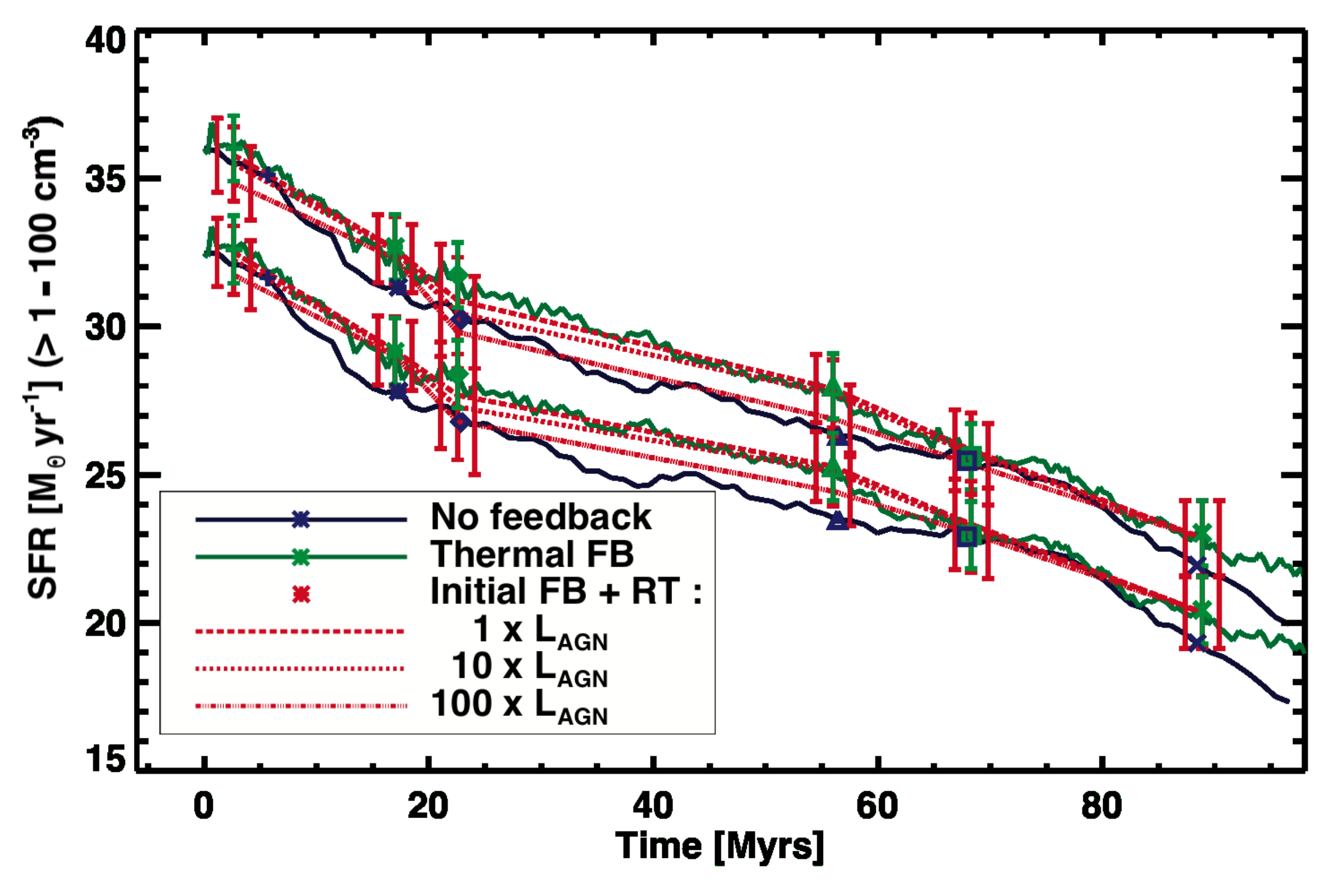}
\caption{Star Formation Rate as a function of time. Same key as Figure~\ref{fig:sfr}. This figure differs from Figure~\ref{fig:sfr} in that we test different SF thresholds: the lower group of curves corresponds to a density threshold for star formation of 100 cm$^{-3}$, and the upper one to a threshold of 1 cm$^{-3}$. The star-forming regions that are cut with a higher threshold are minor contributors to the total SFR along time ($<$ 7 \%).\\} 
\label{fig:sfr1-100Hcc}
\end{SCfigure*}

This weak dependency is also expected from Figure~\ref{fig:pdf}, showing the fraction of initial SFR per density bin: most of the SFR lies in the densest cells, and cutting at 1,~10~or~100 cm$^{-3}$ does not influence the total SFR calculation much. The shape of this SFR-weighted gas density PDF, dominated by high densities, is typical for simulations of standard SFGs \citep{Teyssier2010}. Hence, whatever the value of the threshold (1,~10~or~100 cm$^{-3}$, as long as it is low enough not to include the bulk of the star-forming phase, see Figure~\ref{fig:pdf}), the fraction of missed SFR is very small, and the behaviour is identical. \\

\begin{figure}[htp]
 \begin{minipage} [b]{.65\linewidth}
\includegraphics[trim=2cm 0cm 0cm 0cm, clip=true,width=\linewidth]{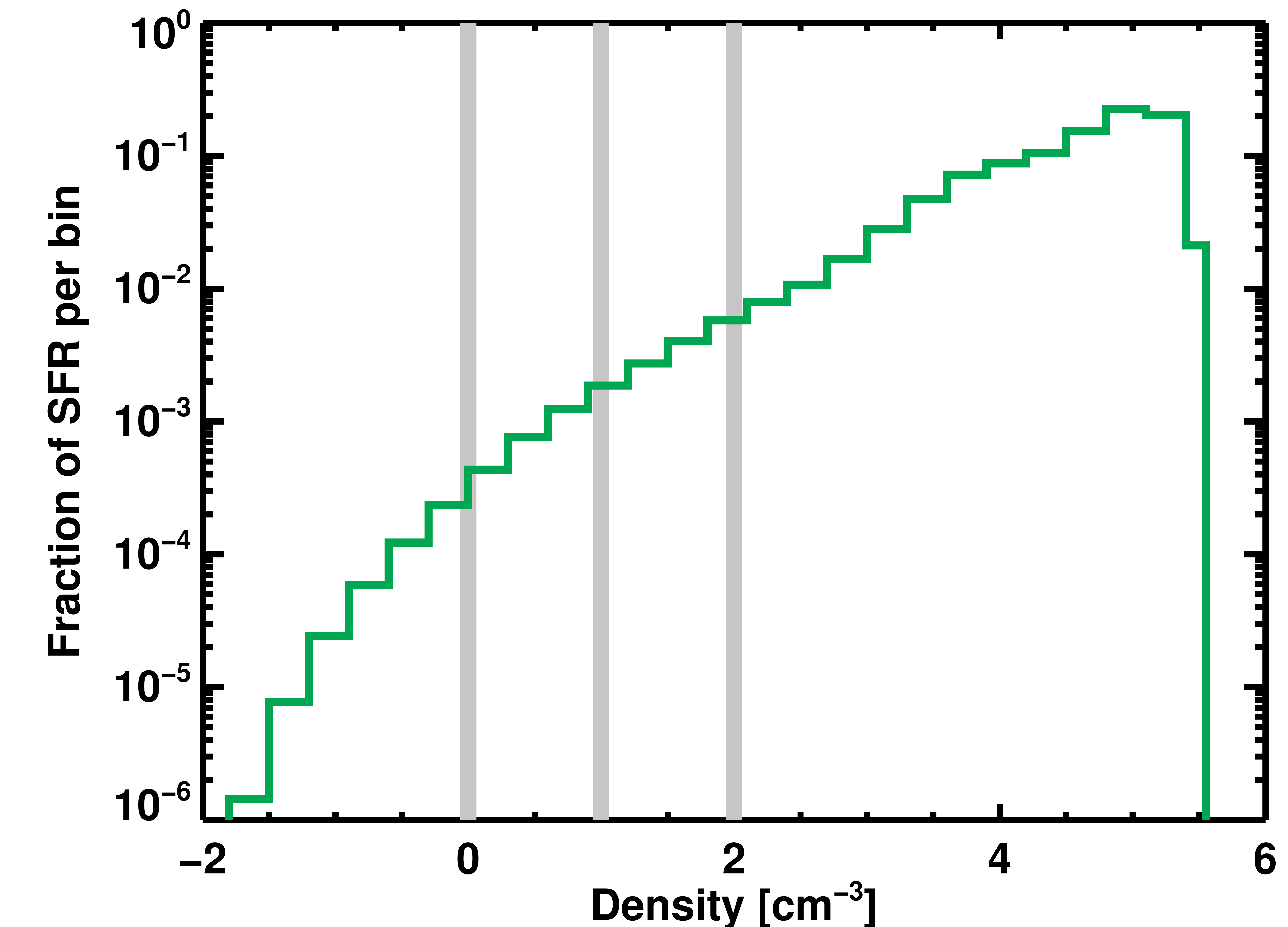}
\caption{Fraction of SFR per density bin before RT. The vertical gray lines are the three varied density thresholds for star formation. The major contributors to the total SFR have densities above $10^3$ cm$^{-3}$ and to cut at 1, 10 or 100 cm$^{-3}$ makes no significant changes.\\\\} 
\label{fig:pdf}
\hfill
\end{minipage}
\hfill
 \begin{minipage} [b]{.25\linewidth}
\includegraphics[trim=4cm 4cm 7.4cm 3.5cm, clip=true,width=\linewidth]{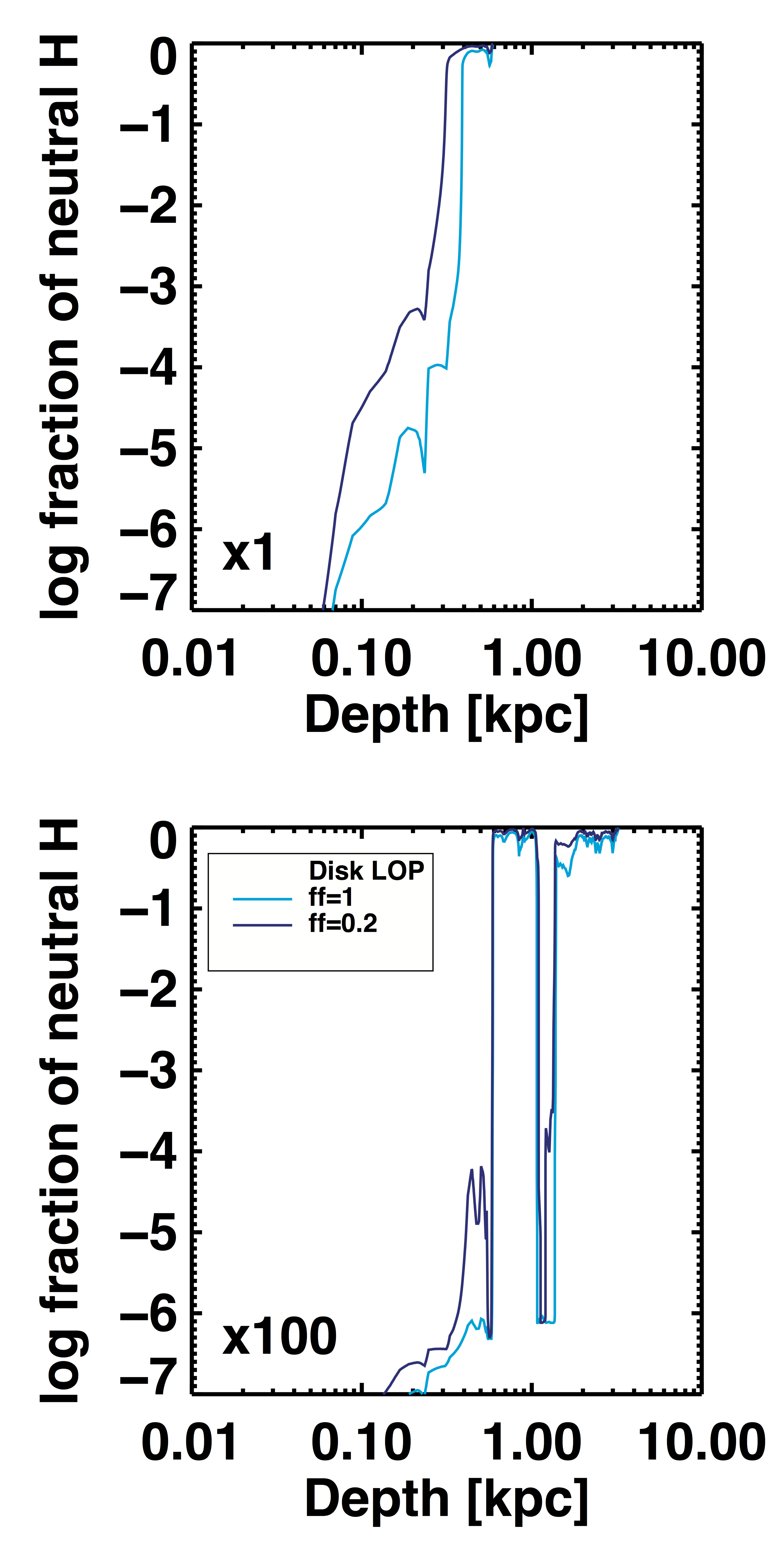}
\caption{Fraction of neutral hydrogen for a typical LOP in the plane fo the disk (as shown in Figure~\ref{fig:lop_example}) with $ff=1$ (\textit{light blue}) and $ff=0.2$ (\textit{dark blue}). Same panels as Figure~\ref{fig:unif}. Changing the filling factor does not affect dense star-forming regions but allows diffuse regions to be ionized to a greater extent.\\} 
\label{fig:comp}

\end{minipage}
\end{figure}

\section{Role of the filling factor}
\label{subSection:fill_fact}

{In the Cloudy calculation, the filling factor is set to account for the clumpiness of the gas in the sphere based on the 1-D profile. When considering a sphere containing a set of gas condensations separated by empty regions, the filling factor $ff$ corresponds to the fraction of volume occupied by the condensations \citep{Osterbrock1959}. The specified density is that in the condensations. The mass of the galaxy is given by:
\begin{equation}
M = \sum_i \rho_i^{sim}\times\textrm{V}_i=\sum_i \left(ff_i\times\rho_i^{cloudy}\right)\times\textrm{V}_i,
\label{eqn:mass}
\end{equation}
where M is the total mass of the galaxy and $ff_i$, $\rho_i^{sim}$ and $V_i$ are respectively the filling factor, original mass density and volume of cell $i$ and $\rho_i^{cloudy}$ is the density specified for Cloudy runs.
To conserve the mass of the galaxy, the density in each cell is divided by the filling factor before the Cloudy computation (see Equation \ref{eqn:mass}). All densities output by Cloudy are multiplied back by the filling factor before performing the next step of the analysis. In the main study, the value of the filling factor is set to 0.2, which is small enough to account for the clumpiness of the ISM, but still in agreement with the mass refinement criterion (i.e. multiplying the density by 5 will not change the level of refinement of the cell).}

\subsection{Constant density profile}

{To study the role of the filling factor, we use the constant density profiles presented in Section \ref{subSection:other} and compare three values of the filling factor (1, 0.2 and $10^{-3}$). We also study the difference between profiles with the original density (conservation of density) versus density divided by the filling factor (conservation of mass):}
\begin{itemize}
\item At constant density, the smaller the filling factor, the further the ionization gets into the disk, favoring the hypothesis that the holes between the clumps are necessary for AGN radiation to propagate past the central kiloparsecs. 

\item At constant mass, the smaller the filling factor, the closer the ionization is stopped in the disk. This can easily be understood by considering the column density of the encountered gas. Indeed, the first cloud encountered is denser and its recombination time is shorter if the filling factor is smaller. The critical column density needed to stop the AGN radiation is thus reached closer to the BH when decreasing the filling factor, even if the condensations are less numerous and the fraction of volume occupied is smaller. 
\end{itemize}

{Both behaviours can also be explained from the theoretical point of view. From \citet{AGN3}, the number of recombination Q for hydrogen is proportional to the proton density $n_p$, the electron density $n_e$ and the filling factor $ff$ :
\begin{equation}
\textrm{Q} \propto ff \cdot n_p \cdot n_e.
\end{equation}
In the conserved density case, the number of recombinations only depends on the filling factor and thus decreases accordingly. This is consistent with the ionization front being located further in the disk. In the conserved mass case, since the densities are divided by $ff$, Q goes with the inverse of the filling factor and is larger for a smaller filling factor. }

\subsection{Typical LOP}

{We also study the effect of the filling factor on a typical profile of a LOP in the disk plane (see Figure~\ref{fig:comp} for $ff = 1$ and 0.2 ; the low filling factor ($10^{-3}$) was not studied because it would conflict with the mass refinement criterion). At all AGN luminosities, the diffuse regions of the LOP ($n<10^1$ cm$^{-3}$ regions located less than 500 pc away from the BH and the ionization spots defined in Section~\ref{discussion:cones}) are ionized to a greater extent when $ff$ is larger, whereas the denser parts show no significant differences. This is easily understood since, in diffuse regions, the density is smaller when $ff$ is larger (since mass is conserved) and the path length needed to reach the column density that stops AGN radiation is larger. Thus AGN ionization goes further in the disk and gas is ionized to a greater degree. In contrast, the density in dense regions is large enough to shield the gas, whatever the filling factor. Therefore, we do not expect the effect of AGN radiation on SF to depend on the filling factor.}

\section{Effect of resolution}
\label{subSection:resolution}

{In our simulations, the GMCs are resolved, which means that some internal structure is visible, and that they  are not at the numerical limit. Furthermore, the SFR of such a simulated galaxy is converged with resolution since the comparison of simulations at $\sim1$ pc resolution \citep{Bournaud2010} and 0.05 pc resolution \citep{Renaud2013} shows no difference in the internal pc-scale substructures of GMCs. Thus, with a higher resolution, we would not get smaller and denser clumps separated by more numerous holes, and the PSD of the ISM would remain the same.}

{However, as suggested by the study of the filling factor, degrading the resolution could induce significant changes by reducing the number of holes between the clumps. We ran the simulation with AGN feedback from $t=0$ to 88 Myrs with a $\sim$ 12 pc resolution (one level of refinement less than the runs used in the main study) and chose a few snapshots, close in time to those of the high-resolution run. The simulation is run over the same period of time as the high resolution one in order to have comparable gas consumptions. However, one has to wait for the GMCs to be disrupted and re-grown at lower resolution, which is why we only study 4 snapshots, close in time to the last 4 snapshots of the high resolution run. The star-forming clumps are thus consistently grown at both resolutions. The PSDs of both high and low resolution snapshots are alike, which means that the intensity of the density fluctuations does not change with resolution. Thus, each pair of high/low resolution snapshots has a similar, though not identical, distribution of gas, and the series of snapshots have to be compared on average.}

\begin{figure}
\center
\includegraphics[trim=0cm 0cm 0cm 0cm, clip=true,width=0.5\linewidth]{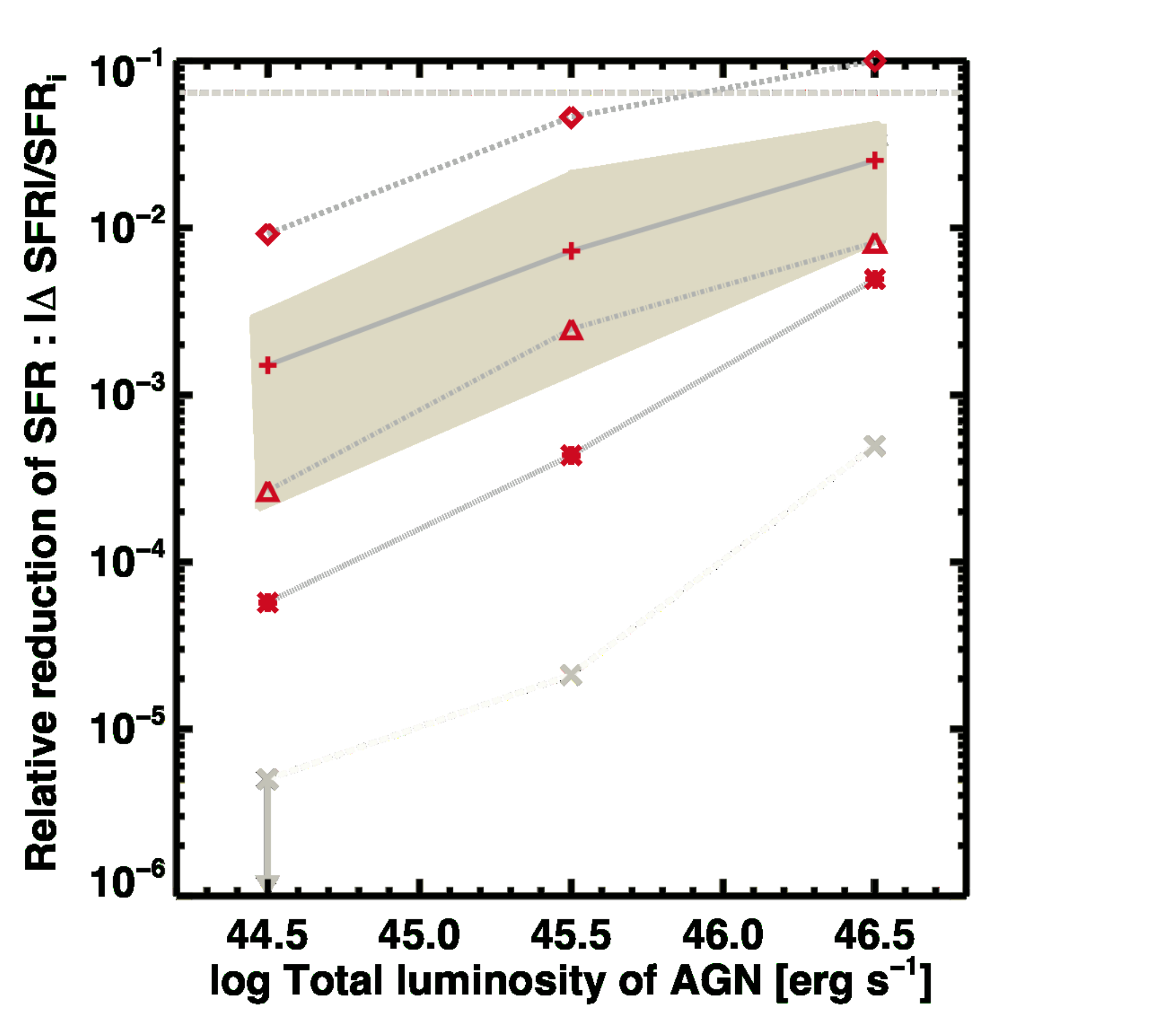}
\caption{Relative reduction of the Star Formation Rate after radiative transfer as a function of AGN luminosity. Each symbol represents a low resolution snapshot. They are linked for clarity. The beige shaded area and curves recall the data for the high resolution snapshots shown in Figure \ref{fig:sfr_ratio}. The median of the relative reduction of the SFR at each luminosity regime is the same as for the high resolution run.\\}
\label{fig:lores}
\end{figure}

{The low resolution simulation lacks very dense clumps ($n\sim 10^6$~cm$^{-3}$) compared to the high resolution run and, as the mass of the galaxy is the same, clouds are more extended. Dense clumps at $n\sim10^{4-5}$~cm$^{-3}$, as that located on the BH in snapshot \#6 of the high resolution run, are present in both high and low resolution runs since the resolution is high enough to reach such densities. However, such configurations are rare, which is why we do not have such a configuration among the low resolution snapshots. }

{The SFR is lower since there are fewer very dense star-forming clumps. However, the relative reduction of the SFR can be compared in both high and low resolution simulations and that of the low resolution run is shown in Figure \ref{fig:lores}. Even though the dispersion is larger for these particular low resolution snapshots compared to the high resolution ones, the median is the same as that of the high resolution snapshots, making both results compatible. This tends to show that degrading the resolution of a factor 2 has no impact on the analysis.}\\

\bibliography{library,pycloudy}

\end{document}